# The Future of Intense Ultrafast Lasers in the U.S.

BLI

Brightest
Light
Initiative

## WORKSHOP REPORT

March 27–29 2019 • OSA Headquarters, Washington, D.C.

SPONSORED BY

U.S. DEPARTMENT OF ENERGY | Office of Science

NSF

NNSA National Nuclear Security Administration

NATIONAL PHOTONICS INITIATIVE

# Brightest Light Initiative Workshop Report

**Chair: Roger Falcone (UC Berkeley)**

**Scientific Research Needs (SRNs) Co-chairs:**

Felicie Albert (LLNL)

Farhat Beg (UC San Diego)

Siegfried Glenzer (SLAC)

**Technical Research Needs (TRNs) Co-chairs:**

Todd Ditmire (UT Austin)

Tom Spinka (LLNL)

Jonathan Zuegel (Univ. Rochester)

This study is based on work supported by the Department of Energy Office of Science under Grant No. DE-SC0019748, by the National Science Foundation under Grant No. PHY-1920984, by the Department of Energy National Nuclear Security Administration under Cooperative Agreement DE-NA0003856, and by the National Photonics Initiative.

# Table of Contents

# Chapter 1 – Executive Summary and Recommendations

This *Brightest Light Initiative (BLI) Workshop Report* captures the important research ideas and recommendations for enabling that work developed by over 100 leading scientists at a community-initiated workshop held March 27-29, 2019 in Washington, DC. Workshop attendees developed an understanding of key opportunities, as well as gaps in current technologies and capabilities, for science enabled by the highest-intensity lasers.

A previous report, *Opportunities in Intense Ultrafast Lasers*: *Reaching for the Brightest Light*, was released by the National Academies of Sciences, Engineering, and Medicine (NASEM) in 2018. It noted important scientific opportunities and applications; a large but fragmented community in pertinent disciplines; recent loss of U.S. leadership; and needs for new programs and facilities. This *BLI Workshop Report* builds on that NASEM study by identifying common themes articulated by the relevant communities, with a goal of informing stakeholders – researchers, funding agency program officers, and government representatives with budgetary responsibility – about the exciting possibilities offered by an expansion of current investments in ultra-intense-laser science, as well as the re-establishment of U.S. leadership.

Since the first demonstration of a laser (the acronym for "light amplification by stimulated emission of radiation") 60 years ago, scientists recognized the unique capability of lasers to provide and utilize high-power light. The physics behind this capability resides in the fundamental properties of a laser, which include stimulated emission and coherence. These properties, first described by Albert Einstein over 100 years ago – before any actual device was envisioned or demonstrated - allow energy stored in a material to be extracted on rapid time scales (limited only by the speed of light) and concentrated to small spatial dimensions (at the length scale of a wavelength of light). A critically important invention in this field – called Chirped Pulse Amplification – was awarded the 2018 Nobel Prize in Physics. The research work that led to this invention was supported by U.S. funding agencies in the 1980s. However, as noted in the NASEM Report, regional European and Asian investments in laser development and laser science research coordination have recently overshadowed related efforts in the U.S., resulting in relative loss of U.S. technological leadership and scientific capabilities.

BLI Workshop participants discussed compelling science and research ideas, which are detailed in the Report and resulted in seven high-level recommendations (not prioritized):

**1. Establish a national, laser research and development program that is broad and diverse; fosters collaborations among universities, national labs, and industry; capitalizes on the strengths of each of these sectors; and has sufficient resources to meet goals.** For example, exawatt-class lasers, petawatt lasers with high repetition rate, and compact and robust short-pulse lasers do not currently exist but would broadly enable new science and applications detailed in this Report. On-going realization of this novel laser technology – including new amplification methods, improved optics, pulse shaping techniques, etc. – requires R&D.

**2. Develop and implement state-of-the-art lasers at the Linac Coherent Light Source (LCLS) free-electron x-ray laser user facility.** Coupling high-power and high-energy lasers with novel x-ray light sources is critical for enabling research and increased understanding in many important fields, such as materials under extreme conditions, plasma physics, high-energy-density science, and laboratory astrophysics. This capability would uniquely enable such leading research and ensure the competitiveness of U.S. facilities for these fields.

**3. Develop ultrahigh-intensity laser technology and build an open-access laser user facility in the U.S. with multiple beamlines at 10-to-100 PW peak powers.** Concepts exist for going beyond state-of-the-art, high-peak-power, short-pulse lasers, and ensuring international leadership in this technology. Such lasers will enable laser-based science to the highest peak powers, and meet scientific challenges in areas of fundamental importance that explore extreme electromagnetic fields, temperature, and pressure.

**4. Develop short-pulse, high-peak power lasers with very high-average power (kilowatts and beyond).** Such advanced lasers would enable key applications in science and technology, industry, and national security, especially those that utilize high-flux secondary sources such as short-wavelength light sources, high-energy particles, and other radiation sources. Such high repetition rate lasers would allow important active feedback for control, as well as high data rates, and are well suited for applying machine learning techniques and statistical and big-data analytical approaches for science and for applications such as quantum materials, energy, biology, and nano-microscopy.

**5. Investigate opportunities to unite traditionally separate scientific fields such as high-energy physics and intense laser physics.** Collocating high-intensity lasers with other scientific infrastructure, such as facilities with relativistic particle beams or other energetic drivers, will enable forefront science in areas such as non-linear quantum electrodynamics, nuclear, plasma and high energy density physics, and astrophysics**.**

**6. Expand the scope and capabilities of LaserNetUS**. This consortium of the high-power laser facilities has recently been launched and is impactful for users. Support for expansion and upgrades of this network, operations and development of scientific programs with grand challenge goals within LaserNetUS, and guidance by a working group (such as recommended below), will foster the important research described in this Report, meet user demand for leadership facilities, and train early-career scientists.

**7. Create a multi-agency/program laser R&D working group, potentially under the National Science and Technology Council and involving all stakeholders.** Research associated with intense lasers is currently supported by several federal agencies that have partially overlapping missions that range from fundamental science to technology development. Greater coordination will facilitate interdisciplinary collaborations, identify evolving scientific opportunities and technological gaps, and result in funding strategies that best support the advanced research and development goals of the agencies and community.

# Chapter 2 – Introduction

## 2.1 Overview of 2018 NAS Report

In early 2018 the National Academies of Science, Engineering, and Medicine (NASEM) issued a consensus report entitled "Opportunities in Intense Ultrafast Lasers: Reaching for the Brightest Light" (www.nap.edu/24939). This report was commissioned by several Department of Defense and Department of Energy agencies out of shared concerns that European and Asian countries were constructing major multi-petawatt laser facilities, but no agency in the U.S. had any plans to do so. The technology for petawatt lasers was developed in the U.S., but the vast majority of these high-intensity laser systems are now overseas.

The agency sponsors requested a survey of high-intensity laser science and related technology, focusing on frontier science opportunities in high-intensity science; the impact of applications associated with high-intensity science; and the status of high-powered laser technology in the United States. They also asked several questions:

- How does high-intensity-science R&D in the U.S. compare to international efforts?
- Is there a national stewardship strategy? If not, what roadmap should the United States follow?
- Is there a case for a large-scale initiative to go well beyond the state of the art?
- Is there a case for forefront U.S. multi-petawatt facilities?
- What parameters or capabilities should be included?

The committee assembled by NASEM was composed of Academy members and other experts in the relevant areas of science, technology, and industry. They drew on published material, visits to high-powered laser facilities in the U.S. and Europe, and discussions with leaders of the laser and applications communities from around the world, with both in-person meetings and biweekly video conferences held over approximately one year. Committee members composed over 35 "white papers" summarizing their findings and then distilled these into a report that was refined through a critical review process directed by the Academy.

The consensus study found that high-intensity laser research has great value for our nation. Petawatt-class lasers can be focused to high intensities to create, diagnose and control extreme plasma conditions found nowhere else on earth. The forces in the plasma can accelerate particles to multi-GeV energies or heat matter to the temperatures and pressures found in the centers of stars. When combined with relativistic electron beams, these laser fields can even exceed the threshold for breakdown of the quantum vacuum. They also deliver applications beyond scientific discovery, in medicine, industry, and the stewardship of the nuclear weapons stockpile. Furthermore, the study concluded that the U.S. has lost its previous dominance in high-intensity lasers and is falling behind in every aspect associated with these unique tools, as reflected in Figure 1.1. The U.S. community is large but fragmented across programs in DOE, NSF, and the defense agencies, and no cross-agency stewardship exists. Collocation with existing infrastructure is essential to realize future opportunities, and cooperation among university, national laboratory, and industry stakeholders is necessary to retain and renew the talent base.

2009 2019

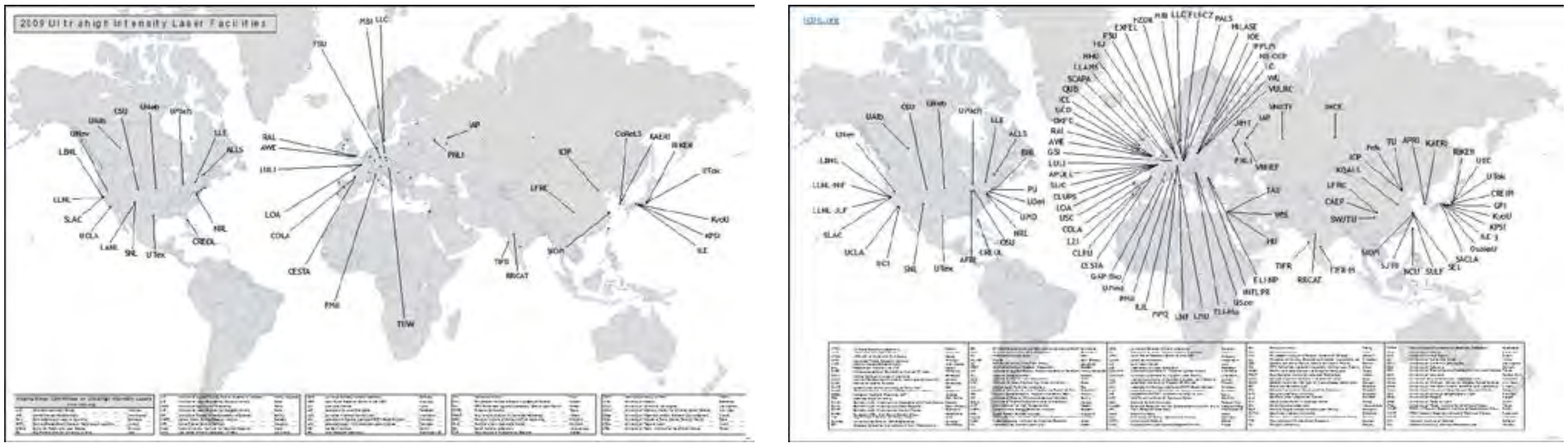

Figure 1.1 – High-peak-power laser research and applications have grown significantly in the last decade; Europe and Asia now dominate a field originally developed in the U.S. (https://www.icuil.org/activities/laser-labs.html)

The study made five major recommendations to the federal sponsors:

(1) develop a National Stewardship Strategy for high-intensity laser science, led by DOE;
(2) create a national coordinated laser facilities network including universities, industry, and national laboratories;
(3) engage the community to define facilities that best serve research needs;
(4) chart a path toward one or more new major petawatt facilities that can meet these needs; and
(5) create broader programs to enable scientists to pursue the best research in these areas.

## 2.2 Community Organization

Conclusion 3 of the 2018 NAS report noted, "*the (US intense laser science) community is large but fragmented. There is a large and talented technical community already, but it is fragmented across different discipline*s." A wide range of independent activities preceding the BLI workshop evidenced the vitality of the U.S. ultra-intense laser research community yet proves this point. Numerous user group meetings, workshops, reports, and planning processes across scientific disciplines and laser facilities have been held and several others are underway, including:

- annual Omega Laser Facility Users Group [1] meetings at UR/LLE and the APS-DPP Annual Meetings since 2009,
- annual Jupiter Laser Facility (JLF) User Group meetings [2] at LLNL since 2012,
- annual High-Power Laser workshops [3] at SLAC National Accelerator Laboratory since 2013,
- 2002 *Science and Applications of Ultrafast, Ultraintense Lasers* (SAUUL) report [4],
- 2003 LaserLab Europe, formed to establish a unified distributed laser research infrastructure that now includes 33 organizations from 16 countries [5],
- 2007 NAS plasma science study – *Advancing Knowledge in the National Interest* [6],
- 2009 *Basic Research Needs for High Energy Density Laboratory Physics* [7],
- May 2010 BES Workshop on Compact Light Sources [8],
- 2011 Extreme Light Infrastructure (ELI) Whitebook [9],
- December 2012 Compact X-Ray Light Source Workshop [10],

1 Omega Laser Facility Users Group, Accessed 3 December 2019, http://www.lle.rochester.edu/about/omega_laser_users_group.php
2 Jupiter Laser Facility, Accessed 3 December 2019, https://jlf.llnl.gov.
3 7th High-Power Laser Workshop, Menlo Park, CA, 26–27 September 2019, https://conf.slac.stanford.edu/hpl-2019/.
4 T. Ditmire and L. DiMauro, "The Science and Applications of Ultrafast, Ultraintense Lasers: Opportunities in Science and Technology Using the Brightest Light Known to Man," *A Report on the SAUUL Workshop held in Washington DC, June 17–19, 2002*, Department of Energy and National Science Foundation, Washington, DC (17–19 June 2002).
5 Laserlab Europe, Accessed 4 December 2019, https://www.laserlab-europe.eu.
6 Plasma 2010 Committee, National Research Council (NRC), "Plasma Physics at High Energy Density," in *Plasma Science: Advancing Knowledge in the National Interest* (The National Academies Press, Washington, DC, 2007), Chap. 3, pp. 75–114.
7 "Basic Research Needs for High Energy Density Laboratory Physics," *Report of the Workshop on High Energy Density Laboratory Physics Research Needs*, Office of Science and National Nuclear Security Administration, U.S. Department of Energy, Washington, DC (15–18 November 2009).
8 W. A. Barletta and M. Borland, *Report of the Basic Energy Sciences Workshop on Compact Light Sorces*, U.S. Department of Energy, Office of Science, Washington, DC (11–12 May 2010).
9 G. A. Mourou *et al.*, eds. *ELI – Extreme Light Infrastructure: Science and Technology with Ultra-Intense Lasers* (THOSS Media GmbH, Berlin, Germany, 2011).
10 *Compact X-Ray Light Source Workshop Report*, U.S. Department of Energy, Washington, DC, Report PNNL-22145 (December 2012).

- January 2013 Workshop on Laser Technology for Accelerators sponsored by DOE/HEP [11],
- 2013 NAS study *Frontiers in High Energy Density Physics: The X-Games of Contemporary Science* [12],
- 2013 NAS study *Optics and Photonics: Essential Technologies for Our Nation* [13],
- February 2014 Basic Energy Sciences Workshop on the Future of Electron Scattering and Diffraction [14],
- 2016–2018 High Energy Density Science Association (HEDSA) community self-organized (CSO) workshops,
- May 2017 Workshop on Laser Technology for k-BELLA and Beyond at LBNL [15],
- January 2018 Texas Petawatt User workshop,
- December 2017 NAS study *Opportunities in Intense Ultrafast Lasers: Reaching for the Brightest Light* [16],
- August 2018 LaserNetUS 1st Annual Meeting [17],
- February 2019 U.S.–ELI Dialogue and September 2019 US-ELI Workshop,
- February 2019 EP-OPAL workshop at UR/LLE,
- May 2019 *Workshop on Opportunities, Challenges, and Best Practices for Basic Plasma Science User Facilities* at the University of Maryland College Park [18],
- a two-year effort initiated in 2018 by Department of Energy Office of Fusion Energy Sciences (DOE-OFES) to establish a long-term strategy for the field [19], and
- 2020 NAS Decadal Assessment of Plasma Science [20] that is underway.

---

11 *Workshop on Laser Technology for Accelerators*, U.S. Department of Energy, Washington, DC (23–25 January 2013).

12 Committee on High Energy Density Plasma Physics, National Research Council (NRC), *Frontiers in High Energy Density Physics: The X-Games of Contemporary Science* (The National Academies Press, Washington, DC, 2003).

13 Committee on Harnessing Light: Capitalizing on Optical Science Trends and Challenges for Future Research, National Research Council (NRC), "Optics and Photonics: Essential Technologies for Our Nation," The National Academies Press, Washington, DC (2013).

14 E. Hall *et al.*, "Future of Electron Scattering and Diffraction," *Report of the Basic Energy Sciences Workshop on the Future of Electron Scattering and Diffraction*, U.S. Department of Energy, Office of Science, Washington, DC (25–26 February 2014).

15 *Report of Workshop on Laser Technology for k-BELLA and Beyond*, Lawrence Berkeley National Laboratory, Berkeley, CA (9–11 May 2017).

16 National Academies of Sciences Engineering and Medicine, *Opportunities in Intense Ultrafast Laser: Reaching for the Brightest Light* (The National Academies Press, Washington, DC, 2018).

17 LaserNetUS 1st Annual Meeting, Lincoln, NE, 20–21 August 2019, https://www.lasernetus.org/event/lasernetus-1st-annual-meeting.

18 H. Milchberg *et al.*, "Workshop on Opportunities, Challenges, and Best Practices for Basic Plasma Science User Facilities," *Workshop Report (October 2019)*, Physics Division, National Science Foundation, University of Maryland, College Park, MD (2019).

19 M. Ambrose, "DOE Fusion Program Developing Long-Range Strategic Plan," *FYI Science Policy News from AIP*, AIP Publishing, Melville, NY, Report 132 (2018).

20 The National Academies of Sciences, Engineering and Medicine, Decadal Assessment of Plasma Science, Accessed 5 December 2019, https://sites.nationalacademies.org/BPA/BPA_188502.

The United States was the leading innovator and dominant user of high-intensity laser technology after its development in the 1990s. The 2018 Nobel Prize in Physics was awarded to scientists for research at the University of Rochester in the late 1980s and later at the University of Michigan for the development of the approach that has enabled high-intensity lasers and the first Petawatt laser exploiting this technology was demonstrated at Lawrence Livermore National Laboratory in the 1990s. Starting at the turn of the millennium, the United States and Europe followed very different tracks in developing science, applications, and technology of intense ultrafast lasers. A brief review below of these two approaches is instructive and illustrates why convening members of the diverse and broad community to organize the U.S. intense ultrafast laser community was a primary goal of the BLI workshop, as described in Sec. 2.3.

A 2002 workshop to assess the *Science and Applications of Ultrafast, Ultraintense Lasers* (SAUUL) and the resulting seminal report [4] highlighted many exciting research areas using ultrahigh intensity lasers, ranging from plasma physics and fusion energy to astrophysics to ultrafast chemistry to structural biology. The report anticipated advances possible with petawatt lasers in five research areas: (1) fusion energy, (2) compact particle accelerators, (3) ultrafast x-ray generation enabling time-resolved structural studies of solids and molecules, (4) creation of extreme states of matter to study astrophysics in the laboratory, and (5) attosecond radiation to study electron dynamics. The SAUUL report identified four central conclusions:

1. Science studied with ultrafast, ultra-intense lasers (UUL's) was one of the **fastest-growing subfields of basic and applied research**;
2. **opportunities for UUL's in many fields of interdisciplinary science had blossomed**;
3. state-of-the-art lasers that make possible these applications were much **more complex and more expensive** than previously; and
4. **a new mode of organization needs to be developed** to maintain the vitality of this research field in the U.S. and to make available the facilities and infrastructure needed to exploit opportunities.

As now, an active community of high-intensity laser scientists existed in 2002 in the U.S., but it was fragmented in single-investigator groups at universities and research scientists at a few DOE national labs. **No single national funding agency had responsibility for this field as a whole**. The SAUUL report recommended organizing and funding a network of institutions devoted to research in UUL science. **Forming a coordinated, cross-agency network** (reproduced in Fig. 2.2.1) would unify the community and greatly enhance its effectiveness, enabling cross-disciplinary interaction among subfields. It would minimize the number of large, expensive facilities and maximize the effectiveness of investments by coordinating activities at next-generation petawatt-peak-power and kilowatt-average power lasers needed to work at the frontier of high-intensity research. Individual nodes of the proposed network would be sponsored by the agency with the greatest interest in the science concentration of the node. Competition for nodes and open access to facilities were essential, so formation of a funded cross-agency body was proposed that would manage the field.

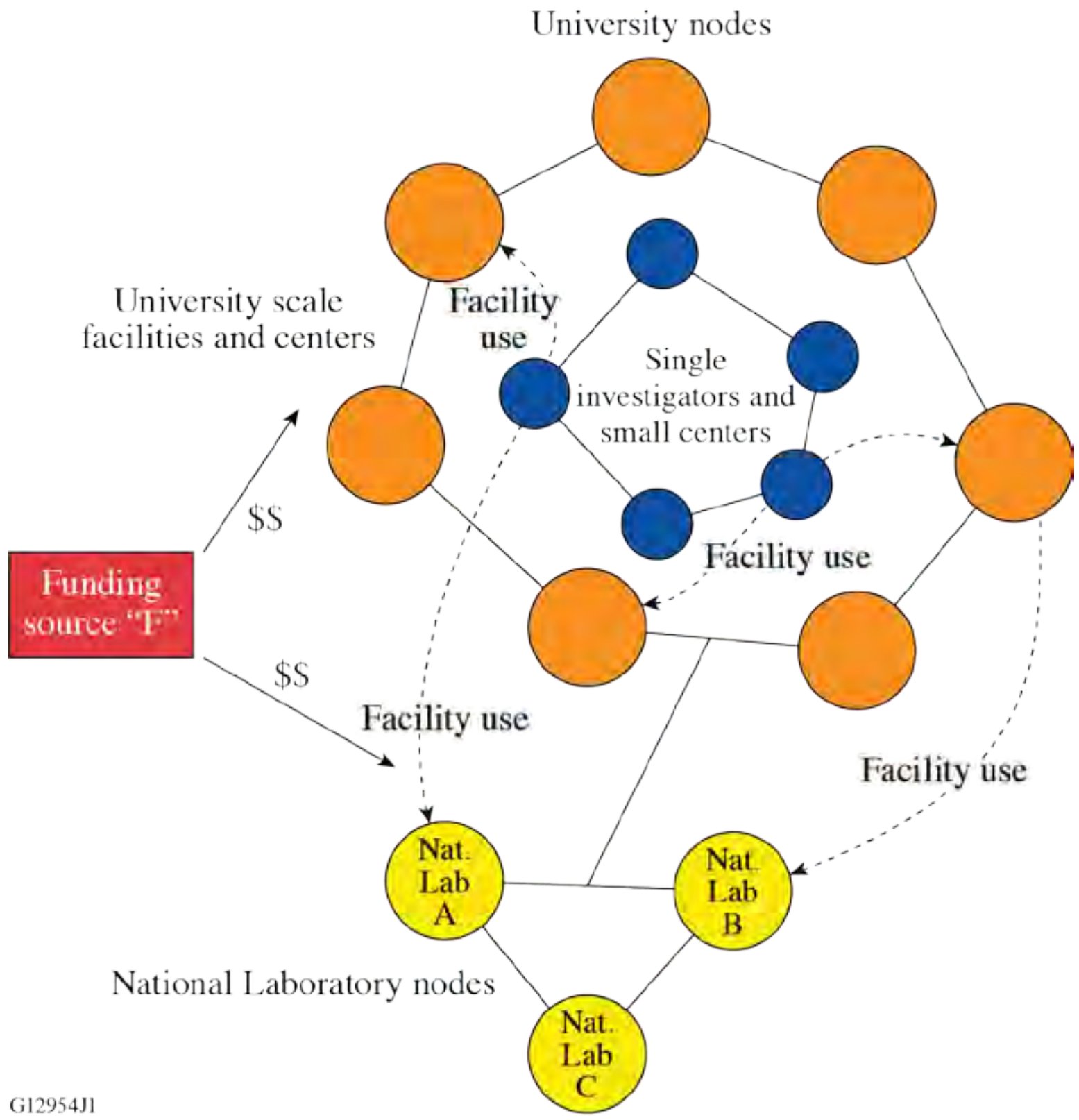

Figure 2.2.1. Coordinated, cross-agency network recommended by 2002 SAUUL report. [4]

Nine possible network node themes were suggested:

1. basic high-field science,
2. computational high-intensity physics,
3. high-energy-density science,
4. laboratory astrophysics,
5. advanced fusion approaches, such as fast ignition,
6. hyperfast (attosecond) x-ray source development and applications,
7. structural dynamics,
8. advanced particle acceleration, and
9. ultrafast nuclear science.

The proposed UUL network and funding scheme proposed by the 2002 SAUUL report came to fruition in part when the DOE Fusion Energy Sciences (FES) program established LaserNetUS [21] in August 2018. LaserNetUS includes ten facilities at universities and national laboratories, shown in Fig. 2.2.2. Two calls for proposals for 2019 and 2020 have been completed. Expanding LaserNetUS and establishing a cross-agency body with dedicated funding would further enable building, upgrading, and operating high-intensity laser infrastructure, developing key technologies, and engaging in research at LaserNetUS and international facilities.

21 "LaserNetUS," https://www.lasernetus.org/.

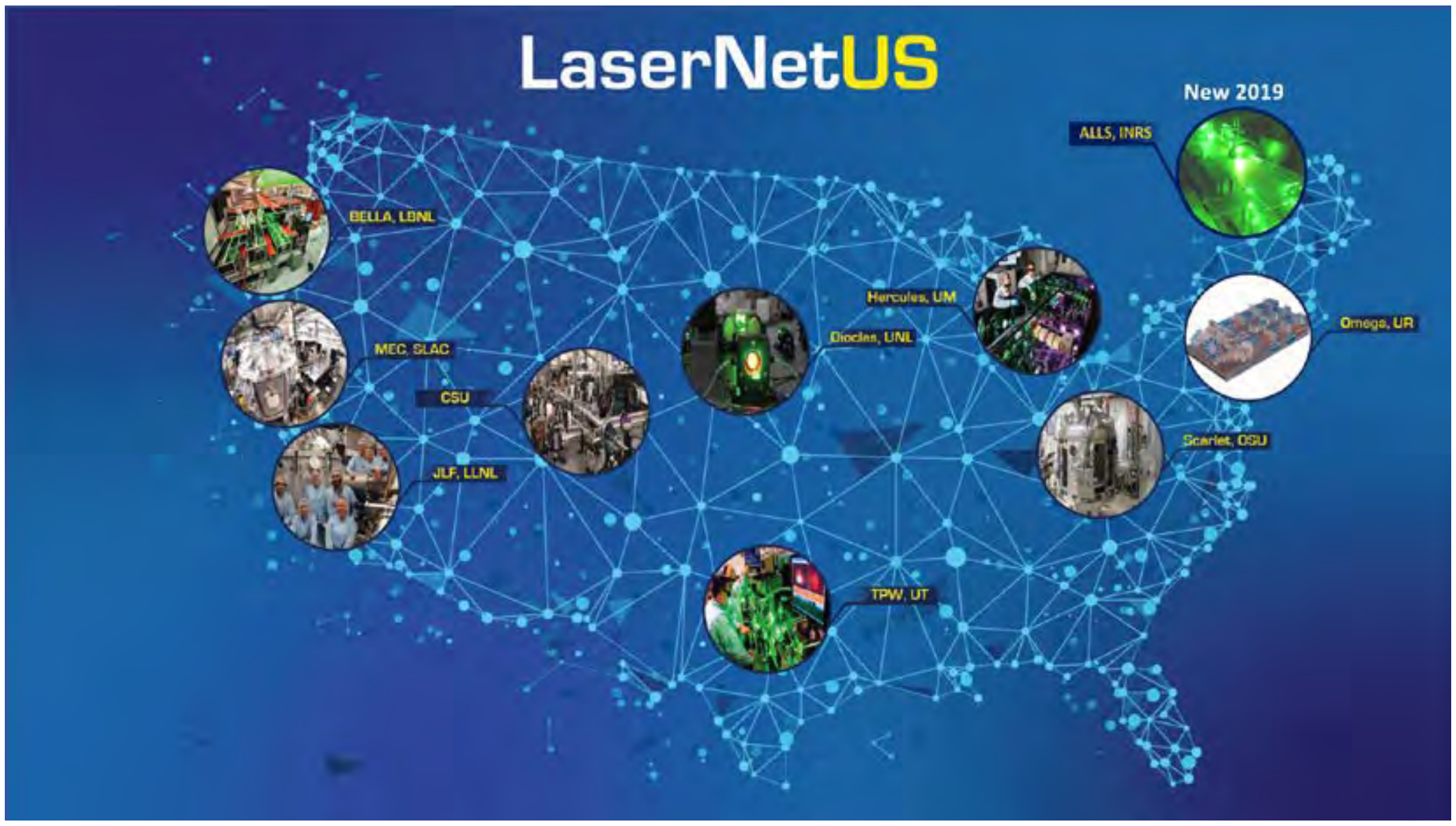

Figure 2.2.2. LaserNetUS network.

At roughly the same time, the SAUUL report was published in 2002, the European Strategy Forum for Research Infrastructures (ESFRI) [22] was established "*to keep Europe at the forefront of rapidly developing science and technology and to increase the capacity needed to meet the needs of the EU and the global scientific community. ESFRI … aims to promote a coherent and strategically driven approach to the design of European research infrastructures and to facilitate multilateral initiatives leading to better use and development of research infrastructures – both within the EU and internationally*."

Laserlab-Europe [5], formed in 2003, is a European consortium of major laser research infrastructures, forming an Integrated Infrastructure Initiative. Laserlab-Europe includes most European Union member states and scientifically covers most areas of laser science and applications with particular emphasis on short pulses and high intensities. Laserlab-Europe supports joint research activities and offers transnational access to European research community members at laser facilities across the unified but distributed laser research infrastructure in a highly coordinated fashion. It also strives to increase laser research and applications by reaching out to other scientific communities.

The exceptional internal coherence the European laser community established by LaserLab-Europe led to the first truly international laser infrastructure, the Extreme Light Infrastructure (ELI), which was initiated as part of a 2005 ESFRI Roadmap. The ELI Preparatory Phase produced the *ELI Whitebook* [10], which described the science, technology basis, and implementation strategy for the Extreme Light Infrastructure, heavily relying on the SAUUL

22 European Strategy Forum for Research Infrastructures (ESFRI), Accessed 29 January 2020, https://www.esfri.eu/.

report [4]. The *ELI Whitebook* also defined three ultra-intense laser user facilities to be implemented under the ELI Delivery Consortium (ELI-DC), which was founded in 2013:

- **ELI-Beamlines** in the Czech Republic, which focuses on short-pulse x-ray generation and acceleration of particles and their applications,
- **ELI Attosecond Light Pulse Source (ELI-ALPS)** in Hungary for the generation and application of the shortest pulses with very high repetition rates, and
- **ELI Nuclear Physics (ELI-NP)**, which uses ultra-intense optical and gamma-ray pulses to investigate fundamental problems in the field of nuclear physics and their practical applications.

Substantial investments by the European Union and the ELI host countries exceeding US$ 1 billion supported construction of these three facilities with world-leading lasers. The ELI project also funded significant technology development activities among both established and relatively new companies, bolstering Europe's commercial leadership in ultrafast laser technology as a whole—most of the laser systems development was done through commercial contracting that resulted in developing and delivering the following advanced laser systems:

- a 100-kHz, <7-fs, 1-mJ carrier-envelope-phase (CEP) stabilized fiber laser system at ELI-ALPS;
- a 100-kHz, <40-fs, 0.15-mJ CEP-stabilized mid-infrared optical parametric chirped-pulse amplifier (OPCPA) system at ELI-ALPS;
- a 1-kHz, <7-fs, >35-mJ CEP-stabilized mid-infrared OPCPA system at ELI-ALPS;
- a 1-kHz, <15-fs, 100-mJ OPCPA system at ELI-Beamlines;
- a 10-Hz, 2-PW (34 J in <17 fs) laser at ELI-ALPS;
- a 10-Hz, 1-PW (30 J in <30 fs) laser at ELI-Beamlines (from LLNL in the U.S.);
- a 10-PW (1500 J in 150 fs), shot per few minutes laser at ELI-Beamlines (from National Energetics in the U.S.); and
- two 10-PW (300 J in 30 fs), shot per minute lasers at ELI-NP.

Commissioning of the three ELI "pillars" was underway at the time of the BLI Workshop and some user operations have started. ELI-DC is coordinating the transition from implementation to the operations phase by leading the creation of a European Research Infrastructure Consortium (ELI-ERIC) [23]. A fourth ultrahigh-intensity pillar is also envisioned that is expected to exceed the current ELI pillars by about one order of magnitude into the sub-exawatt regime, but the location is still to be decided and funding is not secured.

23 ELI ERIC – Moving Ahead, ELI Beamlines, Accessed 6 December 2019, https://eli-laser.eu/news/eli-eric-moving-ahead/.

### 2.3 BLI Workshop Charge, Organization and Process

The charge for the Brightest Light Initiative workshop, sponsored by the DOE Office of Fusion Energy Science, the National Science Foundation, the National Nuclear Security Administration, and the National Photonics Initiative, was to organize the U.S. intense ultrafast laser community and to

- articulate a community response to the 2018 National Academy of Sciences report "Opportunities in Intense Ultrafast Lasers: Reaching for the Brightest Light";
- identify compelling science fundamental and applied research opportunities constituting priority research directions that exploit high-intensity lasers leading to high impact over the next decade and beyond;
- define new and upgraded facility and laser capabilities that will enable compelling science, emphasizing parameters beyond the current state of the art; and
- identify laser research and development to realize both ultrahigh intensities for basic research and high repetition rates, as well as high-average powers needed for applications.

The workshop was organized into two tracks to define (1) the **scientific research needs (SRNs)** for intense ultrafast lasers emphasizing parameters beyond the current state of the art in areas critical to frontier science, and (2) the **technology research needs (TRNs)** to realize these parameters, including peak power, repetition rate, pulse duration, wavelength, and focusable intensity.

SRN topical panels and panel co-leaders who led discussions and drafted material for the workshop report include:

1. **High-energy-density (HED) and basic plasma physics**

   *Bedros Afeyan (Polymath Research) and Sean Finnegan (LANL)*

   Topics: Laser–plasma instabilities (LPIs), plasma photonics, laser-driven magnetic fields, spectroscopy, atomic physics, warm dense matter, and pair plasmas.

2. **Material, planetary, and astrophysical sciences**

   *Frederico Fiuza (SLAC) and Dustin Froula (LLE)*

   Topics: Dynamic material science, shocks, equation of state, opacity, material strength, collisionless shocks, hydrodynamic instabilities, and radiation-induced defects in materials.

3. **Laser wakefield acceleration and applications**

   *Cameron Geddes (LBNL) and Mike Downer (U. Texas, Austin)*

   Topics: Electron acceleration, injection mechanisms, staging, light sources [betatron, Compton, free-electron lasers (FEL), applications (medical, HED, material science], and laser wakefield acceleration (LWFA) diagnostics.

4. **High-field physics, quantum electrodynamics (QED), and attoscience**

   *Alec Thomas (U Michigan), Alexey Arefiev (UCSD), and Lou DiMauro (OSU)*

   Topics: Radiation reaction, quantum effects, high-harmonics generation, Schwinger limit, and high-field science. This panel divided into two subgroups to separately address high-field physics and QED (SRN4A), and high-harmonics generation, and attoscience (SRN4B) led by Margaret Murnane and Henry Kapteyn.

5. **Particle acceleration: neutrons, ions, protons, and positrons**

   *Douglass Schumacher (OSU) and Scott Wilks (LLNL)*

   Topics: Ion acceleration (TNSA, RPA, BOA, shock), neutrons, positrons, and applications (medical, HED science, industry).

6. **Nuclear photonics**

   *Igor Jovanovic (U. Michigan) and Markus Roth (Darmstadt Tech. Univ.)*

   Topics: Nuclear physics with short-pulse lasers and gamma-ray beams: cross sections, nuclear resonances, and applications.

TRN topical panels and panel co-leaders who led discussions and prepared material for the workshop report on challenges related to laser and experimental technologies, theoretical and computational needs, and operational issues associated with four different types of laser capabilities included:

1. **Achieving the highest intensities (beyond $10^{22}$ W/cm$^2$)**
   *Co-leaders: Jake Bromage (LLE) and Erhard Gaul (National Energetics)*

2. **Pushing high-energy (10-J < *E* < 10-kJ), high-intensity lasers to average powers**
   *Co-leaders: Alan Fry (SLAC) and Tom Spinka (LLNL)*

3. **Producing femtosecond systems with the highest average powers (>1 kw per beam)**
   *Co-leaders: Peter Moulton (MIT LL) and Almantas Galvanauskas (U. Michigan)*

4. **Upgrading/extending performance of existing (mid-scale) facilities**
   *Co-leaders: Jorge Rocca (Colorado State Univ.) and Csaba Toth (LBNL)*

Input from the scientific community was solicited before the workshop to guide workshop discussions and provide raw material for a workshop report. Workshop participants and interested community members who were not able to attend uploaded inputs to a collaborative file-sharing site. Those who could not attend arranged to have someone in attendance represent their material during the workshop.

Inputs for compelling science goals with first-day experiments and technology research and development were solicited using templates to provide standardized in two formats:

1. Quad-chart: single-slide “quad charts” summarizing proposals cases were discussed and refined at the workshop.
2. Two-page (max) white paper: white papers provided raw material for preparing the workshop report.

The first day (Wednesday, 27 March) started with a plenary session to kick off the workshop with opening remarks by Roger Falcone (Chair) and a review of the NAS report by Phil Bucksbaum (NAS study group chair). SRN panel leaders presented overviews of their breakout sessions that constituted the remainder of the day. A plenary session at the end of the day reconvened all participants to review “first drafts” of conclusions from these SRN breakout sessions.

The second day (Thursday, 28 March) started with more polished summaries of the first-day SRN breakout sessions by the panel co-leaders. These “second drafts” highlighted capabilities needing attention during the TRN breakout sessions. TRN panel leaders followed with overview presentations of their breakout sessions. A plenary session at the end of the day reconvened all participants to review first drafts of conclusions from these TRN breakout sessions.

The last day (Friday, 29 March) started with more polished summaries of the TRN breakout sessions by the panel co-leaders. These second drafts highlighted technology challenges to feed back for discussion during a final set of SRN breakout sessions. A final plenary session summarized final conclusions and closing remarks.

Workshop chairs and SRN/TRN topical panel co-leaders met after closing the workshop to define the process for preparing the workshop report. Panel co-leaders drafted summary material and forwarded it to their respective track co-chairs for review. The workshop co-chairs met to review the draft summary material and outline the workshop report. Two co-chairs were assigned primary writing and editing duties for each section. Upon completion, all sections were reviewed by all chairs and distributed to all panel co-leaders for their feedback. Chairs resolved any concerns that were identified during this review. A final complete draft report was distributed to external reviewers for comment before finalizing and distributing the final report.

[This page intentionally blank]

# Chapter 3 – Science Research Needs (SRNs) and Priorities

Lasers are the brightest controllable sources of light that are ubiquitous in research and applications to create and probe extreme states of matter. Laser energies available in current state-of-the-art systems routinely allow researchers to produce record material pressures in the laboratory, with pressures in the multi-TPa regime (1 TPa = $10^{12}$ Pa), exceeding pressures that exist at the center of earth and producing conditions found inside giant planets, in brown and white dwarfs, and in the interiors of stars. Thanks to the 2018 Nobel-prize–winning work of Mourou and Strickland [1], lasers are now approaching peak intensities that can produce light pressures in the EPa regime (1 EPa = $10^{18}$ Pa). When interacting with properly chosen targets, such conditions produce copious amounts of radiation and bright beams of energetic particles, including electrons, ions, neutrons, or anti-matter.

Laser capabilities are being vigorously explored with the goal to develop tunable coherent and bright radiation sources that span the range of the electro-magnetic spectrum. This frontier applies a variety of advanced techniques to access the THz regime, optical parametric amplifiers to tune from the infrared through the visible spectral ranges, high-harmonic generation for producing vacuum ultraviolet radiation and x-ray pulses into the attosecond range, and Compton scattering to produce high-energy gamma radiation. Ever brighter sources are needed to selectively excite physical mechanisms and transform materials states, and they can serve as our "eyes" to probe nature's structure and her physical and chemical processes.

Particle acceleration gradients produced in intense laser–matter interactions have surpassed those of conventional accelerators, opening the field for novel acceleration concepts. Such plasma accelerators routinely generate MeV to GeV particles in cm distances, thousands of times shorter than conventional methods. This enables a range of compact accelerator applications and bright, ultrafast electron, ion, and X-ray probe sources. Simulations predict the generation of G-Gauss magnetic fields (G-Gauss = $10^9$ Gauss = $10^5$ Tesla) that exhibit a complex interplay with plasma instabilities and whose study can elucidate the physical mechanisms that lead to cosmic particle acceleration. Such cosmic accelerators produce the highest known particle energies, producing a million times higher particle energies than earthbound accelerators. Studying these particle acceleration processes holds promise for understanding cosmic processes, advances in physics, and future developments for scientific, national security and industrial applications.

This chapter summarizes the input of the community that focused on transformative science goals that make use of bright radiation sources, and has articulated the need for advancing current facilities with lasers that produce higher intensities with better control than currently available. These science applications produce ever more extreme material states, such as the goal to reach the fully developed nonlinear plasma state relevant for cosmic ray physics or producing

1 The Nobel Prize in Physics 2018. www.NobelPrize.org. Nobel Media AB 2019.

unearthly material and plasma states. For example, the latter may be important for understanding exoplanets, whose discovery was recognized by the 2019 Nobel Prize in Physics awarded to Mayor and Queloz [2]. Many of the proposed studies require repetition rates that will enable more and better experiments using active feedback control to fully optimize the laser drive and apply machine learning algorithms to deliver highly accurate measurements that advance science.

The study of extreme material states will greatly benefit from exquisite probing capabilities that have the power to visualize the laser–matter interaction processes and that can be provided by collocating dynamic compression facilities with existing and future x-ray sources, and by multi-beam laser facilities enabling laser-plasma produced probe sources. In addition, bright x-ray and energetic particle probes are needed to measure the structures and phase boundaries of matter in extreme conditions; the same capability is required to resolve the effects on materials that are exposed to harsh radiation environments. Understanding these mechanisms has the potential to greatly advance our understanding of dense plasma physics and radiation damage with the potential to lead to new developments in this research area. Table 3.1.1 summarizes top-level facility requirements to meet the BLI Scientific Research Needs presented in Chapter 3.

Table 3.7.1 – Facility Requirements to Meet BLI Scientific Research Needs

| **Facility Requirement** | **Science case** |
|---|---|
| Synchronized multi-PW (isochoric heating) and multi-kJ long-pulse (compression) | SRN1 + SRN2 |
| Bright x-ray and particle probes | SRN1 + SRN2 + SRN6 |
| Multi-PW lasers/multi-10 GeV electrons | SRN2 + SRN4A |
| High rep-rate (≥ kHz), efficient, multi-J lasers with feedback control | SRN3 + SRN4B |
| Higher photon energies → hard x-rays (~6-10 keV and 30-100 keV) | SRN4B |
| 100-kHz/kW mid-IR lasers to drive higher flux HHG sources for applications | SRN4B |
| Ultra-relativistic (> $10^{22}$ W/cm$^2$) intensities (150 to 1500 J; ≥ 100 fs to multiple ps) with ultrahigh temporal contrast (≥ $10^{12}$) + femtosecond pulse shaping + large uniform focal spots | SRN5 |
| 50-300 MeV electrons + mJ/ps laser pulses for tunable MeV Compton sources | SRN6 |
| Energetic (multi-MeV) and high-flux ($10^{12}$/s) ion and neutron sources | SRN6 |

2 The Nobel Prize in Physics 2019. www.NobelPrize.org. Nobel Media AB 2019.

## 3.1 SRN-1 – Basic and High-Energy-Density (HED) Plasma Physics

An important aspect of High Energy Density (HED) plasma physics is concerned with the generation, study and control of self-organization, far from equilibrium, through the action of intense fields. It is also often more generally defined by a range of pressure and temperature that greatly exceed normal conditions of materials on the surface of the Earth. Research in plasmas at HED has emerged in the past decade as a frontier area for discovery, not only in plasma science but throughout broader areas of science, as well as in commercial and national-security applications. First-generation technical capabilities and resulting fundamental discoveries have transformed the field of plasma science and its applications towards deeper maturity.

HED plasma physics is inherently multi-disciplinary and multi-scale. Fundamental questions of many-body interactions can be answered by studying and manipulating the behavior of HED plasmas. In particular, understanding self-organization of matter under long-range forces is important, both in the small amplitude limit (linear regime), and the turbulent limit (ergodic regime). The potential for discoveries lies between these two extremes, motivated during the workshop by questions, such as:

- How large can nonlinear kinetic structures be made and maintained in HEDP?
- How far from equilibrium can structures in HEDP dynamical systems be pushed, and for how long can they be sustained?
- How does one bridge the gap of efficiently modeling weakly to strongly coupled plasmas?
- How do self-generated magnetic fields influence multidimensional plasma dynamics?
- How can one characterize and control HED plasma turbulence including kinetic effects?

Answering these questions will transform understandings of HED plasmas in the laboratory and the cosmos at large, and open new avenues for potent breakthrough application. This section emphasizes foundational scientific advances in basic and HED plasma physics that can be realized in the coming decades along with the laser hardware capabilities required to facilitate their success.

*Fundamental questions in High-Energy Density Science can now be addressed by investing in powerful optical and x-ray laser facilities and by coupling experimental optimization and data analysis with advanced machine-learning tools.*

### 3.1.1 Opportunities for Frontier Science

Implementing high repetition-rate light sources, both coherent x-ray light sources and high-average-power optical lasers, and advancing machine learning for data analysis will revolutionize HED plasma physics. Exploiting these capabilities represents the future of scientific discovery and enable two major opportunities for frontier science.

#### 3.1.1.1 Long-time behavior of parametric instabilities

Parametric instabilities play a crucial role in governing the energetics in laser-plasma interactions (LPI), significantly affecting instability growth, saturation, and energy coupling to the plasma. The capability to control and manipulate LPI will increase abilities to create an extended range of HED conditions and realize the full potential of laser-driven inertial fusion platforms. First experiments that could test these concepts and methods could focus a diffraction-limited, sub-picosecond probe laser with flexible spatio-temporal parameters into a plasma with well-controlled density, temperature and flow conditions driven by a high energy laser. Radiation from the probe scattered off driven plasma waves via stimulated Raman scattering (SRS) or stimulated Brillouin scattering (SBS), correlated with the back-reflected light signals, provides information about the plasma conditions. These measurements require 100-fs time resolution over 10s to 100s ps time records. In addition, transverse cross-communication, self-organization and magnetic interconnectivity triggered by anisotropic distribution functions benefit from having a second laser hot spot nearby.

Specifically, by increasing the pulse width of the single hot spot from 100 fs up to 100 ps, in small steps and with variable delays in between subsequent spikes, we can test the efficacy of STUD pulses [3], spike trains of uneven duration and delay, plasma recurrence and memory build-up. In addition, transverse cross-communication, self-organization and magnetic interconnectivity triggered by anisotropic distribution functions could also benefit from having a second witness speckle or laser hot spot present in the vicinity of the first. Additionally, increasing the delay between spikes we can study the healing-time effects of undriven waves which are then re-driven by uncorrelated, secondary spikes. Furthermore, such a system would allow for studying the effects of nearby spikes cross-communicating statically or in the local, dynamically scrambled environment.

#### 3.1.1.2 Kinetic plasma turbulence

Perhaps the most cross-cutting grand challenge in plasma physics is developing a complete understanding for the dynamics of fully-evolved plasma turbulence including the interplay between self-generated fields and the hydrodynamic evolution [4], as well as nature of coherent structures [5]. More specifically, in search of self-consistent entanglement, the key role of coherent structures, and dimensional reduction in plasma and magnetic turbulence. Multi-point, multi-field measurements, along with theory and state-of-the-art computation are critical to answering once and for all profound question that remain such as: *Is there a direct persistent causal link between large scale coherent structures and fine scale turbulence? What mechanisms lead to long-lived coherent structures (why do they persist), and govern their creation, destruction, and regulate*

---

3 B. Afeyan and S. Hüller, "Optimal Control of Laser Plasma Instabilities Using Spike Trains of Uneven Duration and Delay (STUD Pulses) for ICF and IFE," EPJ Web Conf. **59**, 05009 (2013).

4 P. Tzeferacos *et al.*, "Laboratory Evidence of Dynamo Amplification of Magnetic Fields on a Turbulent Plasma," Nat. Commun. **9** (1), 591 (2018).)

5 F. Pegoraro *et al.*, "Coherent Magnetic Structures in Self-Organized Plasmas," Plasma Phys. Control. Fusion **61** (4), 044003 (2019).

*their energy dissipation?* Resolving these questions requires measuring the velocity field, viscosity, density, temperature, and conductivity all at the same time in order to adequately constrain models.

Developing a complete understanding for the dynamics of fully evolved plasma turbulence, including the interplay between self-generated fields and hydrodynamic evolution, as well as the nature of coherent structures, represents a cross-cutting grand challenge in plasma physics. Multi-point, multi-field measurements along with new theory and state-of-the-art computation are critical to answering questions, such as: *Is there a direct persistent causal link between large scale coherent structures and fine scale turbulence? What mechanisms lead to long-lived coherent structures (why do they persist), and govern their creation, destruction, and regulate their energy dissipation?* Resolving these questions requires measuring the velocity field, viscosity, density, temperature, and conductivity simultaneously to constrain models. An example of such an experiment addressing the complex transport regimes is shown in Fig. 3.1.1.

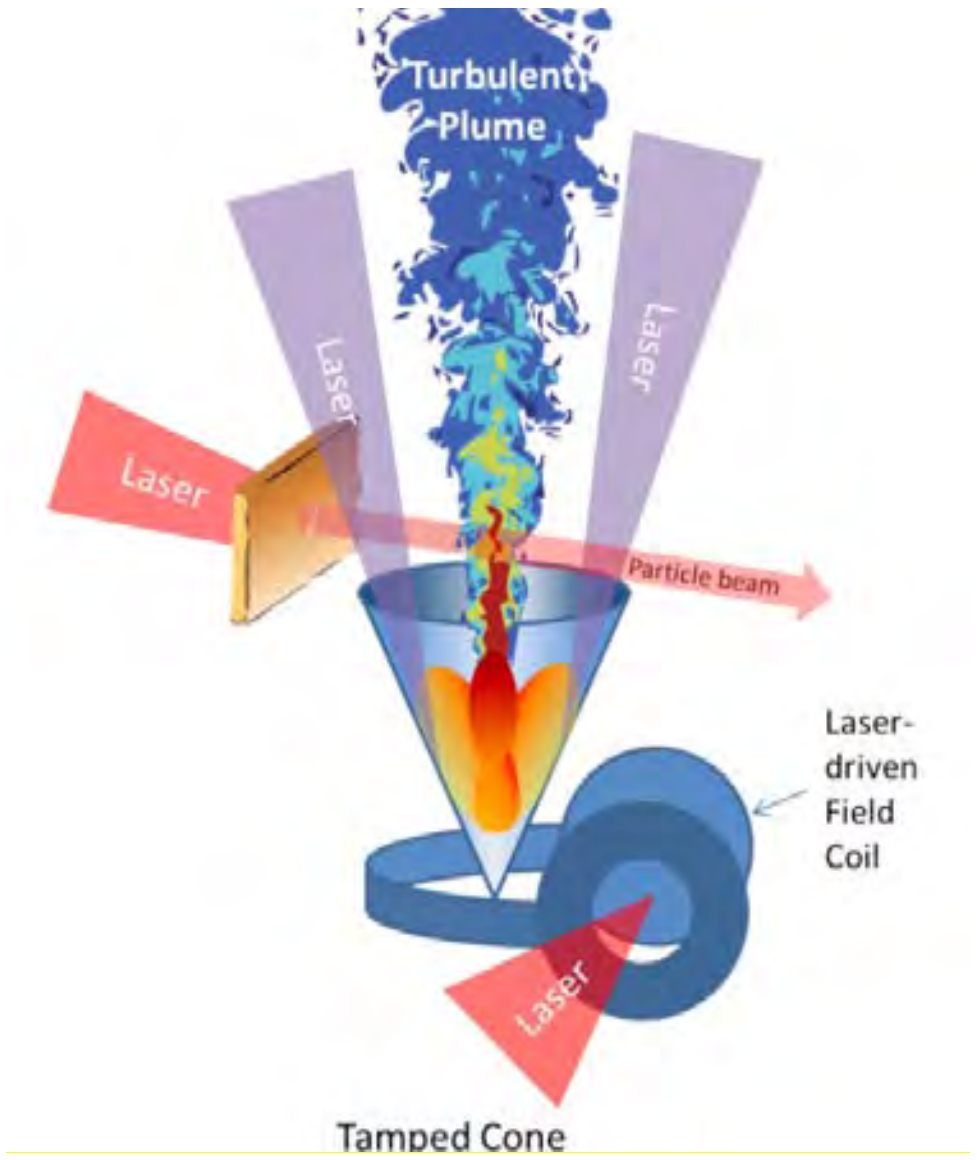

Figure 3.1.1. A laser-driven experiment to study plasma dynamo evolution schematically shows how plasma and magnetic field turbulence in the kinetic regime creates complex energy transport. [6]

### 3.1.2 Underlying Physics

The BLI workshop identified four grand challenges for discovery in basic and HED plasma sciences that would be enabled by multiple, collocated ultraintense and high-energy laser platforms with repetition rates higher than currently available. These challenges exemplify unifying themes for all of plasma science, and represent the future of the field.

6 Courtesy of K. Flippo (LANL).

#### 3.1.2.1 Self-organization in HED plasmas

Coherent structures (i.e., self-organized patterns) are complex and fascinating phenomena that occur in a wide variety of biological, chemical, and physical systems. Their occurrence is associated with the supply of energy to an open dissipative system, leading to instability, symmetry breaking, bifurcation, and ultimately to the formation of self-sustaining coherent structures. Coherent structures can be observed in a wide diversity of space and laboratory plasmas and high-intensity beams [7]. These interactions are caused by the self-consistent electromagnetic fields generated by the charged particles that make up plasma structures. These electromagnetic fields can transfer energy from fast particles to slow particles even in the absence of collisions in a way that frustrates and invalidates fluid modeling of plasmas. Consequently, coherent structure formation in plasmas depends crucially on the details of the velocity distribution of the particles in phase space. The current level of advanced diagnostics, as well as recent progress in numerical and analytical modeling, now make it possible to understand the mechanisms of coherent structure formation and to manipulate them with unprecedented accuracy on the particle-kinetic level by measuring, sculpting, and simulating the evolution of velocity distribution functions in phase space. Although much knowledge has been acquired about specific nonlinear structures that occur in particular contexts, there remains a need to develop a comprehensive picture that can explain and manipulate such nonlinear coherent plasma pattern formation synthesizing all these tools and concepts.

#### 3.1.2.2 Energy flows in HED plasmas

Self-organization processes, such as those found in turbulent flow generation or dynamo action, involve the nonlinear transfer of energy across spatiotemporal (and energy) scales and the formation of coherent structures at intermediate scales between the micro-turbulence and the global system scale. These processes transform energy from one form into another (e.g., kinetic energy to magnetic field energy). Important challenges include: understanding transformations of energy from particle flows to electric and magnetic fields, magnetic field energy to plasma flows and energetic particles; and magnetically constraining the systematic complexity of HED plasma. It is crucial to obtain a deep understanding of the interaction and interplay between plasma turbulence, self-organization, turbulent transport and subsequent system-scale behavior for scientifically and technologically interesting plasma systems, such as the mitigating effect of magnetic fields on Rayleigh-Taylor unstable inertial confinement fusion plasmas illustrated in Fig. 3.1.2.

Scientific progress requires an approach that links theory, multi-point measurements, and state-of-the-art computation. Multi-point, multi-field measurements of turbulence provide the relevant turbulent fluxes (e.g., particle, momentum and heat flux for fluid-like processes) through space; together with background measurements these can provide flux-gradient relationships (or the equivalent for fundamental kinetically driven processes) which can be used to build reduced models that can then explore a wide range of parametric conditions to trigger further precision and synthesis. The highly nonlinear nature of the transport can often lead to distinctly different regions in parameter space that are bounded by phase transitions and bifurcations that need to be clearly identified and understood.

---

7 B. Eliasson and P. K. Shukla, "Formation and Dynamics of Coherent Structures Involving Phase-Space Vortices in Plasmas," Phys. Rep. **422** (6), 225–290 (2006).

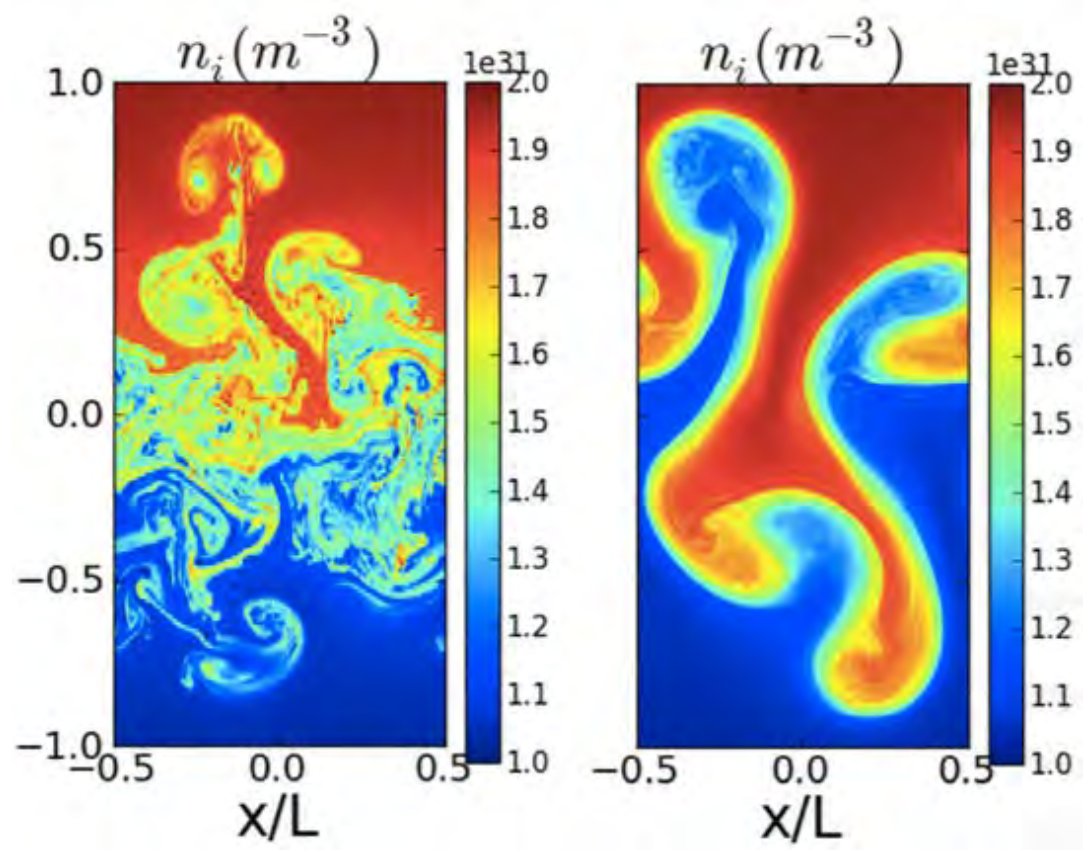

Figure 3.1.2. Numerical MHD simulations of the constraint of turbulent scales via external magnetic fields. Magnetic fields severely limit the speed and complexity of turbulent mixing. [8]

#### 3.1.2.3 Relativistic pair plasma in the laboratory

Research with trapped positrons has been at the forefront of discovery in plasma physics for several decades. This has led to applications in materials characterization[9] and probes of molecular structure[10], and there continues to be great focus on improving the ability to accumulate and store positrons, studying the properties and pushing the limits of these unusual systems, and on finding new applications. Nonlinear kinetic structures in pair plasmas are interesting since they involve electrons and positrons on an equal footing. Both phase spaces have holes and vortices of comparable size evolving on the same time scale. Only a drift is needed to trigger them and a temperature difference, if any. A much richer collection of kinetic behavior is expected since particles trapped in one well will also be strongly affected by the wells in the drifting wells as well. The behavior in magnetized plasmas is equally compelling. Then at relativistic energies along with strong gamma ray fields, a new arena of strong-field plasma physics results. Similar effects have been predicted for the astrophysically relevant case of relativistic pair plasmas, such as those expected at the poles of neutron stars. In spite of extensive theoretical work on pair plasmas, they have not yet been studied in the laboratory. Progress in the accumulation and manipulation of antimatter made over the past two decades is now at the cusp of making experimental studies of pair plasmas possible.

---

8 B. Srinivasan and X.-Z. Tang, "The Mitigating Effect of Magnetic Fields on Rayleigh–Taylor Unstable Inertial Confinement Fusion Plasmas," Phys. Plasmas **20** (5), 056307 (2013).

9 A. E. Hughes, "Probing Materials with Positrons," Mater. Des. **2** (1), 34–40 (1980).

10 K. Chen and X. Chen, "Design and Development of Molecular Imaging Probes", Curr. Top. Med. Chem. **10** (12), 1227–1236 (2010).

#### 3.1.2.4 1,2,3 … Infinity – From weakly to strongly coupled plasmas in HED systems

George Gamow described primitive cultures where bartering was done by counting to three and lumping the rest under "many" [11]. Similarly, in plasma kinetic theory, as the plasma coupling constant increases, one-body and two-body correlation function descriptions quickly give way to the too-many-to-count limit, also known as the strong coupling limit where dynamics are simple and far from violent, but stochastic, diffusive and very high dimensional. This inherent duality between trading low-dimensional deterministic chaotic dynamical complexity for very high-dimensional, stochastic, small-amplitude dynamics represents a fascinating feature to explore and exploit.

From high-dimensional stochastic dynamics of strongly coupled plasmas to the violent intermittent behavior of weakly coupled HED systems. HED plasmas can be highly nonlinear in the weak coupling regime where long-range collective interactions dominate their response, while in the strongly coupled regime they tend to be rigid, inherently high-dimensional and stochastic in their response and may also include quantum correlations. The unifying theme is how coherent, external, well-controlled radiation can help steer the flow of energy to self-organize and self-sustain states far from equilibrium. The goal is to exploit nonlinear, kinetic plasma-wave structures created by optical mixing techniques with modern laser pulse sequences and used for pressing applications, beyond the fundamental many-body physics questions that arise along the way. Whether it is the steering of high energy (x-ray) photons or accelerating particles in controllable electric field gradient structures, high-impact applications await the manipulation of HED plasmas on the 100-fs time scale lasting for picoseconds. Features that are highly complex can be sculpted by controlled regiments of crossing laser pulses creating nonlinear, kinetic, phase space barriers that can steer high energy or shorter wavelength radiation on the 10-fs time scale effectively and disposably. Both electrostatic and electromagnetic interactions must be deployed and nonlinear plasma waves, the Weibel, Harris and Buneman instabilities – all three exploiting velocity space anisotropy and inter-species drifts – as well as trapping physics, sidebands and merging vortices will play a role.

11 G. Gamow, *One, Two, Three… Infinity: Facts and Speculations of Science*, (Viking Press, New York, NY, 1961).

### 3.1.3 Meeting the challenges – technology requirements

The workshop clearly identified the need for high repetition-rate light sources (coherent x-ray light sources and optical lasers) and machine learning for data analysis. Advancing frontiers in basic and HED plasma science will require collocation of multiple capabilities:

- Collocation of highly energetic (100s J up to 10-100 kJ) nanosecond, 150+ J (150 fs), and 30+ J (30 fs) optical lasers to serve as both pump and probe beams for driving particle and radiation diagnostic sources, while simultaneously driving matter on hydrodynamic time-scales. Energetic drivers are required to drive mm-scale targets to sufficiently large Reynolds number for turbulence to develop.
- 50-100 MeV protons beams (or 2-5 MeV electrons) coupled with tunable x-rays to penetrate four-times solid density, mm-scale, 10-eV HED plasmas, and measure 50 kG to 500 kG fields. Tunable x-ray probes will deliver precision radiography for quantifying structural turbulent features.
- High-resolution spatial and temporal diagnostics to capture long-time behavior (10s of ns) with multiple frames within a single shot (~1 frame/ns of a 30 ns experiment) and high-dynamic range to capture these measurements over four to five orders of magnitude.

Experimental lines of research with the potential to generate breakthroughs in basic and HED plasma physics, and their associated technical requirements, are identified in Table 3.1.1.

### 3.1.4 Summary and Recommendations

High-energy-density physics explores the extremes of matter and radiation inextricably entangled that challenge our understanding of nature. In particular, HEDP strives to generate, study and control nonlinear self-organization far from equilibrium through the action of intense electromagnetic fields. Bringing this mission to future experimental facilities that produce precision data at significantly improved repetition rates will allow coupling these methods with novel data analysis techniques and advanced machine-learning algorithms. This will deliver new insights and a deeper understanding of plasma self-organization, turbulence and multi-scale kinetic system behaviors that are needed to make predictions of laboratory experiments addressing classical and quantum many-body dynamics, astrophysical processes and laboratory inertial fusion.

A near-term priority is developing a facility with multiple, high-repetition-rate and synchronized laser drivers with modern state-of-the art diagnostics to study ultrafast plasma processes. In parallel, a program needs to support the development of theory and modern simulation algorithms, suited for university and single-investigator groups, to take full advantage of future experimental capabilities. Societal benefits span a wide range of applications from national security to diagnostic and therapeutic medicine, and will enable new technologies for the energy security of mankind.

**Table 3.1.1.** Required technical capabilities for basic and HED plasma physics

| | Research Topic | 1-10kJ (ns) | 150+ J (150 fs) | 30+ J (30 fs) | Rep Rate > 0.01 Hz | x-ray source |
|---|---|---|---|---|---|---|
| **Self-organization in HED plasmas and energy flows in HED plasmas** | Creating controlled incoherence in crossing laser beams | ● | ● | ● | ● | |
| | Magnetic interactions influencing SRS | ● | ● | | ● | |
| | Deciphering microphysics of turbulent magnetic dynamos | ● | ● | | ● | ● |
| | Femtosecond plasma phase-space photonics | ● | ● | ● | ● | ● |
| | Constraining plasma turbulence using external magnetic fields | ● | ● | | ● | ● |
| **Relativistic pair plasmas in the laboratory** | | | | ● | ● | |
| **1,2,3 … Infinity – From weakly to strongly coupled plasmas in HED systems** | Plasma kinetic diffusion beyond the Fickian limit at high-Z/low-Z interfaces | ● | ● | | ● | ● |
| | Memory accumulation effects in successively and multiply shocked interfacial dynamics | ● | ● | | ● | ● |
| | Uniformly heating large samples for WDM studies | | ● | ● | ● | ● |
| | DC conductivity of WDM using novel THz radiation | ● | ● | | ● | ● |

## 3.2 SRN-2 – Materials and Laboratory Astrophysics

Compelling science exists in producing, observing and controlling materials and plasmas in extreme conditions. The vast majority of condensed matter in the universe exists at pressures currently inaccessible to earthbound experiments. The core pressures of confirmed exoplanets generally exceed current experimental capabilities to probe their material properties. These extreme states are also important to dynamic systems vital to the security of our nation. At lower densities, the universe is magnetized and filled with cosmic rays, energetic particles with energies up to eight orders of magnitude higher than those produced in terrestrial/laboratory accelerators such as the LHC at CERN. The mechanisms behind magnetic field generation, amplification, and annihilation, as well as particle acceleration remain a research frontier.

Beyond its relevance to understanding the extreme universe, the study of particle acceleration in plasmas has a significant impact on generating new laboratory accelerators for a variety of applications. It is known that the magnetic field dynamics and particle acceleration is tightly associated with collision-less shocks, magnetic reconnection, turbulence, and plasma instabilities, through complex and nonlinear processes. Combining spacecraft measurements and laboratory experiments has led to significant progress in the study of nonrelativistic collisionless plasmas relevant to our solar system; however, relativistic plasma regimes relevant to high-energy astrophysical environments have been beyond the reach of *in situ* studies that require new lasers with higher energies and intensities, and high-precision ultrafast diagnostics to characterize these extreme states of matter based on x-ray lasers or energetic particle beams.

> *Well-characterized compression experiments enabled by cutting-edge lasers can help resolve important questions about astrophysical processes like the source of cosmic rays and the habitability of exoplanets, as well as synthesize novel materials not found on earth of interest for applications.*

### 3.2.1 Opportunities for Frontier Science

#### 3.2.1.1 Planetary, dwarf star, and solar interiors

A wide range of opportunities would use lasers to compress matter isentropically to produce novel material states with new properties in strength, hardness, or conductivity. Structural characterization of these high-density states requires bright ultrafast x-rays for imaging and spectroscopy, simultaneous measurements of material properties, for example, with THz probing and high shot rates.

*Required: 200 J to > 10 kJ shaped nanosecond laser pulses with repetition rates to map phase boundaries and to control the path through the HED matter phase space. Collocation with intense short pulse lasers and coherent x-rays is needed for characterizing material properties.*

#### 3.2.1.2 Relativistic shocks, magnetic reconnection, and plasma instabilities

This area of research requires ultra-intense lasers that produce and drive the target to relativistic temperatures and for sufficiently long time that allow the generation of strong magnetic fields via current-driven instabilities and onset of magnetic reconnection or collisionless shocks in relativistic environments with subsequent particle acceleration.

<u>Required</u>: *Multi-PW laser, focal intensity in the range $10^{21}$ to $10^{23}$ W/cm$^2$ with high temporal contrast (> $10^{12}$), and Strehl ratios approaching unity.*

#### 3.2.1.3 Jets and pair plasmas

Similar to 3.2.1.2, but now the focus is generating large scale-length plasma jets.

<u>Required</u>: *In addition to the requirements in 3.2.1.2, diagnostics with high temporal resolution and long record lengths.*

### 3.2.2 Underlying Physics Enabled by Next-Generation Lasers

The control of matter in extreme conditions allows the exploration of new properties, which can lead to transformative new materials that impact society and our economy. The past few years have seen a revolution in material science by applying precise pulse shaping of the output of high-power laser systems, such as NIF and OMEGA, to compress, in a shock-free, quasi-isentropic manner, solid-state and warm dense matter into new pressure regimes. From the perspective of fundamental science, the novel states obtained are directly relevant to understanding the composition of the plethora of exoplanets that have been discovered over the past two decades and are still being discovered on an almost daily basis. Closer to home, access to as-yet-unexplored high-pressure regions of the phase diagram of all of matter affords the opportunity to address the grand challenge of designing and engineering new metastable phases, with the ultimate aim of recovering them to ambient conditions. Just as one of the most industrially impactful metastable materials—diamond—was formed under high pressure on our own planet, many other metastable materials are known. Density functional theory predicts that certain materials subjected to even higher pressures than those within the earth’s core could be metastable and recoverable to standard pressures and temperature with unique and useful properties (e.g., the BC8 form of diamond is predicted to be harder than the natural forms).

#### 3.2.2.1 High-energy, spatially smoothed, and temporally shaped laser pulses

**Dynamic compression**. High-energy-density—in particular, warm dense matter with densities comparable to or above solid density—is a very challenging research subject since such matter combines strong interactions of the particles with a distinct quantum behavior including bound states and degeneracy. One crucial point for reduced modeling is the effective ion–ion potential, which is often strongly modified from potentials applicable for ambient material states. Since this potential is determined by the charge distribution, electron screening, and the interaction of bound and free electrons in this parameter region, it captures the complex interactions in the matter under investigation.

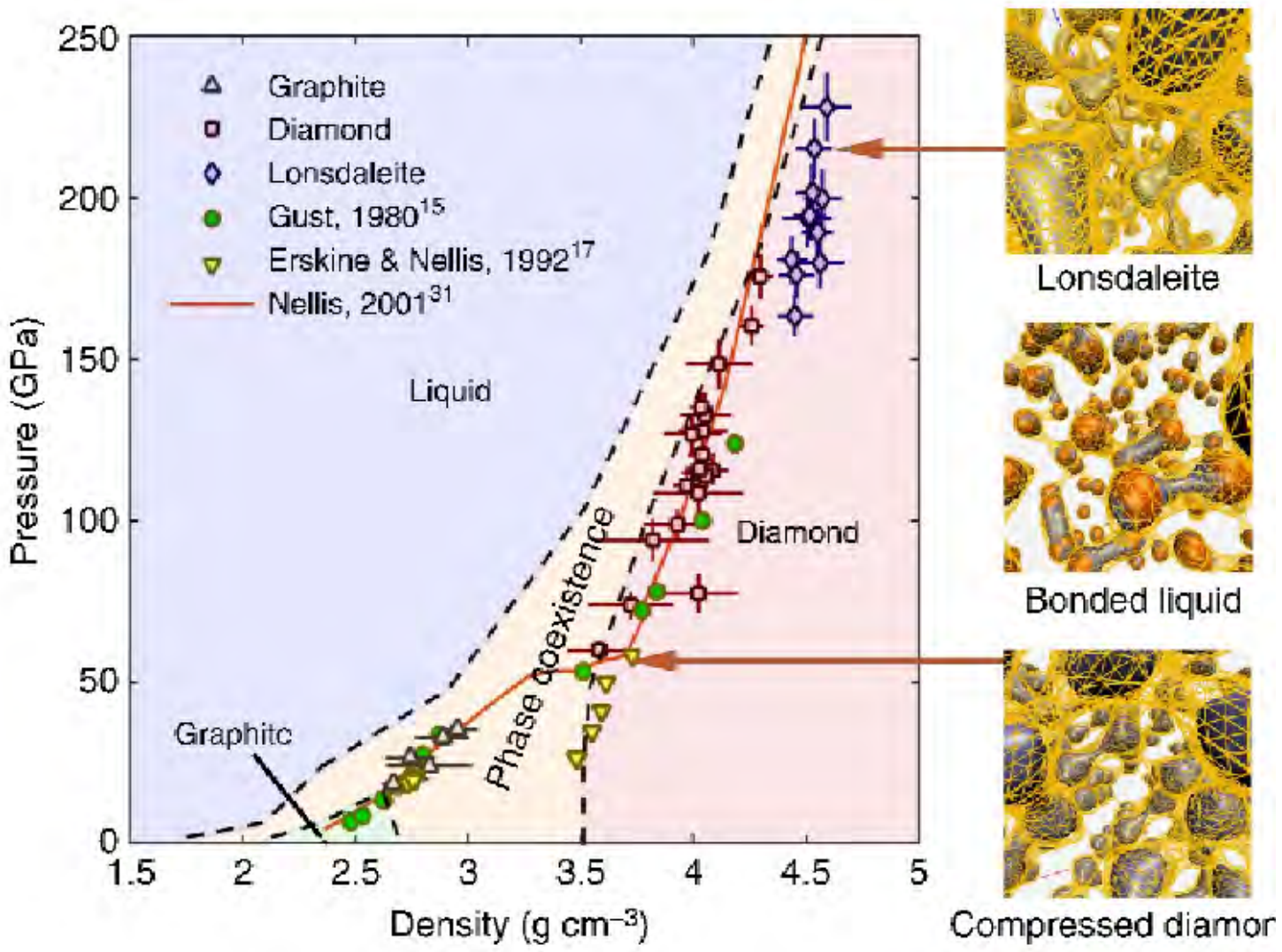

**Figure 3.2.1.** (a) New phase diagram for graphite for pressures up to 250 GPa (2.5 Mbar) showing the transition from graphite to compressed diamond to the bonded liquid and to the elusive lonsdaleite phase. This study is enabled by the collocation of an energetic laser driver with the exquisite probing capabilities of an x-ray free electron laser. [12]

Experimental investigations of dense matter cannot rely on usual plasma physics techniques because the high density makes the matter opaque in the visible spectrum and beyond. Further, the remaining atomic structure can change absorption considerably, and only transient states can be created in the laboratory. Most existing techniques therefore focus on the macroscopic behavior like the equation of state or absorption. Over the last decade, x-ray scattering and diffraction technique have been developed as suitable techniques to investigate the microphysics of these extreme states [12,13,14,15]. The x-rays or energetic electron beams can penetrate dense matter and couple to the electrons or ions, respectively, in the matter probed. Since part of the electron density follows the ion motion, both methods can observe the ionic structure. Recent proof-of-principle experiments have demonstrated this approach with great success [16]. Figure 3.2.1 shows recent data together with the results of theories indicating phase boundaries and the first production of the hexagonal diamond phase in the laboratory.

12 D. Kraus *et al.*, “Nanosecond Formation of Diamond and Lonsdaleite by Shock Compression of Graphite,” Nat. Commun. **7**, 10970 (2016).
13 A. E. Gleason *et al.*, “Ultrafast Visualization of Crystallization and Grain Growth in Shock-Compressed $SiO_2$,” Nat. Commun. **6**, 8191 (2015); 8709(E) (2015).
14 M. G. Gorman *et al.*, “Direct Observation of Melting in Shock-Compressed Bismuth With Femtosecond X-Ray Diffraction,” Phys. Rev. Lett. **115** (9), 095701 (2015).
15 D. Kraus *et al.*, “Formation of Diamonds in Laser-Compressed Hydrocarbons at Planetary Interior Conditions,” Nat. Astron. **1** (9), 606–611 (2017).
16 M. Z. Mo *et al.*, “Heterogeneous to Homogeneous Melting Transition Visualized with Ultrafast Electron Diffraction,” Science **360** (6396), 1451–1455 (2018).

### 3.2.2.2 High-intensity, high-energy laser pulses

**Particle acceleration in relativistic plasmas**. The fast progress in high-power laser technology enables advancing the study of relativistic plasmas from theoretical investigations and astrophysical observations into the realm of laboratory studies. In particular, the interaction of relativistic high-intensity lasers with dense plasmas promises the possibility of driving collisionless shocks, turbulence, and magnetic reconnection at relativistic temperatures [17,18]. External magnetic fields can be applied to control the initial plasma magnetization. The small spatial (microns) and short temporal (femtosecond) scales associated with collisionless relativistic plasma phenomena requires ultrashort, coherent x-ray sources to probe these processes in detail using a combination of x-ray Thomson scattering, phase-contrast imaging, small-angle x-ray scattering, and x-ray Faraday rotation diagnostics. Developing such well-diagnosed laser-driven plasma experiments and integrated fully kinetic simulations that can model realistic laboratory configurations will guide important developments in high-energy plasma astrophysical models [19,20], such as shown in Fig. 3.2.2.

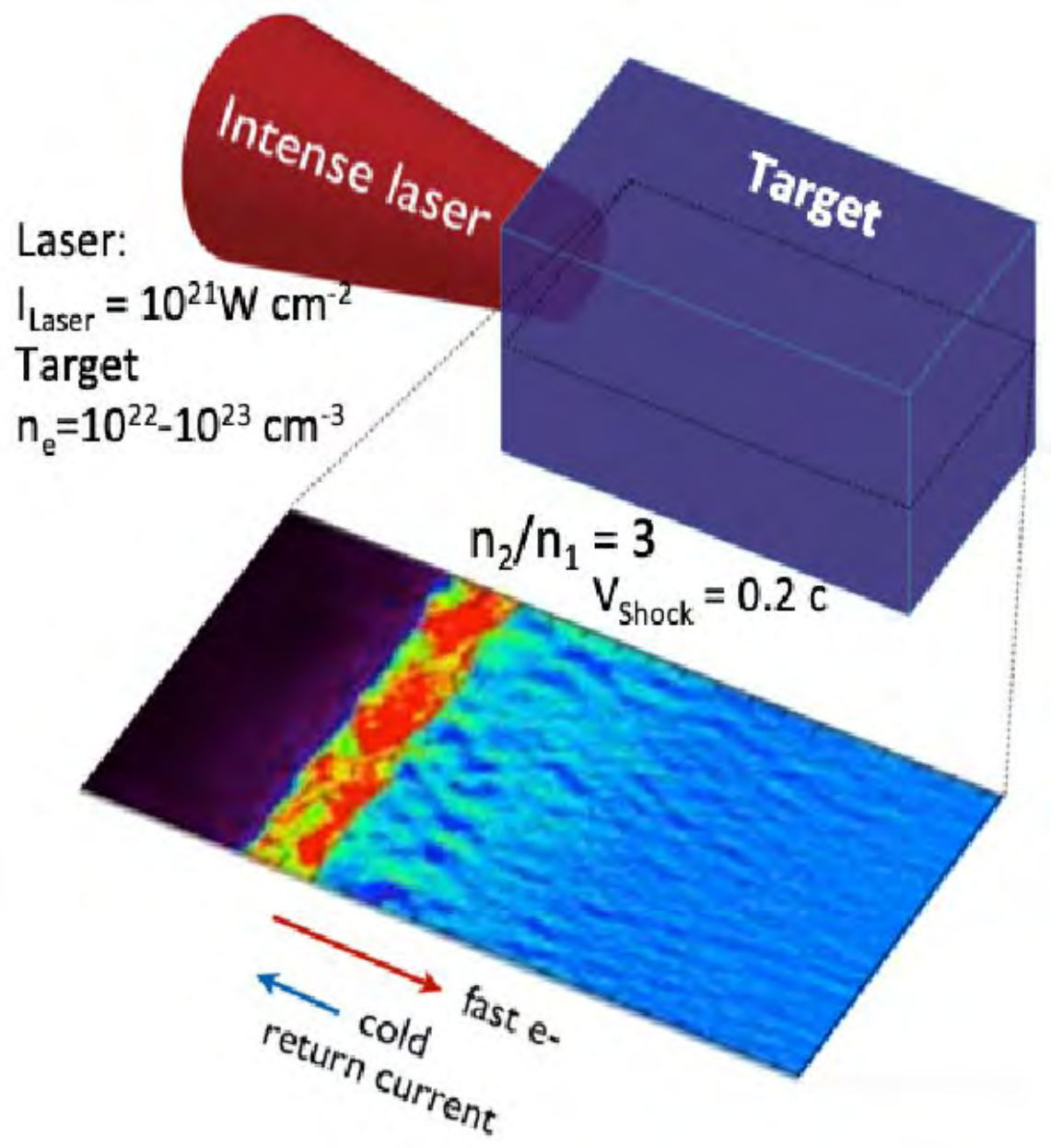

**Figure 3.2.2.** Particle-In-Cell (PIC) simulations show the formation of a collision-less shock wave by irradiation of a dense target with a multi-PW laser. High laser intensities are required for sufficiently long times to drive the instability [19]. The Weibel filaments form due to counter-propagating currents induced by fast laser accelerated electrons in the forward direction and a cold return current. The filaments coalesce into a shock with density compression by a factor of three which can be observed with ultrafast x-ray and particle imaging techniques to validate this physics picture.

Experiments play a critical role by (1) characterizing the dominant plasma processes that lead to magnetic-field amplification and turbulence in relativistic counter-streaming plasmas, (2) characterizing shock formation and particle acceleration in these systems, and (3) benchmarking the codes that are being used to explain observational data and develop astrophysical models.

---

17 M. Ackermann *et al.*, "Detection of the Characteristic Pion-Decay Signature in Supernova Remnants," Science **339** (6121), 807–811 (2013); A. R. Bell, "The Acceleration of Cosmic Rays in Shock Fronts – I," Mon. Not. R. Astron. Soc. **182** (2), 147–156 (1978); A. Marcowith *et al.*, "The Microphysics of Collisionless Shock Waves," Rep. Prog. Phys. **79** (4), 046901 (2016).

18 R. Blandford and J. P. Ostriker, "Particle Acceleration by Astrophysical Shocks," Astrophys. J **221**, L29–L32 (1978).

19 F. Fiuza *et al.*, "Weibel-Instability-Mediated Collisionless Shocks in the Laboratory with Ultraintense Lasers," Phys. Rev. Lett. **108** (23), 235004 (2012).

20 A. Spitkovsky, "Particle Acceleration in Relativistic Collisionless Shocks: Fermi Process at Last?" Astrophys. J. **682** (1), L5–L8 (2008).

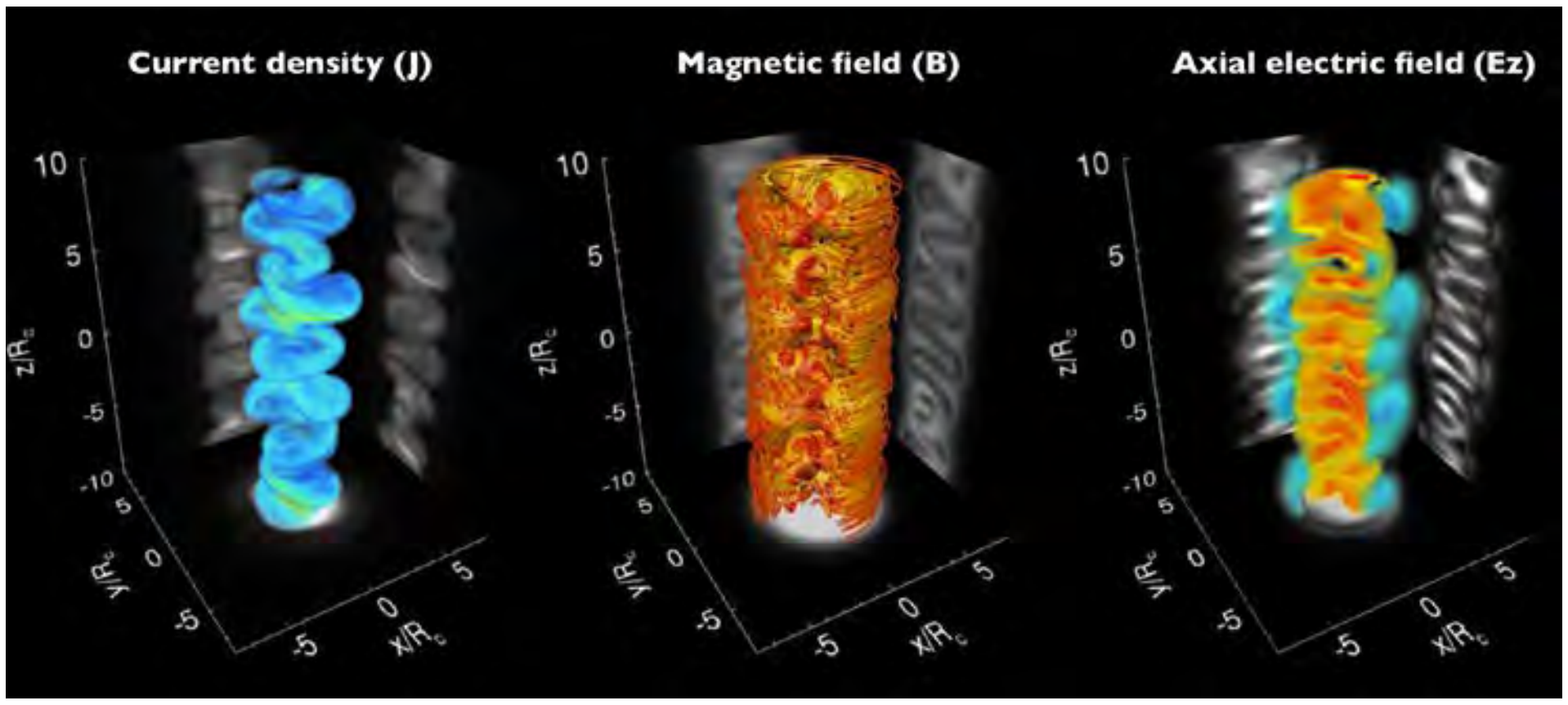

**Figure 3.2.3.** Particle In Cell (PIC) simulations show the formation of the kink instability in electron-positron jets that tangles the magnetic field lines and produces axial electric fields that produce acceleration gradients consistent with those required to explain the very high-energy cosmic rays. [24]

**Plasma jet instabilities**. Large-scale three-dimensional Particle-In-Cell (PIC) modeling is now routinely applied to predict extreme plasma states and their nonlinear evolution. These simulations show how magnetic instabilities in relativistic plasmas can efficiently accelerate high-energy particles, which is central to understanding the particle acceleration and radiation emission mechanisms of extreme astrophysical environments like the relativistic magnetized jets launched by active galactic nuclei (AGN). AGN jets are among the most powerful cosmic accelerators [21], but their particle acceleration mechanisms remain a mystery. While global magneto hydrodynamic (MHD) simulations have made great progress in identifying the dominant competing instabilities experienced by AGN jets during their propagation [22,16], the MHD framework is unable to determine if and how the development of such instabilities can produce the high-energy particles required to explain the observed radiation emission. A significant advance revealed that a nonlinear development of the kink instability (a fundamental instability in jets) [23] can efficiently convert the copious magnetic energy stored within the jet into high-energy radiating particles [24], as shown in Fig. 3.2.3. The particle spectra predicted by PIC simulations have been shown to be consistent with the measured non-thermal emission spectra in the optical and x-ray bands. Simulations further show that the kink instability can accelerate ions to the confinement energy of the jet, which can reach ~$10^{20}$ eV in some jets, such as in Knot A in the M87 giant elliptical galaxy.

21 A. Celotti and G. Ghisellini, “The Power of Blazar Jets,” Mon. Not. R. Astron. Soc. **385** (1), 283–300 (2008).
22 D. Giannios and H. C. Spruit, “The Role of Kink Instability in Poynting-Flux Dominated Jets,” Astron. Astrophys. **450** (3), 887–898 (2006); O. Bromberg and A. Tchekhovskoy, “Relativistic MHD Simulations of Core-Collapse GRB Jets: 3D Instabilities and Magnetic Dissipation,” Mon. Not. R. Astron. Soc. **456** (2), 1739–1760 (2016).
23 M. C. Begelman, “Instability of Toroidal Magnetic Field in Jets and Plerions,” Astrophys. J. **493** (1), 291–300 (1998).
24 E. P. Alves, J. Zrake and F. Fiuza, “Efficient Nonthermal Particle Acceleration by the Kink Instability in Relativistic Jets,” Phys. Rev. Lett. **121** (24), 245101 (2018).

Producing and observing these instabilities in the laboratory will be important to advance our understanding of the physics.

### 3.2.3 Meeting the challenges – technology requirements

This research area requires the capability to produce material pressures of several TPa and light pressures exceeding EPa. These laser driver capabilities will need to be matched with ultrafast high-precision x-ray and optical diagnostics to characterize phase boundaries, discover new materials, and measure the nonlinear evolution of plasma and material states.

#### 3.2.3.1 Increased repetition rate

The discovery of novel materials will require the capability to produce and probe plasmas over a wide range of parameters in P,T,V space (using tailored pressure and, consequently, laser pulses that can combine shock and ramp compression): a few shots per day is insufficient. In practice this means shaped nanosecond multi-kJ-class lasers operating at up to 10 Hz. Active feedback loops will need to tailor the laser pulse shape in real time with target handling solutions that support fast and precise delivery with target debris mitigation.

#### 3.2.3.2 Collocation with x-ray lasers

The compression laser must be sited alongside a source of coherent x-rays so that the material phases produced can be accurately diagnosed in terms of structure, density, and pressure, and new structures can be indexed. The x-ray source should be capable of producing photon energies in excess of 20 keV to allow sufficient coverage of reciprocal space (needed for structural determination) and be developed to have seeded and monochromatic ($\Delta E/E < 10^{-6}$) pulses to enable density and temperature measurements via inelastic scattering from plasmons and phonons. Additional capabilities to probe and dynamically perturb the states of matter require laser systems with 100-J to 1-kJ pulse energies, 100 fs to 1-ps pulse durations.

#### 3.2.3.3 Laser pulse shaping and drive uniformity

The driver requires flexible and high-fidelity laser pulse shaping over a range of time durations and intensity ranges to achieve near-isentropic laser ablation pressures over large target samples of millimeter scale. This requirement implies a laser driver with frequency doubling and tripling multiple kilojoules of pulse energy to compress targets with sufficient mass. Laser beam smoothing will be required to produce uniform drives with extremely small pressure nonuniformities.

### 3.2.4 Summary and Recommendations

Just a tiny fraction of the phase space of matter has been explored. Coupling cutting-edge high-repetition-rate, high-power optical lasers with an x-ray laser will enable creating and probing the structure of extreme states of matter in unprecedented detail and applying newfound knowledge to search for new metastable phases of matter that potentially can be recovered to ambient conditions. The high data-collection rates required will in turn drive the development of novel new sample delivery methods and tools for real-time analysis. The challenges are substantial, but the payoffs for both fundamental and applied material science are enormous.

In the astrophysical context, high-intensity lasers creating exciting opportunities to probe for the first time the processes associated with magnetic-field dynamics and particle acceleration of relativistic collisionless plasmas in the laboratory. In addition, new studies on magnetic reconnection and pair-plasma production will be of great importance and enabled by atomic scale spatial and femtosecond temporal resolution of x-ray lasers. This will require the combination of a high-energy, short-pulse laser that can drive targets on time scales up to picoseconds at ultrahigh laser intensities.

## 3.3 SRN-3 – Laser-plasma wakefield acceleration and related physics

New and novel particle acceleration mechanisms, whereby large numbers of high-energy particles ($\gg$ MeV to GeV) can be obtained over short distances (typically anywhere from < 1 $\mu$m to a few cm) are a cornerstone of short-pulse-laser science. This section covers science opportunities related to the laser wakefield acceleration (LWFA) of particles in underdense plasmas [25]. Almost all of the major advances in this field have been achieved with close integration of theory and experiment; this achievement is expected to continue in the next decade. Progress has been driven by a broad array of facilities at varying scales from smaller university systems to larger national laboratory devices that provide user access, support training, and enable innovation. This research has the potential to advance plasma science as well as impact many scientific disciplines that rely on accelerators as scientific tools, such as colliders and light sources, with applications spanning diverse areas from medicine to national security.

*Extreme acceleration gradients in laser-plasma accelerators can be leveraged for future applications and light sources that need low-emittance, high-brilliance beams by investing in short-pulse laser systems with kHz to MHz repetition rates.*

### 3.3.1 Opportunities for Frontier Science

#### 3.3.1.1 High-brilliance, low-emittance, >> GeV-class electron beams

While many important "firsts" have been demonstrated, including high-energy beams with low-energy spread and high transverse quality [26], laser-accelerated beam performance is still far from what is theoretically achievable. The grand challenge going forward is generating efficient phase-space–shaped particle beams and manipulating ultrabright beams well beyond the GeV. This needs to be done in various regimes scaling laser wavelength, intensity, and other parameters optimized to enable unique physics, bright injectors, and applications such as colliders, plasma-based XFELs, and high-field quantum electrodynamics (QED) studies.

#### 3.3.1.2 High-flux, high-repetition-rate, ~GeV electron beams

Compact high-energy electron beams have the potential to enable broad societal benefits in medicine, security, and industry by bringing the precision realized at very large accelerator

25 T. Tajima and J. M. Dawson, "Laser Electron Accelerator," Phys. Rev. Lett. **43** (4), 267–270 (1979); E. Esarey, C. B. Schroeder, and W. P. Leemans, "Physics of Laser-Driven Plasma-Based Electron Accelerators," Rev. Mod. Phys. **81** (3), 1229–1285 (2009).

26 S. P. D. Mangles *et al.*, "Monoenergetic Beams of Relativistic Electrons from Intense Laser–Plasma Interactions," Nature **431** (7008), 535–538 (2004); C. G. R. Geddes *et al.*, "High-Quality Electron Beams from a Laser Wakefield Accelerator Using Plasma-Channel Guiding," Nature **431** (7008), 538–541 (2004); J. Faure *et al.*, "A Laser–Plasma Accelerator Producing Monoenergetic Electron Beams," Nature **431** (7008), 541–544 (2004); J. Faure *et al.*, "Controlled Injection and Acceleration of Electrons in Plasma Wakefields by Colliding Laser Pulses," Nature **444** (7120), 737–739 (2006).

facilities to a variety of everyday applications. Bright beams with high charge, high repetition rate, and at modest energies well below those required for colliders or QED applications will enable a new class of photon sources.

#### 3.3.1.3 New photon sources from THz to gamma rays

Secondary sources, produced by GeV-class electrons from LWFA, have the potential to span the entire electromagnetic spectrum [27]. Processes including betatron emission, Thomson/Compton scattering, bremsstrahlung, free-electron laser generation, and coherent transition/THz radiation are all outcomes of LWFA providing near-term applications [28]. A particularly attractive feature of LWFA is the intrinsic synchronization between the drive laser beam, the generated particle bunch, and any secondary radiation. High-energy-density science and other areas of plasma physics need but do not have routine access to advanced photon probes that could offer greater resolution for precision in the spectral, temporal, and spatial domains.

The major classes of systems derived from these challenges and corresponding electron energy are summarized in Table 3.3.1.1.

**Table 3.3.1.1.** Primary accelerator drive laser parameter ranges of interest.

| **Electron energy** | **Laser energy** | **Pulse duration** | **Repetition Rate** | **Applications** | **Timeline** |
|---|---|---|---|---|---|
| 20 MeV | 5-20 mJ | 4 fs | kHz | Ultrafast electron diffraction, keV Thomson sources | 1-5 years |
| 1 GeV | 2-4 J | 30 fs | kHz | High performance LWFA, photon sources, 10 GeV collider stages, HED science | |
| 10 GeV | 7-80 J | 100 fs | 50 kHz | | 5-10 years |
| 50 GeV | 0.2 – 1 kJ | 250 fs | Hz | High field Physics | |
| Sub-GeV | 1 kJ | 50 fs | < Hz | HEDS, dynamic exp. | |

### 3.3.2 Underlying Physics

LWFA occurs through resonant excitation of large-amplitude electron plasma waves, which can be achieved by propagating relativistic laser pulses with durations comparable to the plasma wave period. The laser pulse displaces electrons with respect to the background ions by means of the ponderomotive force (radiation pressure). Because the transit time of the laser is fast compared to the ion plasma frequency, the ions remain immobile during the interaction. The resulting laser-excited plasma waves can be driven from linear to highly nonlinear regimes, depending on the laser intensity. The phase velocity of the plasma wave is near the group velocity

27 S. Corde *et al.*, "Femtosecond X Rays from Laser-Plasma Accelerators," Rev. Mod. Phys. **85** (1), 1–48 (2013).

28 F. Albert and A. G. R. Thomas, "Applications of Laser Wakefield Accelerator-Based Light Sources," Plasma Phys. Control. Fusion **58** (10), 103001 (2016).

of the laser in the underdense plasma, which allows such plasma waves to trap and accelerate particles to relativistic energies (Fig. 3.3.1). Plasma-based accelerators have attracted considerable attention as a potential next-generation accelerator technology, because the accelerating electric fields of the plasma wave can be orders of magnitude greater than conventional linear accelerators based on radio-frequency waves in metallic structures.

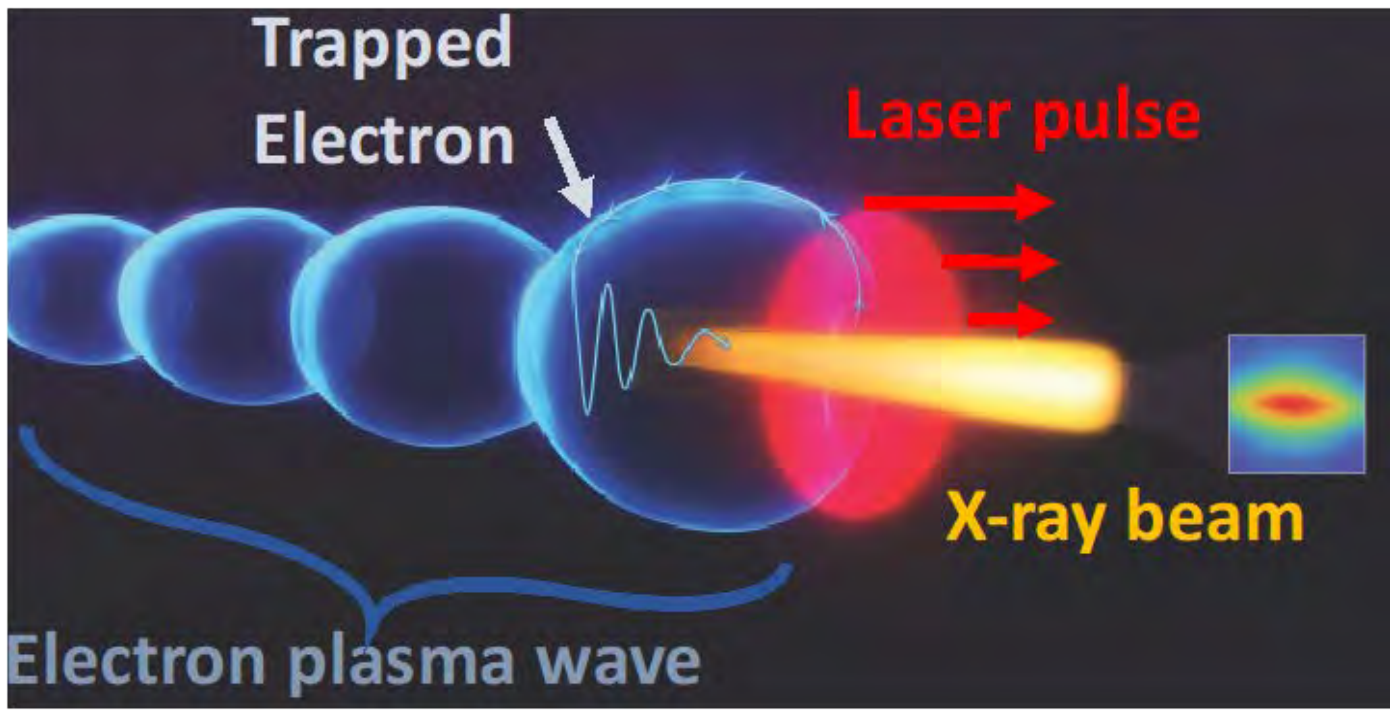

**Figure 3.3.1.** The figure illustrates the principle of laser-wakefield acceleration. An intense laser pulse propagates through a plasma and drives an electron-plasma wave. Electrons are pushed by the laser ponderomotive force, trapped in the plasma wave and accelerated to relativistic energies. Electrons trapped off the laser axis also oscillate, which causes them to emit synchrotron-like radiation. [29]

Over the past 15 years there has been tremendous progress in the field of LWFA. An important experimental milestone was achieved in 2004 when three laboratories reported the acceleration of background plasma electrons to produce quasi-monoenergetic electron beams at energies of approximately 100 MeV [26]. In 2006, 1-GeV electron beams were produced using 100 TW-class laser pulses in a few-cm-long plasma channel [30]. Today, many laboratories around the world can achieve laser-plasma–accelerated electron beams with energy ranging from 100 MeV to 8 GeV with 10s to 100s of pC per bunch [31]. The beam-energy gain in a laser-plasma accelerator is determined by the amplitude of the excited plasma wave and the interaction distance. The quality of the accelerated electron beam is determined by how background plasma electrons are trapped in the accelerating and focusing bucket of the plasma wave.

One of the most promising applications of LWFA is the generation of novel sources of photons, with applications in medicine, high-energy-density and material sciences, and national security [27]. Five types of light sources, spanning a large range of photon energies (from THz to gamma rays) can be produced by LWFA. **Betatron motion, Compton scattering, and undulators** produce x-rays or gamma rays when relativistic electrons from the wakefield oscillate behind the laser pulse, in a counter-propagating laser field, or a magnetic undulator, respectively. **Bremsstrahlung radiation** from the interaction of relativistic LWFA electrons with a solid target, and **transition radiation** at plasma/vacuum interfaces also produce photons in the gamma-ray and terahertz energy range, respectively. As noted elsewhere in this report, such sources have broad impact across plasma science and societal applications. In the short term, multi-beam laser facilities can offer ultrafast, brilliant incoherent probes via these mechanisms to complement x-ray lasers or to support experiments at facilities where such coherent sources are not available. In the mid to longer term, plasma-driven FELs and XFELs present exciting opportunities.

---

29 F. Albert, "Laser Wakefield Accelerators: Next-Generation Light Sources," Optics and Photonics News, **29** (1), 42–49 (2018).

30 W. P. Leemans *et al.*, "GeV Electron Beams from a Centimeter-Scale Accelerator," Nat. Phys. **2** (10), 696–699 (2006).

31 A. J. Gonsalves *et al.*, "Petawatt Laser Guiding and Electron Beam Acceleration to 8 GeV in a Laser-Heated Capillary Discharge Waveguide," Phys. Rev. Lett. **122** (8), 084801 (2019).

Even with great progress in using lasers for LWFA to achieve up to 8-GeV electrons [28], the most exciting opportunities still lie ahead. Advancing the field to applications requires precision that will push laser capabilities in pulse shaping, beam control, repetition rate, and efficiency. The grand challenge for laser wakefield acceleration will be phase-space–shaped particle beam generation and manipulation of ultrabright beams and that enable applications.

#### 3.3.2.1 Electron trapping

The quality of the accelerated electron beam is determined by how plasma electrons are trapped in the accelerating and focusing bucket of the plasma wave. Several methods of initiating electron trapping have been proposed that offer the potential to improve the beam quality (e.g., beam emittance and brightness) significantly beyond state-of-the-art particle sources. This represents an important area of plasma physics research over the next decade to improve beam brightness. Methods include (i) millimeter plasma bubbles to be used for sub-1%-energy-spread GeV electron acceleration with external electron injection, (ii) controlled multi-pulse injection, and (iii) methods relying on ionization injection, using multipulse, multicolor, and colliding laser pulse schemes.

#### 3.3.2.2 Large-amplitude nonlinear electron plasma waves over long distances

Sustaining large-amplitude nonlinear electron plasma waves over long distances requires a greater understanding of nonlinear laser–plasma interactions, including energy deposition and laser propagation physics, relativistic self-focusing, ponderomotive self-channeling, interactions with preformed plasma channels, and short-pulse laser–plasma instabilities, such as laser self-modulation and hosing. Methods center on correcting for laser fluctuations by tailoring pulses from current PW and TW lasers, applying LWFA active feedback to achieve spatial and temporal laser pulse shaping, controlling carrier-envelope phase effects, and spatially tapering plasma densities.

#### 3.3.2.3 Staging and guiding

Preserving high beam quality over many LWFA stages and tens of meters of plasma, requires critical attention to beam instabilities and alignment. Collider applications require studies to map the electron beam phase space in a plasma using radiation generated by the electrons and to mitigate hosing and beam break-up.

LWFA regimes identified for both colliders and photon sources require external guiding by shaping the plasma density into a channel, in addition to self-guiding provided by the high laser intensity, to achieve the required efficiency. Optical Field Ionization (OFI) is one possible route to generate long, low-density plasma channels at high repetition rates; discharge capillaries and shaped flows are other approaches.

#### 3.3.2.4 New regimes of underdense laser-plasma acceleration

Novel regimes of laser pulse duration or wavelength, or plasma density may offer new capabilities. For example, longer laser wavelengths reduce the plasma density for a given electron dephasing energy, which raises the effective stage charge. New research are needed to study: LWFA at longer wavelengths or pulse durations; the use of temporally structured laser drivers, single-stage 40–GeV LWFA in channels, large plasma bubbles for easier external injection, and direct laser acceleration.

#### 3.3.2.5 Light source development and applications

The development of new light sources and their key parameters (emittance, energy, photon flux, spatial and temporal resolution) must be closely coordinated with scientific applications in high-energy-density science; planetary, material, and astrophysical sciences; and nuclear photonics to enable new discoveries.

### 3.3.3 Meeting the Challenges – Laser Capability Requirements

Proof-of-principle experiments can be realized at low repetition rates, but most research directions ultimately require a high repetition rate (kHz and above), to enable both applications and active laser feedback for precision control. This defines a core accelerator development track: establishing a few-joule, multi-kHz system enabling high-precision LWFA with feedback for stabilization that could lead to a future 10-J/50-kHz collider stage driver, as well as next-generation light sources. Additional laser beams are desired for injection and guiding control but they are typically lower energy and do not therefore drive overall laser development. There are three significant areas with distinct laser needs:

- For colliders, while the main stages must operate at high efficiency (and have charge and laser energy set by interaction point physics), the injector stage may benefit from different regimes of operation such as long-wavelength drivers to enable advanced injectors.
- For low-energy applications, such as medicine and ultrafast electron diffraction, few-fs laser systems are needed at modest energies in the 10-mJ class.
- Stages producing tens of GeV electron beam energies are required for high-field physics, and very high charge electron beams are of interest for HEDS photon probes, both of which motivate higher peak powers.

### 3.3.4 Summary and Recommendations

Laser-plasma acceleration has made great progress; advancing this technology to realize applications requires high precision that will push laser capabilities in pulse shaping, beam control, repetition rate, and efficiency. The range of needs motivates a range of facilities addressing different capabilities and approaches. A near-term priority is developing a few-joule, kHz laser system for laser wakefield acceleration with closed-loop stabilization that will enable light sources and ultimately lead toward a future 10-J/50-kHz collider stage driver. Additional lasers for electron injection and guiding control will play significant roles and affect LWFA facility configurations. Long-wavelength or high-energy long-pulse (ns) systems will provide LWFA electron probe and heating sources, respectively, for the broad range of science described in this report.

## 3.4 SRN-4A – High-Field Physics and Quantum Electrodynamics (QED)

As peak laser intensity continues to increase since the invention of chirped-pulse amplification, and understanding the collective properties of plasmas advances, laser–plasma interaction studies enter a new regime where the physics of relativistic plasmas is strongly affected by strong-field quantum electrodynamics (QED) processes, including hard photon emission and electron–positron ($e^+e^-$) pair production [32]. Moreover, nonlinear quantum effects emerge at field strengths much lower than the QED critical field strength. This coupling of quantum emission processes and relativistic collective particle dynamics can result in dramatically new plasma physics phenomena, such as generating dense $e^+e^-$ pair plasmas from near vacuum, or even stopping ultrarelativistic electron beams using a hair's breadth of laser light that would otherwise penetrate a centimeter of lead. It is crucial to study this new regime with the next generation of ultra-high-intensity laser–matter experiments and to develop new applications, such as high–energy ion and photon sources for medical radiotherapy or next-generation radiography for homeland security and industry.

> *New ultra high-intensity lasers will allow reaching novel physics regimes dominated by strong-field quantum electrodynamics processes that are characterized by high-energy gamma emission and electron-positron pair plasmas that may be tailored for applications.*

### 3.4.1 Opportunities for Frontier Science

Strong-field QED (SF-QED) describes the physics of strong electromagnetic (EM) fields and is broadly characterized by environments where the strength of electric fields is large relative to the QED critical field $E_{cr} = m_e^2c^3/e\hbar$, which is the field that *classically* would accelerate an electron to its rest mass energy in a Compton length ($h/mc$ = $2.4\times10^{-12}$ m), the wavelength of a photon whose energy is the same as the rest mass of the electron [33]. At fields of this strength, QED processes are highly nonlinear and cannot be described by straightforward perturbation theories. Strong-field QED processes have for a long time been believed to be the domain of elementary particle physics, which has made tremendous progress in the last 100 years from the formulation of basic laws and the construction of the first particle accelerators to the creation of the elaborate Standard Model and its experimental verification at grand-scale experimental facilities, such as LEP, LEP II, SLAC, Tevatron, and LHC. At the beginning of the 21st century, there appeared a demand to understand the cooperative behavior of relativistic quantum systems

32 G. A. Mourou, T. Tajima, and S. V. Bulanov, "Optics in the Relativistic Regime," Rev. Mod. Phys. **78** (2), 309–371 (2006); M. Marklund and P. K. Shukla, "Nonlinear Collective Effects in Photon-Photon and Photon-Plasma Interactions," Rev. Mod. Phys. **78** (2), 591–640 (2006); A. Di Piazza *et al.*, "Extremely High-Intensity Laser Interactions with Fundamental Quantum Systems," Rev. Mod. Phys. **84** (3), 1177–1228 (2012).

33 J. Schwinger, "On Gauage Invariance and Vacuum Polarization," Phys. Rev. **82** (5), 664–679 (1951).

and the basic properties of the quantum vacuum. It was realized that one of the possible ways of studying these phenomena applies high-power laser systems and collective properties of plasmas. Consequently, the study of the physics of laser–plasma interactions in completely new regimes may contribute to studies of the most fundamental properties of nature and lead to new sources of photons and charged particles.

The highest laser intensities demonstrated to date are orders of magnitude lower than necessary to reach the critical field $E_{cr}$. However, since the electric field is not Lorentz invariant, a subcritical field may be boosted to the critical field strength in the rest frame of an ultrarelativistic particle; therefore, SF-QED processes, such as multiphoton Compton emission and Breit–Wheeler pair production, can occur at significantly lower field strengths than $E_{cr}$. Realizing that laser–plasma interactions receive a boost from the fact that the laser accelerates the particles to relativistic energies and provides the strong-field strengths led to seminal theoretical breakthrough [34] predicting prolific pair-plasma production at fields orders of magnitude below $E_{cr}$.

As has been known for many years, electrons reach their rest mass energy in laser field strengths exceeding $I\lambda^2 = 10^{18}$ Wcm$^{-2}\mu$m$^2$, and the plasma becomes relativistic. More recently, it was discovered that the interaction between charged particles and EM fields enters a radiation-dominated regime at laser intensities above $I\lambda^2 = 10^{23}$ Wcm$^{-2}\mu$m$^2$ [35], which completely changes the dynamics of individual particles, as well as the collective properties of plasmas. At still higher laser intensities, interactions become highly nonlinear and other quantum effects come into play until eventually vacuum breakdown becomes possible. These effects represent part of an emerging branch of physics that employs both strong-field effects, and both classical and quantum electrodynamics [36] that is largely unexplored both theoretically and experimentally. The study of strong-field effects is generally important for understanding nonperturbative quantum field theories. It opens up the possibility of a controllable study of matter in extreme conditions with applications to high-energy hadron interactions and the creation of quark-gluon plasmas. Strong-field effects in connection with plasmas are also crucial to understanding astrophysical phenomena involving neutron stars, magnetars, and black holes, and will also be important for future lepton colliders.

Figure 3.4.1 illustrates the rich physics of SF-QED effects connected with laser–plasma interactions parameterized in terms of the electron density $n_e$ and the field strength (classical nonlinearity parameter) $a = eE/mc\omega$, where $\omega$ is the laser frequency. The following thresholds are shown:

---

34 A. R. Bell and J. G. Kirk, "Possibility of Prolific Pair Production with High-Power Lasers," Phys. Rev. Lett. **101** (20), 200403 (2008).

35 S. V. Bulanov *et al.*, "On the Design of Experiments for the Study of Extreme Field Limits in the Interaction of Laser with Ultrarelativistic Electron Beam," Nucl. Instrum. Methods Phys. Res. A **660** (1), 31–42 (2011).

36 S. S. Bulanov *et al.*, "Electromagnetic Cascade in High-Energy Electron, Positron, and Photon Interactions with Intense Laser Pulses," Phys. Rev. A **87** (6), 062110 (2013).

- Radiation dominated regime: This is where the radiation reaction effective force approaches the same order strength as the Lorentz force.
- Quantum effects begin to dominate as the field strength approaches the QED critical field for an electron in a rotating electric field (the standing wave formed by colliding circularly polarized laser fields or a circularly polarized laser reflecting from a dense target) is boosted to the critical field limit in its rest frame.
- $e^+e^-$ pair cascades start at values of a~$10^3$ where the electric field experienced by the electrons/generated positrons in their rest frame is many times the critical field.
- The Ritus-Narozhny $\alpha(\gamma a/aS)^{\frac{2}{3}}$= 1 threshold, where the field is strong enough that strong-field perturbation theory breaks down, is reached at a = 2.5x$10^4$.
- The field in the laboratory frame reaches the critical field strength, E=$E_{cr}$, at a = 4.1x$10^5$.
- The laser field is depleted due to the QED-interaction with dense electron/positron plasma.

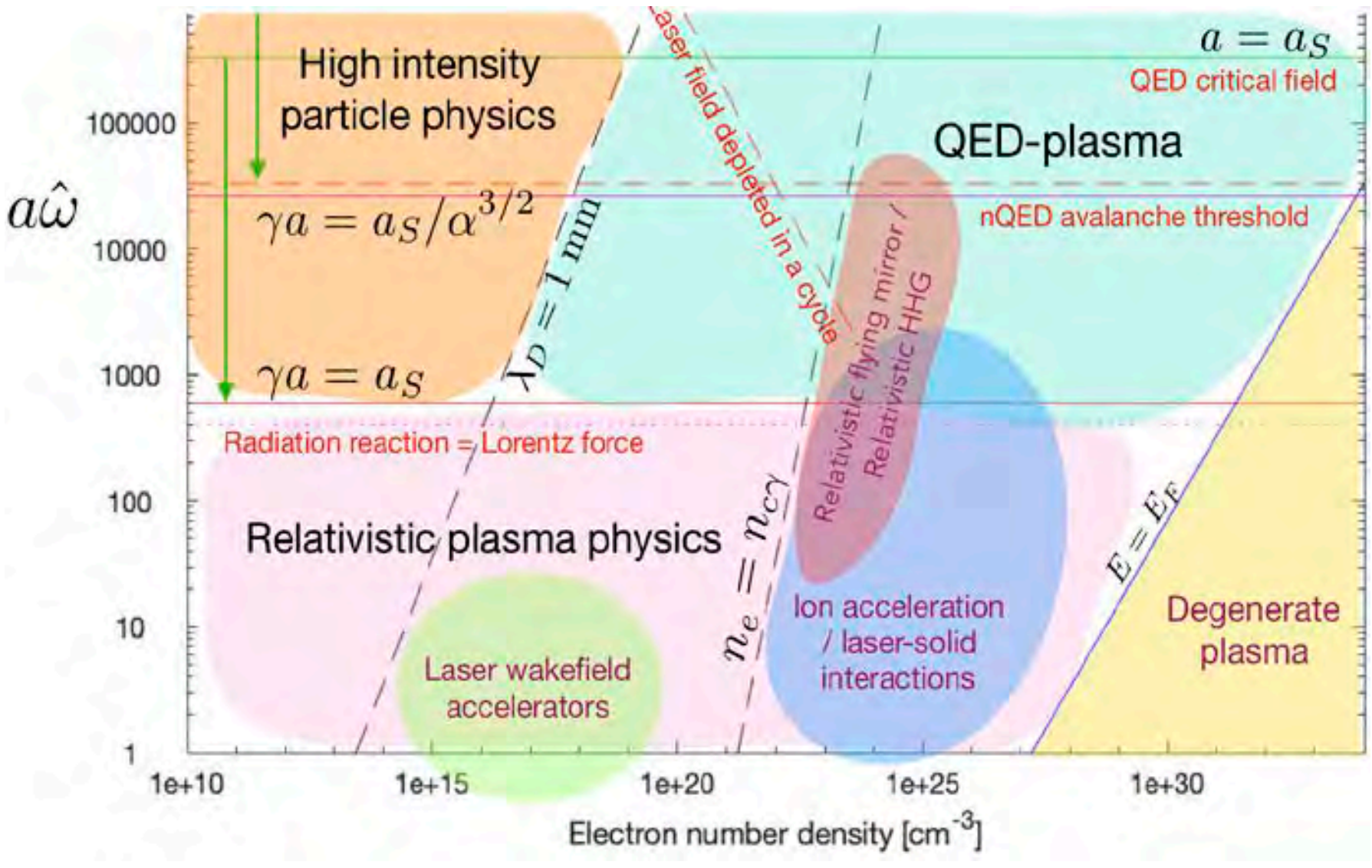

**Figure 3.4.1.** Regimes of strong field QED for plasma with densities ranging over 24 orders of magnitude. $a$=$eE/mc\omega$ is the normalized field strength and $\hat{\omega}$ is the laser frequency normalized to that of a 1 $\mu$m wavelength laser.

The ($n_e$, $a$) plane is subdivided into distinct regimes of behavior: "relativistic plasma physics," which is mostly the domain of classical electrodynamics and relativistic plasma physics; "high-intensity particle physics," which is the domain of SF-QED noncollective interactions; "QED plasma," which is the domain where plasma physics and SF-QED couple together to result in new phenomena that is bounded by the "degenerate plasma" state where optical lasers do not propagate.

Several regions correspond to important applications of laser–plasma interactions, some of which may be affected by the SF-QED effects: laser wakefield acceleration, ion acceleration/laser solid interactions, relativistic flying mirrors, and relativistic high-order harmonic generation. A significant part of the plot is *"terra incognita"* with the QED-plasma domain occupying a significant portion. Even a part of high-intensity particle physics is bordered by $\alpha\chi^{2/3} = 1$ [37] and laser depletion thresholds [38].

### 3.4.2 Underlying Physics and State of the Art

In the last 10 years, the study of SF-QED has greatly advanced understanding of charged-particle interactions with intense EM fields. Particle-in-Cell (PIC) methods [39] rely on the separation of scales, where the characteristic scale of the SF-QED emission processes is much smaller than the scales of the EM field or plasma phenomena. QED-PIC codes are extensively used to study the interactions of high-intensity EM fields with energetic beams of charged particles, photons, and plasmas of different composition and density. Apart from either enhancing or suppressing particle acceleration (see, e.g., [40]), the strong fields were found to modify the trajectories of charged particles in the radiation-dominated regime, trapping $e^+$ and $e^-$ on stable or quasi-stable trajectories inside the EM fields [41].

As new results piled up, an understanding developed that the models used are not capable of describing new interaction regimes. A good example is the local-constant-field approximation (LCFA), which is the backbone of almost all numerical tools employed. Recent studies identify the parameter region where LCFA predictions significantly differ from the full QED calculations. A number of solutions to this problem were proposed by modifying the LCFA [42]. Another example is the interaction of charged particles with super-strong EM fields, where perturbation theory no longer applies, since contribution of second-order processes become comparable with the first-order ones of the so-called "$\alpha\chi^{2/3} > 1$" regime [37]. Proper treatment of spin and polarization effects on plasma dynamics also merits study [43].

---

37 V. I. Ritus, "Radiative Corrections in Quantum Electrodynamics with Intense Field and Their Analytical Properties," Ann. Phys. **69** (2), 555–582 (1972); N. B. Narozhny, "Expansion Parameter of Perturbation Theory in Intense-Field Quantum Electrodynamics," Phys. Rev. D **21** (4), 1176–1183 (1980).

38 D. Seipt *et al.*, "Depletion of Intense Fields," Phys. Rev. Lett. **118** (15), 154803 (2017).

39 C. P. Ridgers *et al.*, "Modelling Gamma-Ray Photon Emission and Pair Production in High-Intensity Laser–Matter interactions," J. Comp. Phys. **260**, 273–285 (2014); A. Gonoskov *et al.*, "Extended Particle-in-Cell Schemes for Physics in Ultrastrong Laser Fields: Review and Developments," Phys. Rev. E **92** (2), 023305 (2015).

40 P. Zhang *et al.*, "The Effect of Nonlinear Quantum Electrodynamics on Relativistic Transparency and Laser Absorption in Ultra-Relativistic Plasmas," New J. Phys. **17**, 043051 (2015).

41 A. Gonoskov *et al.*, "Anomalous Radiative Trapping in Laser Fields of Extreme Intensity," Phys. Rev. Lett. **113** (1), 014801 (2014); T. Zh. Esirkepov *et al.*, "Attractors and Chaos of Electron Dynamics in Electromagnetic Standing Wave," Phys. Lett. A **379** (36), 2044–2054 (2015); J. G. Kirk, "Radiative Trapping in Intense Laser Beams," Plasma Phys. and Contr. Fusion **58** (8), 085005 (2016).

42 T. G. Blackburn *et al.*, "Benchmarking Semiclassical Approaches to Strong-Field QED: Nonlinear Compton Scattering in Intense Laser Pulses," Phys. Plasmas **25** (8), 083108 (2018); A. Di Piazza *et al.*, "Implementing Nonlinear Compton Scattering Beyond the Local-Constant-Field Approximation," Phys. Rev. A **98** (1), 012134 (2018); A. Ilderton, B. King, and D. Seipt, "High-Intensity QED beyond the Locally Constant Field Approximation," High Energy Physics Archive, available at https://arxiv.org/abs/1808.10339v1 (2018).

43 D. Del Sorbo *et al.*, "Spin Polarization of Electrons by Ultraintense Lasers," Phys. Rev. A **96** (4), 043407 (2017).

#### 3.4.2.1 Beyond the plane-wave approximation

Most researchers analyzing different processes in strong laser fields simulate a laser pulse field by a plane monochromatic wave. In this case, the Dirac equation for the electron in the field of the plane EM wave has exact solutions, which enable one to obtain quite easily analytic formulas for the probabilities of quantum processes. However, there are processes that cannot be described by the plane-wave model and a self-consistent treatment of these processes in the case of a non-plane wave is still missing.

#### 3.4.2.2 Beyond the local-constant-field approximation

Present-day laser systems achieve very high intensities localized in a small space–time volume with a characteristic size of a wavelength, and the EM field has a complicated, 3-D structure. If the formation length or time of a process is much smaller than the respective spatial and time inhomogeneities of the laser pulse, the local probability of the process can be calculated in the framework of a plane-wave or constant-field model, which is almost always the case with the present-day calculations; otherwise, finite-size effects must be taken into account explicitly.

#### 3.4.2.3 Beyond the external-field approximation

One of the open questions is how back reactions affect pair production or photon emission in the intense EM field. Usually these processes are considered using an external field approximation; however, the creation of new particles can lead to the depletion of the EM field energy, which invalidates the approximation of the external field.

#### 3.4.2.4 Multistage processes

The most-straightforward examples of such processes are avalanche and shower cascades [44], which are fascinating phenomena of fast transformation of laser and/or charged-particle beam energy into high-energy photons and $e^+e^-$ pairs, respectively. Theoretical and simulation studies of the cascades up to now have relied on the fact that formation length/time is much smaller than the respective spatial and time inhomogeneities of the electromagnetic field.

### 3.4.3 Meeting the Challenges—Facility and Laser Capability Requirements

The above unanswered questions together with the need to develop *numerical tools* make a strong case for further theoretical and numerical studies of SF-QED to advance understanding of EM field interactions with plasmas at highest intensities. Such development requires a concentrated experimental effort to test theoretical and numerical models, to pave the way for future applications, and to explore new phenomena. Realizing the facilities to study the SF-QED physics regime requires a staged approach.

---

44 A. A. Mironov, N. B. Narozhny, and A. M. Fedotov, "Collapse and Revival of Electromagnetic Cascades in Focused Intense Laser Pulses," Phys. Lett. A **378** (44), 3254–3257 (2014).

A first stage will study the basic quantum processes of strong-field QED in the high-intensity particle physics regime together with relativistic plasma physics phenomena. This can be carried out at a PW-class laser facility featuring an additional colliding electron beam that could be provided by two laser beamlines with one of them for particle acceleration, or with a laser and a conventional electron beam. The main laser beamline with power $P_{laser}$ with near diffraction-limited focusing should satisfy $E_{beam}[GeV]\sqrt{P_{laser}[PW]}\,/\,\lambda_{laser}[\mu m] >> 1$, where $E_{beam}$ is the electron beam energy.

A later stage to study the QED-plasma regime with the ultimate goal of "producing plasma from light" requires multi-10 PW to EW-class laser facilities able to deliver multiple laser pulses to the interaction point at extreme intensities. Assuming these lasers achieve nearly diffraction-limited focusing, the laser power should satisfy $P_{laser}[PW]\,/\,\lambda_{laser}[\mu m] >> 10$ to fully enter this regime.

### 3.4.3.1 Near-term experiments

Multi-beam PW-class laser facilities are needed to study the basic properties of the electron, positron, and photon interaction with intense laser pulses (multiphoton Compton and Breit–Wheeler processes) by colliding intense lasers with electron beams produced by plasma acceleration. Different theoretical and numerical models must be tested, and the limits of their applicability must be established. This work will require two petawatt-class beamlines. The first U.S. laser facility with 0.5-PW and 2.5-PW beamlines (NSF ZEUS) is under construction and funded through the NSF mid-scale infrastructure initiative.

### 3.4.3.2 Far-term experiments

Multibeam-PW- to EW-class laser facilities are needed to study the development of electromagnetic cascades and unexplored QED–plasma interactions, as well as to test the theoretical models going beyond the plane wave, the external field, and local constant-field approximations. The scientific case for new high-power laser facilities is often the acceleration of particles and generation of new sources of radiation; since the regimes of these applications will be affected or even dominated by SF-QED–plasma interactions, it is of paramount importance that such studies become an integral part of the scientific program of such facility concepts described in Secs. 4.2.3.2 and 4.2.4.

### 3.4.4 Summary and Recommendations

Strong-field quantum electrodynamics in plasma is a new field of research where new phenomena result from coupling quantum emission processes with relativistic plasma dynamics. It has seen significant theoretical excitement but few experiments so far, and none that have pushed into the regime where $\chi > 1$. The frontier experiments exploring this new state of matter will require substantial upgrades to facilities. The entry level to this field is to couple current state-of-the art PW-class lasers to ultrarelativistic electron beams to explore the quantum processes, pushing the laser capabilities in stability and repetition rate to enable *precision* experiments. The electron beam could leverage existing conventional particle accelerators or employ laser plasma wakefield accelerators. Multi-10 PW near-IR lasers will be needed to initiate experimental studies of avalanche-type cascades of positron-electron plasma, which is one of the more dramatic predictions of SF-QED theory. Moderate repetition rates are desirable, and extremely high pointing, temporal and phase stability will be required to overlap multiple tightly focused laser and particle beams. Strong-field QED also requires dedicated experimental efforts coupled to developing new computational tools and simulation capabilities to benchmark experimental results against new models.

## 3.5 SRN-4B – High Harmonic Generation and Attosecond science

Intense ultrafast lasers can be used for powerful, table-top-scale sources of extreme ultraviolet and soft x-ray radiation through the process of high-order harmonic generation (HHG). Such HHG sources enable the study of atomic, molecular, and solid-state material systems with unique chemical, spatial, and temporal resolution, and yield insights into how these systems function. Taking HHG sources to the highest temporal resolution has enabled important new science in the attosecond time domain associated with fundamental electron motion in matter [45].

Advancing science that takes advantage of this technology requires a research model that integrates advances in laser technology and HHG development while addressing the scientific challenges. This synergy of technology and scientific application can be accomplished by multiple approaches, which can exploit single investigator laboratories, mid-scale centers that support multidisciplinary collaborative projects and teams, or large user facilities that provide broad access to the scientific community. Examples of the latter two approaches include the mid-scale, NSF NeXUS facility that recently started construction at the Ohio State University, which is dedicated to HHG/attosecond science, and the ELI-ALPS facility in Hungary, a large facility that will push the frontiers technology and support a wide range of users.

HHG sources complement those of synchrotron and XFEL light source user facilities, with each having unique characteristics. Synchrotrons provide many portals of light extending from the infrared, through the soft x-ray, to the hard x-ray regime, with high average flux. XFELs are also accelerator-based sources and yield a smaller number of beams but with extremely intense, short duration bursts of light and increasingly high average flux. HHG sources are more compact, are expected to continue to yield the shortest, controlled pulses, and a novel capability to sculpt and control short wavelength light. They are also amenable to use in the laboratories of individual investigators.

*Novel fully coherent light sources that provide bright pulses for applications in the x-ray regime and with short attosecond durations require investments in new high-average-power, few-cycle lasers in the near- to mid-infrared wavelength ranges.*

### 3.5.1 Opportunities for Frontier Light Science

#### 3.5.1.1 High Harmonic Generation and Attosecond science

The interaction of intense light with matter, or strong-field science, is an active field of physics, made possible by the development of intense lasers. HHG is an example of such highly non-linear processes, which makes it possible to efficiently upshift intense visible and mid-infrared

45 P. Agostini and L. F. Dimauro, "The Physics of Attosecond Light Pulses," Rep. Prog. Phys. **67** (6), 813–855 (2004).

(mid-IR) laser light to shorter wavelengths. It has been known for decades that scaling laws of laser-matter interaction favor the use of mid-IR laser wavelengths to generate high-order harmonics: specifically, the ponderomotive energy of an electron, the HHG single-atom cutoff, and the critical power for self-focusing all scale as $\lambda^2$. Recent work has leveraged this scaling law to demonstrate efficient, phase-matched generation of >keV harmonics driven by a 4-µm intense femtosecond laser. Extrapolating performance shown in Figure 3.5.1 suggests that using ultrashort-pulse lasers in the far-IR (15- to 20-µm wavelength) might produce coherent *hard* x-ray sources that would create exciting opportunities, such as probing time-scales of a zeptosecond (a thousandth of a billionth of a billionth of a second), potentially relevant to nuclear processes ).

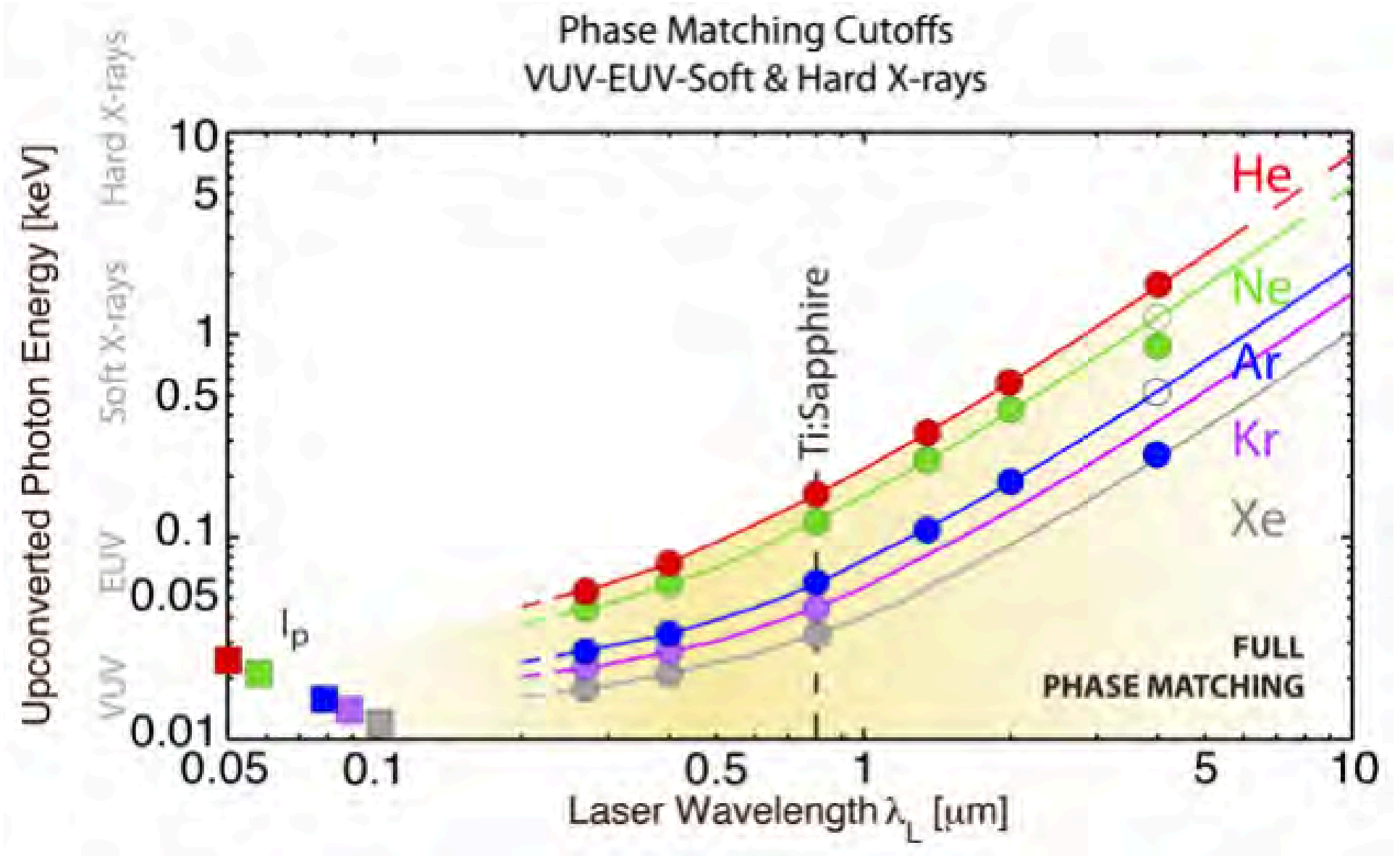

**Figure 3.5.1** Phase matching of high harmonic generation: experiment [46] confirms predictions on wavelength scaling of HHG [47]. The continuous curves show theoretical scaling of HHG spectral cutoff vs driving laser wavelength, while the dots correspond to experimental confirmations. Most remarkably, the use of mid-IR lasers—with lower photon energy— results in the generation of higher energy soft x-rays. In experiment, coherent >keV energy x-ray photons result each from the coherent combination of *>5000* photons from the laser.

Fundamental questions remain for extending current HHG sources to higher photon energies. The recollision process requires maintaining coherent entanglement of the electron and ion for ~100 fs in a dense plasma/gas environment. Maintaining quantum coherence has been demonstrated for HHG in the keV photon regime with a ~4 µm laser. Simple calculations show that the free-electron excursion passes many neighboring atoms in the high-pressure gas medium,

46 P. Colosimo *et al.*, "Scaling Strong-Field Interactions Towards the Classical Limit," Nat. Phys. **4** (5), 386–389 (2008).

47 J. Tate *et al.*, "Scaling of Wave-Packet Dynamics in an Intense Midinfrared Field," Phys. Rev. Lett. **98** (1), 013901 (2007).

so quantum coherence in high energy-density plasmas seems problematic. Additionally, the interplay of nonlinear beam propagation and high harmonic generation plays a critical role for HHG, so understanding stable propagation with self-focusing and phase matching will be needed.

Finally, it is important to understand that strong field physics supports going beyond the predictions of Figure 3.5. 1. Other recent work has shown that the use of ultra-intense pulses in the deep-UV can also be used effectively for HHG [48]. This corresponds to a new regime of much higher intensity interactions where the nonlinear response of the medium alters the driving laser propagation, but the detailed physics are far from understood or optimized.

### 3.5.1.2 High harmonic source polarization and waveform control

In recent years, it has become apparent that HHG allows for coherent manipulation of short-wavelength light with the highest level of control. Examples include the generation and characterization of complex polarization states of HHG light, [49] and the recent generation of light pulses of time-varying orbital angular momentum—a new "self-torque" property of light (Figure 3.5.2). These new states of light promise to find application in imaging and manipulating magnetic, chiral, and topological materials that may be the basis for new computing technologies.

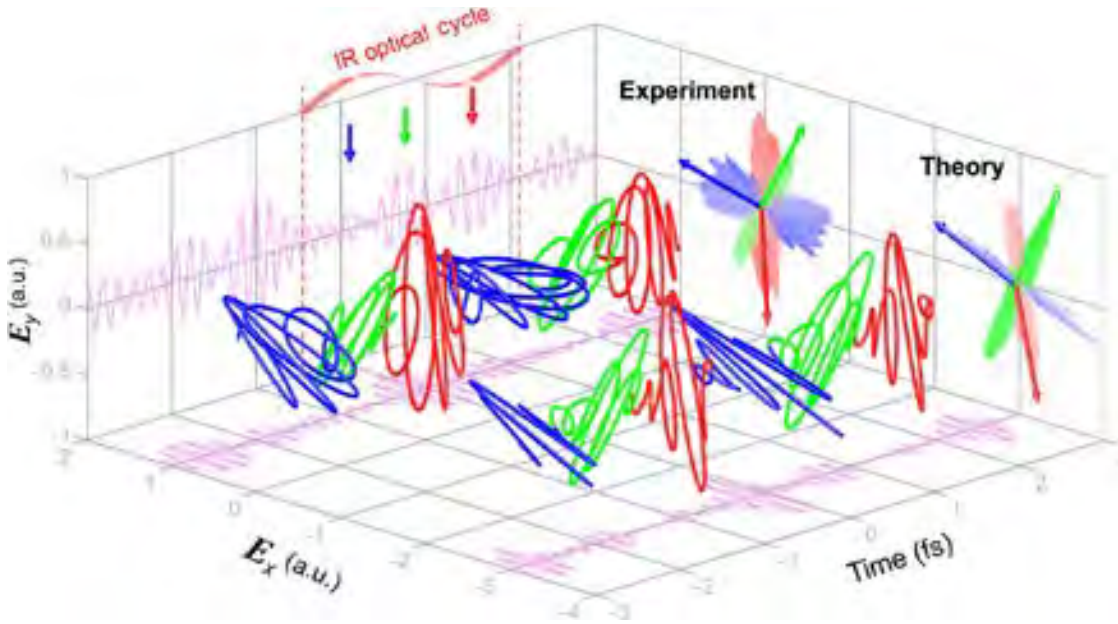

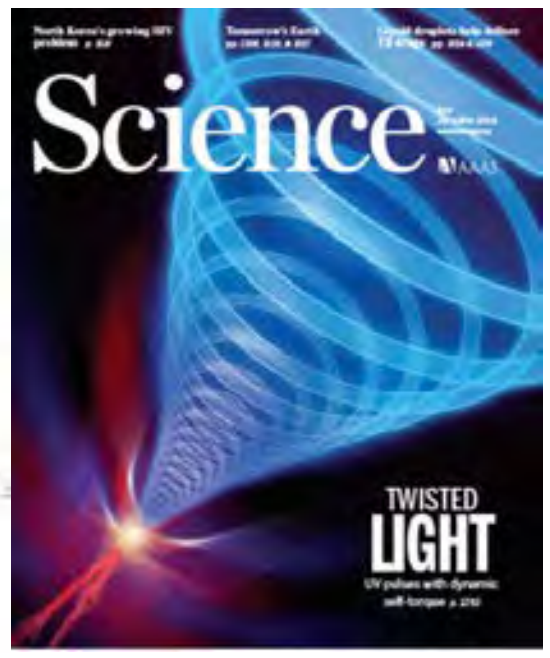

Figure 3.5.2: High-harmonic generation can realize new properties for light fields. Left: tomographic reconstruction of the time-varying polarization state of a complex attosecond pulse train. Right: Orbital angular momentum (OAM) states can enable new imaging modalities. The coherent nature of the HHG process makes it possible to generate short-wavelength light with controllable, as well as "accelerating" OAM, or self-torque of light. [52]

Fundamental questions in the science of coherent HHG include cycle-to-cycle control of pulse shape and polarization of HHG properties. The concept of controlling the detailed shape of a few-cycle light pulse to sculpt the spectral characteristics of HHG is the earliest experimental

48 D. Popmintchev *et al.*,"Ultraviolet Surprise: Efficient Soft X-Ray High-Harmonic Generation in Multiply Ionized Plasmas," Science **350** (6265), 1225–1231 (2015).

49 C. Chen *et al.*, "Tomographic Reconstruction of Circularly Polarized High-Harmonic Fields: 3D Attosecond Metrology," Sci. Adv. **2** (2), e1501333 (2016); L. Rego *et al.*, "Generation of Extreme-Ultraviolet Beams with Time-Varying Orbital Angular Momentum," Science **28** (6447), eaaw9486 (2019).

demonstration establishing the concept of "attosecond science" [50] along with the use of very short-duration pulses for isolated attosecond bursts, [51] and with polarization control. [52] Each of these concepts is still being actively developed, with current questions primarily concerning the effects of high intensity laser propagation on the emission characteristics.

### 3.5.2 Applications of New HHG Light Sources

HHG sources are becoming more widely used and are extremely useful for addressing challenging research. Currently active areas of research include:

#### 3.5.2.1 Dynamics and nanoscale imaging of magnetic and quantum systems

The use of HHG sources to capture the dynamics of magnetic systems results in new understandings of magnetism and quantum materials. Using EUV light, it is possible to probe *element selective* dynamics through the magneto-optic Kerr effect at core-level absorption edges, and to use angle-resolved photoemission spectroscopy (ARPES). Applying HHG for these techniques allows for femtosecond and even attosecond time-resolved studies. [53] This work to-date has identified that light can directly manipulate spins on few-femtosecond timescales [54], and can drive unprecedented spin-polarized currents [55]. Experiments to-date have been done using uniform substrates. Spin-based logic is a leading contender for truly practical large-scale energy-efficient and quantum information processing applications, some of which are speculative and others that are contemplated for incorporation into manufacturing [56,57]. Stroboscopic *imaging* of magnetic dynamics with nm-scale spatial resolution and femtosecond to attosecond time resolution can accelerate this work that can benefit from next-generation HHG sources.

#### 3.5.2.2 Water-window soft x-ray imaging using a stand-alone, tabletop-scale microscope

Sources throughout the soft x-ray spectral region can be implemented using mid-IR ultrafast lasers to drive the HHG process that are beginning to be used for time-resolved spectroscopic applications [58] for both XANES [59] and XAFS [60]. Extending the CDI techniques illustrated in Fig. 3.5.3 into the 2.5 to 4.5 nm spectral region would enable tabletop implementations of a "water window" soft x-ray microscope allowing for ultrahigh 5- to 10-nm resolution imaging of hydrated single cells, viruses, etc.—for both bioscience and eventual clinical application.

---

50 R. Bartels *et al.*, "Shaped-Pulse Optimization of Coherent Emission of High-Harmonic Soft X-Rays," Nature **406** (6792), 164–166 (2000).

51 A. Baltuška *et al.*, "Attosecond Control of Electronic Processes by Intense Light Fields," Nature **421** (6923), 611–615 (2003).

52 T. Fan *et al.*, "Bright Circularly Polarized Soft X-Ray High Harmonics for X-Ray Magnetic Circular Dichroism," Proc. Nat. Acad. Sci. **112** (46), 14,206–14,211 (2015).

53 Z. Tao *et al.*, "Direct Time-Domain Observation of Attosecond Final-State Lifetimes in Photoemission from Solids," Science **353** (6294), 62–67 (2016).

54 P. Tengdin *et al.*, "Critical Behavior Within 20 fs Drives the Out-of-Equilibrium Laser-Induced Magnetic Phase Transition in Nickel," Sci. Adv. **4** (3), eaap9744 (2018).

55 D. Rudolf *et al.*, "Element Selective Investigation of Spin Dynamics in Magnetic Multilayers," in *Ultrafast Magnetism I.* Springer Proceedings in Physics, Vol. 159, edited by J. Y. Bigot *et al.* (Springer, Cham, Switzerland, 2015), pp. 307–309.

56 I. Žutić, J. Fabian, and S. Das Sarma, "Spintronics: Fundamentals and Applications," Rev. Mod. Phys. **76** (2), 323–410 (2004).

57 A Vaysset *et al.*, "Wide Operating Window Spin-Torque Majority Gate Towards Large-Scale Integration of Logic Circuits," AIP Adv. **8** (5), 055920 (2018).

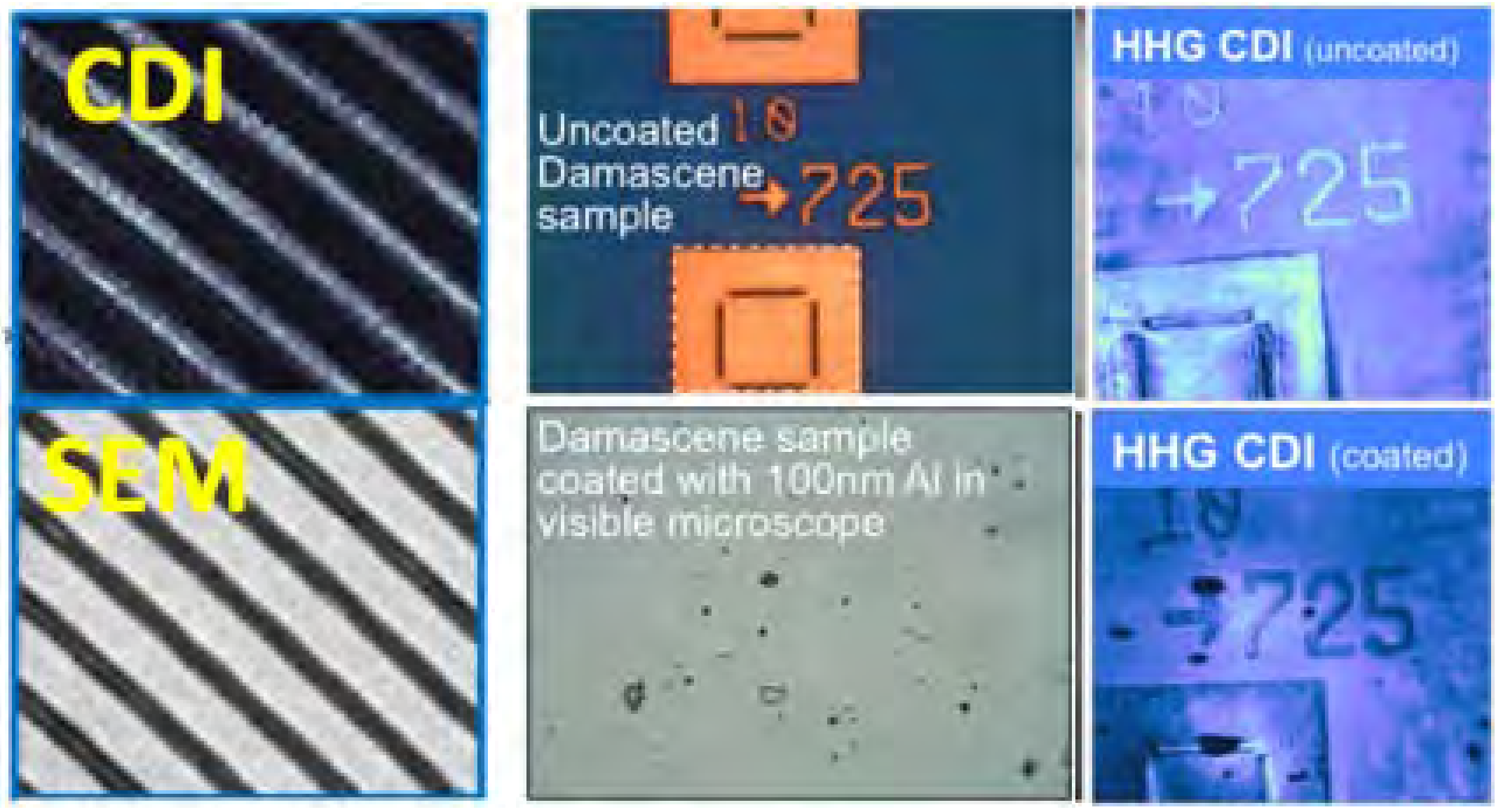

**Figure 3.5.3.** Using HHG as a source for new technological capabilities. Coherent Diffractive Imaging data (left) and comparison of Atomic Force Microscope data (AFM) with imaging of buried interfaces (right) [61].

### 3.5.2.3 Fully-coherent tabletop-scale hard x-ray sources

Very recent theory shows that advances in laser technology for very long wavelengths, fully-coherent HHG sources in the 6 to 10 keV spectral region may be possible. [62] The applications of such a source are quite rich: x-ray CDI imaging with atomic-resolution on a tabletop, medical imaging using the coherence of the source for imaging with a "quantum limited" low dose, and ultralow-dose x-ray backscatter security screening at standoff distances that can be deployed in a wide variety of environments. The timeline for these advances depend on developing new technologies for high-power, few-cycle far-infrared ($\lambda \geq 10$ μm) sources.

---

58 T. Popmintchev *et al.*, "Bright Coherent Ultrahigh Harmonics in the keV X-Ray Regime from Mid-Infrared Femtosecond Lasers," Science **336** (6086), 1287–1291 (2012).

59 A. R. Attar *et al.*, "Femtosecond X-Ray Spectroscopy of an Electrocyclic Ring-Opening Reaction," Science **356** (6333), 54–58 (2017).

60 D. Popmintchev *et al.*, "Near- and Extended-Edge X-Ray-Absorption Fine-Structure Spectroscopy Using Ultrafast Coherent High-Order Harmonic Supercontinua," Phys. Rev. Lett. **120** (9), 093002 (2018).

61 D. F. Gardner *et al.*, "Subwavelength Coherent Imaging of Periodic Samples Using a 13.5 nm Tabletop High-Harmonic Light Source," Nat. Photonics **11** (4), 259–263 (2017); E. R. Shanblatt *et al.*, "Quantitative Chemically Specific Coherent Diffractive Imaging of Reactions at Buried Interfaces with Few-Nanometer Precision," Nano Lett. **16** (9), 5444−5450 (2016).

62 B. R. Galloway *et al.*, "Lorentz Drift Compensation in High Harmonic Generation in the Soft and Hard X-Ray Regions of the Spectrum," Opt. Express **24** (19), 21,818–21,832 (2016).

#### 3.5.2.4 High-average-power light sources in the VUV for industrial applications

The semiconductor industry currently uses deep-UV light sources for a variety of defect inspection tasks, but conventional nonlinear optical harmonic generation with crystals has reached its practical limit at ~200 nm signal wavelength. Gas-phase nonlinear optics can circumvent this limitation, and provide a path forward for defect inspection with light in the 50- to 200-nm spectral region, which is likely the shortest wavelength useful for reflection-based scatterometry. [63] This application requires ~kW class ultrafast lasers operating at ~100 MHz repetition-rate to generate watt-level coherent VUV light sources. The first kW-class laser-driven HHG user facility (NSF National Extreme Ultrafast Science, NeXUS) is under construction and funded through the NSF mid-scale infrastructure initiative.

### 3.5.3 Meeting the Challenges – Facility and Laser Capability Requirements

Progress in HHG sources has been driven historically by advances in laser technology; close linkage of technology development efforts to HHG advances has been extremely effective. In light-science and strong-field physics, laser technology needs focus on refining few- to 10-cycle laser technologies to permit more precise experiments, and developing intense driving lasers that can span the spectrum from the ultraviolet, to the mid-infrared and even the THz spectral regions. For scientific and industrial applications, high-average-power ultrafast laser sources are needed with similar pulse durations and wavelengths in the near-to-mid-infrared. Obtaining the requisite peak powers, short pulses, and focusability is far from routine at wavelengths other than 800 nm. The pulse energy required to drive the HHG process at longer mid-infrared wavelengths increases since diffraction sets the minimum practical spot size.

#### 3.5.3.1 New technologies for intense ultrafast lasers in the mid-infrared spectral region

HHG can produce soft x-rays and potentially hard x-rays. Robust, high-average-power ultrafast lasers are needed f, especially in the mid-infrared spectral region. Mid-IR ultrafast lasers are typically implemented through parametric down conversion from shorter wavelengths. Peak intensity requirements are approximately constant while the spot size for a given confocal parameter of the laser increases with wavelength. Furthermore, the density-length product required for optimal conversion rapidly increases. All these parameters increase pulse energy requirements when scaling from EUV to x-ray HHG. Developing low-loss waveguides that can confine the gas

63 D. Winters *et al.*, “1 MHz Ultrafast Cascaded VUV Generation in Negative Curvature Hollow Fibers,” in *Conference on Lasers and Electro-Optics*, OSA Technical Digest (online) (Optical Society of America, Washington, DC, 2018), Paper STh1N.2.

and guide MIR laser pulses [64] offers an approach relieve this demand and improve HHG efficiency.

#### 3.5.3.2 Pulse compression and pulse shaping at high average power

Robust technologies for pulse shaping and manipulation, especially in the mid-IR, are needed. Pulse shaping will enable optimizing dispersion compensation in amplifier systems to obtain the maximum peak intensity, as well as optimizing specific source characteristics [50]. Techniques for pulse compression at very high average powers and higher pulse energies are needed. These technologies are improving, but all current implementations suffer experimental limitations, including: alignment sensitivity, high loss, high cost, limited fidelity, and limited dynamic range.

#### 3.5.3.3 Improved ultrafast pulse diagnostics and detectors

Ultrafast laser pulses are complex spatio-temporal concentrations of energy that can have complex spatio-temporal aberrations. Most current measurement approaches cannot accurately predict peak focal intensity. New techniques for full spatiotemporal characterization of ultrafast pulses promise to address this issue; however, these techniques must be benchmarked against direct, verifiable peak intensity measurements. The HHG cutoff photon energy, when properly characterized, relates directly to this important parameter. Relevant detector technologies are mainly developed at facility-scale sources, like synchrotrons and FELs, which can also benefit the science using HHG sources.

#### 3.5.3.4 High-average-power, few-cycle lasers in the NIR/visible

Light can be upconverted into the VUV using gas-phase harmonic generation and four-wave mixing that are useful for inspection applications with repetition-rates of 10-100 MHz. The pulse energy requirements are in the ~1-10 μJ range with conversion efficiency approaching 0.1% in guided-wave geometries. High-power (100-1000W) Yb-fiber ultrafast laser technology is well suited for this application.

#### 3.5.3.5 High-average-power laser user facility

The U.S. needs user facilities for developing and applying high average power lasers. In principle, a single ultrafast driver laser can generate high-brightness coherent waveforms spanning the THz, IR, visible, VUV, EUV and x-ray regions of the spectrum that are well synchronized, while also generating ultrafast electron beams. Such a facility could serve as a technology development testbed

64 F. Yu, W. J. Wadsworth, and J. C. Knight, “Low Loss Silica Hollow Core Fibers for 3-4 $\mu$m Spectral Region,” Opt. Express **20** (10), 11,153–11,158 (2012)

#### 3.5.3.6 Other R&D needs

Developing powerful predictive capabilities for HHG sources would significantly advance the field. Theory and modelling of HHG requires calculations that are an elegant combination of atomic quantum dynamics with macroscopic electromagnetic propagation, and scale lengths spanning many orders of magnitude. Nonlinear effects, such as filamentation and self-focusing will become more significant at higher powers. There is a clear potential for computation to predict "sweet-spot" regimes where filamentation and HHG phase-matching coincide. Easy-to-use simulation tools for spatio-temporal pulse propagation, dispersion, and focusing are needed, along with standardized data formats to share and compare results and characteristics of commonly used components, such as mirror dispersion.

### 3.5.4 Summary and Recommendations

Progress using high-harmonic light sources over the past two decades has been extraordinary. Early work simply observing the harmonic spectrum has progressed to using these sources for advanced imaging and spectroscopy to provide new insights into materials and quantum systems. While coherent x-ray science and technology has primarily originated in facility-scale projects such as synchrotrons and XFELs [65], the ability to generate coherent x-rays on a tabletop is enhancing new discoveries and applications by making them accessible to investigators in their home laboratories.

---

65 S. H. Glenzer *et al.*, "Matter Under Extreme Conditions Experiments at the Linac Coherent Light Source," J. Phys. B: At. Mol. Opt. Phys. **49** (9), 092001 (2016).

## 3.6 SRN-5 – Particle Acceleration (electrons, ions, neutrons, and positrons)

Short-pulse lasers can produce extraordinarily short bursts of light that are capable of producing extremely high intensities, temperatures, and pressures. Novel laser-based particle acceleration techniques generate large numbers of high-energy particles ($\gg$ MeV) over short distances (typically from less than 1 μm to 100 μm). Strong nonlinear coupling of the laser to plasma electrons produces highly energetic electron populations in large densities ($> 10^{21}$ $cm^{-3}$) on the time scale of the laser pulse length. An effort to study these complex interactions with experiments using the most advanced lasers, simulations, and theory can maximize the energy coupled to the electrons or other secondary particles, like protons, ions, neutrons, and positrons. Targets are becoming increasingly sophisticated to maximize and isolate various acceleration mechanisms of specific secondary particles. The pathway to dramatic increases in particle production and energy is becoming clear and requires next-generation lasers. Higher secondary particle energies and higher fluxes produced at high repetition rate will enable many exciting applications.

Decisively breaking the ~100 MeV/nucleon barrier for accelerated ions represents a grand challenge. This is best achieved with a quasi-mono-energetic distribution in the 100-300 MeV range or a broad energy distribution with a high yield in this energy range. Mechanisms that achieve higher energies also have better conversion efficiency and higher fluxes. Some applications that drive the capabilities of future facilities are described below.

*Collocating laser-based particle beams with energetic drivers and x-ray laser probes will enable exploring the rich science of the ultrafast interaction of energetic short-pulse laser accelerated particle beams with matter and electromagnetic fields.*

### 3.6.1 Opportunities for Frontier Science and Applications

#### 3.6.1.1 Neutron generation

A wide range of applications use neutrons for imaging and spectroscopy with both national security and commercial impact, including: detection of nuclear materials in shipping containers, stockpile stewardship, and nondestructive inspections of structures and defects deep inside solid-density materials. Laser-accelerated ions can drive different nuclear reactions to produce neutrons, like deuteron-deuteron (d-d) or $^{10}$B-d reactions. Increasing laser-accelerated ion yield through higher power lasers and repetition rates provides a path to pursue these areas of physics and applications.

*<u>Required</u>: A wide range of requirements exists depending on the specific approach with considerable overlap on the requirements for protons and ions, described below.*

#### 3.6.1.2 Generating near-solid-density reactive-ion distributions

Nuclear synthesis for the higher-*Z* elements remains an unsolved problem. About half of the elements heavier than Fe and all elements heavier than Bi are produced in nature via the rapid neutron-capture process (r process) [66]. Current data on neutron-capture cross sections and isotope half-lives is limited to cold material conditions. One can directly measure cross sections relevant to nuclear synthesis under hot dense plasma conditions using high-power lasers and suitable targets [67]. A laser-based high-flux neutron source would allow one to study these r-process reactions under the plasma conditions relevant to astrophysical scenarios. These experiments can achieve high density and simultaneously operate at high temperature so that the effect of the plasma on the bare reaction is measured.

*Required: Minimum 50-J, sub-100-fs pulses, at the highest possible repetition rate (≥ 1 Hz).*

#### 3.6.1.3 Anomalous mix dynamics of dense plasmas and warm dense matter

Matter isochorically heated by intense laser-driven ion beams enables directly studying plasma mix in cases where kinetic effects, strong coupling, or collective effects matter. This has broad applicability to space and planetary science (Sec. 3.3), inertial confinement fusion, industrial plasmas, and national defense, but existing capabilities must be significantly improved.

*Required: >200-J/pulse, high-contrast pulses required for thin targets, high repetition rate for collecting sufficient data, and collocation with an x-ray FEL for diagnostics.*

#### 3.6.1.4 Isochoric heating of dense plasmas and dynamic materials

Warm dense matter (WDM) is found in planetary interiors and astrophysical environments, and it is a transient state encountered in many proposed applications. Isochoric heating can generate homogeneous WDM samples that allow studying the equation of state, transport, and opacity using arbitrary loading paths for dynamic materials [68]. Combined with a shock-compression driver, off-Hugoniot areas in the phase space diagram become accessible.

*Required: In addition to 3.6.1.3, collocation with a kJ,* pulse-shaped driver laser.

#### 3.6.1.5 Dynamic thermometry of materials and dense plasmas

A revolutionary thermometry diagnostic that is volumetric and time- and space-resolved can be established using neutron resonance spectroscopy (NRS) for dense matter [69]. This would provide the ability to measure thick solid-density samples inaccessible with x-ray probes.

---

66 M. E. Burbidge *et al.*, "Synthesis of the Elements in Stars," Rev. Mod. Phys. **29** (4), 547–650 (1957).

67 A. J. Kemp, "Generating keV Ion Distributions for Nuclear Reactions at Near Solid-Density Using Intense Short-Pulse Lasers," Nat. Commun. **10**, 4156 (2019).

68 W. Bang *et al.*, "Visualization of Expanding Warm Dense Gold and Diamond Heated Rapidly by Laser Generated Ion Beams," Sci. Rep. **5**, 14318 (2015).

69 J.C. Fernández *et al.*, "Requirements and Sensitivity Analysis for Temporally- and Spatially-Resolved Thermometry Using Neutron Resonance Spectroscopy," Rev. Sci. Instrum. **90** (9), 094901 (2019).

*<u>Required</u>: Lasers produced >200-J, high-contrast pulses for thin targets collocated with a kJ/ns shaped-pulse laser.*

#### 3.6.1.6 Multi-MeV gamma-rays and GeV electrons from laser-driven magnetic fields

Multi-MeV gamma-ray and GeV-electron beams generated by laser-driven magnetic fields would facilitate fundamental high-field studies and excitation of nuclear isomer states.

*<u>Required</u>: Multi-PW, 150-fs laser with $10^{23}$ W/cm$^2$ focal intensity, and structured targets.*

#### 3.6.1.7 Protons for diagnosing electric and magnetic fields

Laser-generated proton beams serve as probes of transient space and time-varying magnetic fields in fundamental studies of high-energy-density plasmas and space physics (Sec. 3.2). Higher-energy, higher-yield proton beams are needed to access new regimes of field strength [70].

*<u>Required</u>: Sub-ps pulses with tens of kJ energies.*

#### 3.6.1.8 Understanding the properties of matter exposed to extreme radiation conditions

The transport and structural properties of matter are affected by defects from exposure to long-term, high radiation under fusion energy conditions. A copious source of secondary particles is required for testing that could ultimately lead to developing accurate materials modeling.

The typical range of neutron energies required for neutron-induced damage studies spans the fission spectrum up to 14-MeV for D-T fusion. These studies use in-core reactor irradiation, large accelerator-based sources, and high-flux neutron generators based on the DT reaction. A complementary approach uses heavy ions that simulate the energy deposition by mimicking the recoil nuclei, but limits studies to approximately micron-thick layers on material surfaces. High demand for access to test facilities limits progress, so establishing laser-based neutron sources could make a considerable impact on this field by increasing the community access. The high peak flux available from laser-based sources would enable studies of potential collective and transient damage phenomena in conditions that resemble those that would occur in inertial confinement fusion systems and in defense applications.

A special application of a laser-based neutron source that interfaces with a state-of-the-art FEL, such as LCLS, has been proposed with several motivations: engineering new materials suitable for extreme conditions encountered in fusion experiments, space or nuclear environments; determining transport and structural properties of matter affected by defects exposed to extreme radiation conditions; and testing inter-atomic potentials in widely used physics molecular dynamics models to make long-term predictions of material responses and properties.

---

70 L. Yin *et al.*, "GeV Laser Ion Acceleration from Ultrathin Targets: The Laser Break-Out Afterburner," Laser Part. Beams **24** (2), 291–298 (2006); L. Yin *et al.*, "Monoenergetic and GeV Ion Acceleration from the Laser Breakout Afterburner Using Ultrathin Targets," Phys. Plasmas **14** (5), 056706 (2007).

*<u>Required</u>: Ultrashort pulse laser producing 150- to 1500-J pulses at high repetition rate collocated with x-ray FEL.*

### 3.6.2 Underlying Physics – Outlook for Acceleration Mechanisms

Despite great progress in using lasers for accelerating secondary particles to relatively high energies (< 100 MeV for protons), the most exciting opportunities still lie ahead. New particle acceleration mechanisms have been proposed using both next-generation lasers and tailored targets to achieve extraordinary control of the relativistic laser interaction resulting in higher particle energies and fluxes. Current state-of-the-art ion acceleration focuses intense laser pulses on very thin, flat targets (typically < 100 nm), but this excludes a key degree-of-freedom in the laser-plasma interaction (LPI). Ion energies from 100 MeV to perhaps as high as 1 GeV may be possible by controlling the LPI using tailored laser pulses and/or tailored targets. An order of magnitude increase is required in both laser energy and repetition rate to realize these advances. Specifically, energetic laser pulses (>100 J) at relativistic intensities with longer durations (~100 fs) and extremely high temporal contrast ($10^{12}$ or better) over nanosecond timescales are needed.

#### 3.6.2.1 Relativistically induced transparency (RIT)

A particularly compelling scientific development would be observing laser plasmas in the RIT regime [71] and its use for laser-driven ion acceleration [72]. In this regime, the full electron population is driven to relativistic temperatures (i.e., a relativistic factor $\gamma_e >> 1$) by intense laser heating before or at the peak of the laser pulse. Although the plasma may be classically over-dense (*e.g.*, solid density) in such conditions, the relativistically corrected plasma frequency (~1/ $\gamma_\varepsilon$) drops below the laser frequency and the laser can traverse the plasma, leading to a volumetric interaction. This regime has been realized in the past decade by the confluence of several technological advances: temporal pulse-cleaning techniques delivering unprecedented temporal contrast on intense laser pulses that enable using nanofoil laser targets as thin as 10 nm. An example of an unexpected acceleration regime, "breakout afterburner (BOA)" [70], was identified in simulations and demonstrated that generates ~100-MeV protons and the possibility of quasi-monoenergetic ions.

*<u>Required</u>: Intensities >$10^{22}$ W/cm$^2$ with pulse durations up to ~ps duration, and spatial and temporal pulse shaping to reduce variations in laser intensity.*

#### 3.6.2.2 Relativistically induced transparency in channels

Recent three-dimensional kinetic simulations show that in the RIT regime, a high-intensity laser pulse drives a strong mega-ampere-level electron current that generates and sustains quasi-

71 S. Palaniyappan *et al.*, "Dynamics of Relativistic Transparency and Optical Shuttering in Expanding Overdense Plasmas," Nat. Phys. **8** (10), 763–769 (2012).

72 A. Henig *et al.*, "Enhanced laser-driven ion acceleration in the relativistic transparency regime," Phys. Rev. Lett. **103**, 045002 (2009).

static magnetic fields in the mega-tesla range [73]. Such magnetic fields are known to exist in astrophysics but they are not currently accessible in laboratory conditions. Future laser drivers producing such extreme quasi-static magnetic fields would alter the electron dynamics of the propagating laser pulse enabling:

- electron acceleration to GeV energies over just tens of laser wavelengths;
- efficient and directed gamma-ray emission at intensities as low as $5 \times 10^{22}$ W/cm$^2$, with over $10^{12}$ multi-MeV photons in a 30-fs bunch; and
- proton-accelerating structures that generate dense monoenergetic beams with 200 MeV in energy and tens of nC of charge [74].

An example of a novel laser-driven ion-acceleration mechanism that can be enabled by a high-intensity laser pulse and a strong plasma magnetic field uses a structured target, where the channel material is hydrogen and the bulk material is carbon. Relativistic transparency enables a volumetric laser–matter interaction at solid densities that generates monoenergetic proton beams with energies reaching several hundred MeV. Fully kinetic relativistic three-dimensional PIC simulations, shown in Fig. 3.6.1, predict a large beam charge up to tens of nC, and relatively low angular divergence below 10 degrees produced by a 1-PW laser pulse with a peak intensity of $5 \times 10^{22}$ W/cm$^2$.

*<u>Required</u>: 300-J pulses with a 150-fs pulse length.*

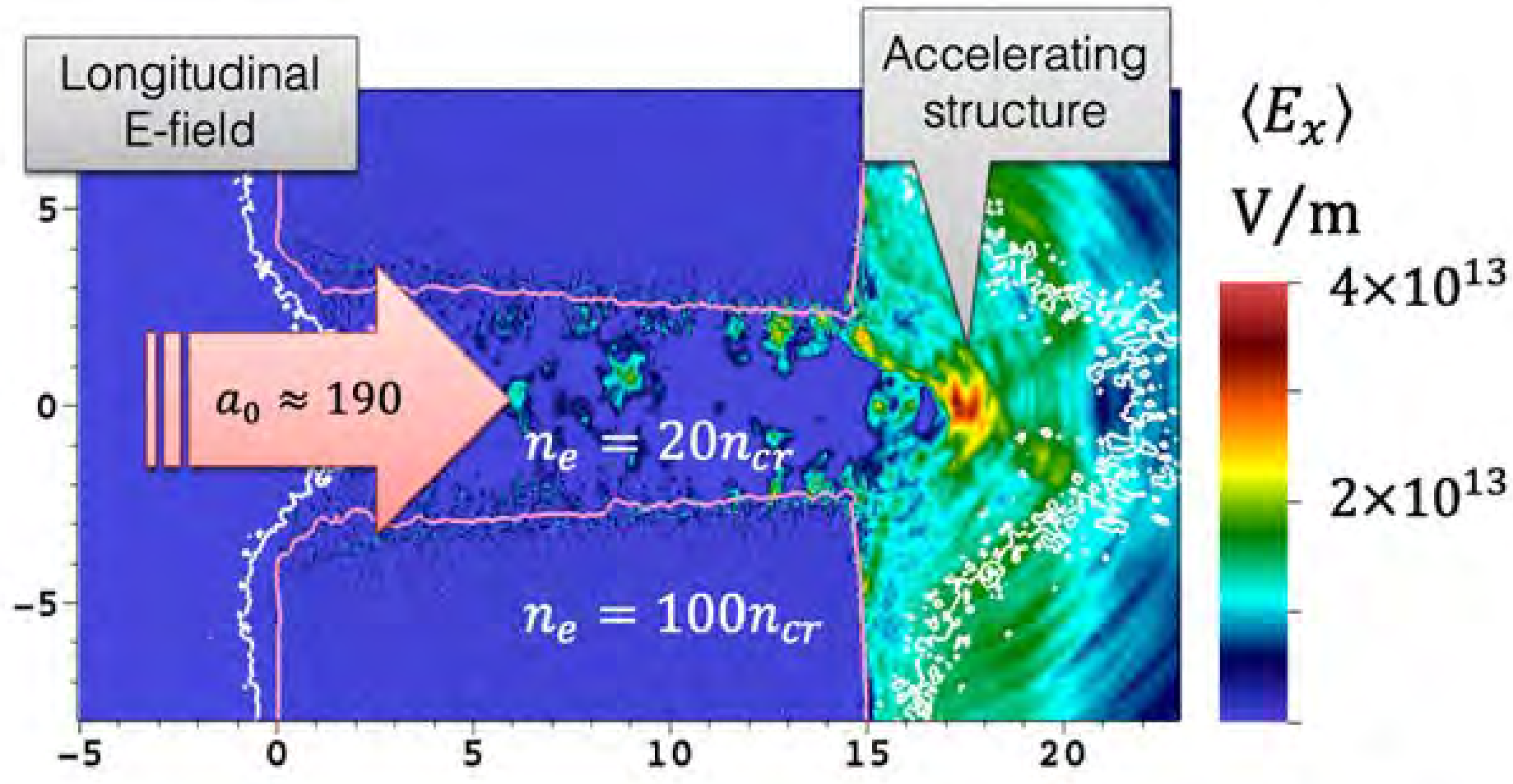

**Figure 3.6.1.** Ion acceleration scheme based on a relativistically induced transparent channel and subsequent magnetic field behind the target.

73 D. J. Stark, T. Toncian, and A. V. Arefiev, "Enhanced Multi-MeV Photon Emission by a Laser-Driven Electron Beam in a Self-Generated Magnetic Field," Phys. Rev. Lett. **116** (18), 185003 (2016).

74 A. V. Arefiev, Z. Gong, T. Toncian, and S. S. Bulanov, "Laser Ion-Shotgun Acceleration," Plasma Physics Archive, available at https://arxiv.org/abs/1807.07629 (2018).

### 3.6.2.3 Laser-ion lensing and acceleration

Figure 3.6.2 illustrates combining a next-generation high-power laser with a tailored target involves driving a plano-convex lens–shaped target (laser incident on the flat side) to both accelerate the target and drive a transverse collapse. The resulting plasma then collides with a second, lower-density plasma that is stationary [18]. A collisionless shock occurs via the Weibel instability in a process similar to that in astrophysical environments, and the shock accelerates ions to high energies. Monoenergetic protons beams with energies exceeding 150 MeV are predicted.

*Required: 35 PW, 50 fs, 1600 J laser.*

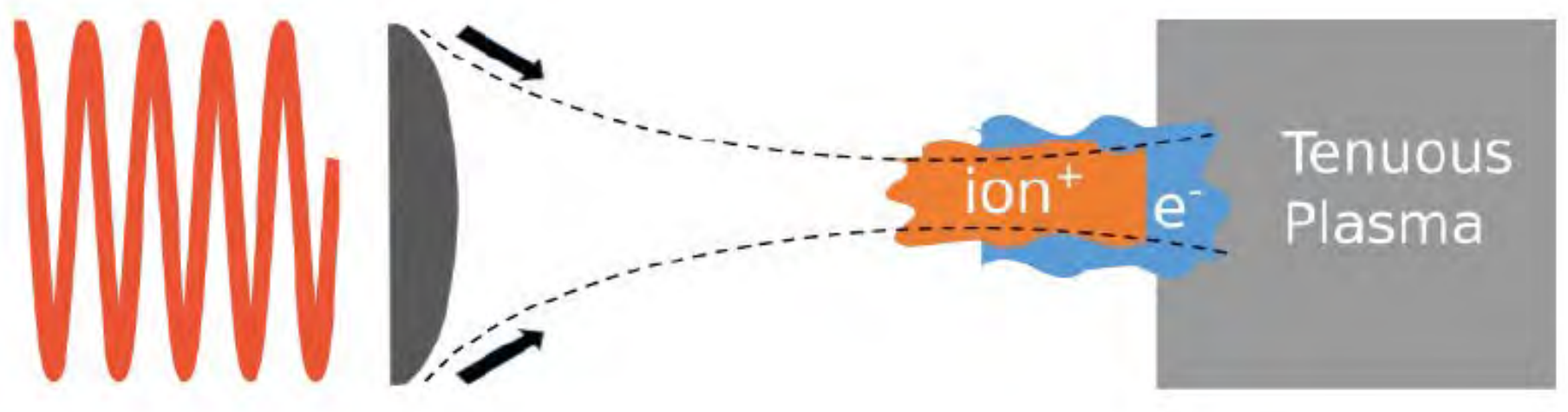

**Figure 3.6.2.** A shaped target is simultaneously focused and accelerated resulting in transverse compression and longitudinal expansion.

### 3.6.2.4 Superponderomotive electron-driven target normal sheath acceleration (TNSA)

Proton acceleration schemes would benefit from superponderomotive electrons in the NIF Advanced Radiographic Capability (ARC) regime at higher intensity. Recent experiments [75] demonstrated a very robust and useful proton source, and opened up an entirely new regime of multi-ps, short-pulse particle acceleration. A future facility providing 10-ps, 5-kJ pulses focused to $D_{80\%}$ = 50 $\mu$m could achieve a highly relativistic intensity of $3 \times 10^{19}$ W/cm$^2$, while retaining the quasi one-dimensional and multi-picosecond conditions that have proven to produce much higher maximum proton energies than those predicted for femtosecond pulses. By focusing to a spot size of $D_{80\%}$ = 16 $\mu$m, one could achieve a current state-of-the-art intensity of [76], $3 \times 10^{20}$ W/cm$^2$, but with an 11× longer pulse duration. This is predicted to result in dense proton beams with peak energies of several hundreds of MeV.

---

75 D. Mariscal *et al.*, "First Demonstration of ARC-Accelerated Proton Beams at the National Ignition Facility," Phys. Plasmas **26** (4), 043110 (2019).

76 A. Higginson *et al.*, "Near-100 MeV Protons via a Laser-Driven Transparency-Enhanced Hybrid Acceleration Scheme," Nat. Commun. **9**, 724 (2018).

### 3.6.3 Meeting the challenges – technology requirements

Much research can be done using existing facilities before next generation facilities will come on-line. Realistic upgrades to existing facilities can greatly increase their utility in studying secondary particle generation and acceleration and developing the new approaches. In particular, studies are required of the properties and relative interplay of multiple mechanisms, including hole boring, relativistic transparency, radiation pressure acceleration, shock acceleration, magnetic vortex acceleration and more.

#### 3.6.3.1 Temporal pulse shaping

The ability to shape short laser pulses or string together separate laser pulses in time could open a whole new area of phase space in plasma physics by providing exquisite control of LPI and particle acceleration. For example, recent research [77] shows that adding a "picket" laser pulse in front of a 10-ps laser pulse doubles the maximum proton energy. Simulations with advanced pulse shapes, coupled with structured targets show peak proton energies over 300 MeV are possible. Future lasers with 5-kJ pulse energy could produce copious protons with peak energies approaching 1 GeV.

#### 3.6.3.2 Increased pulse contrast

High temporal pulse contrast limits target expansion before arrival of the main laser pulse, and it preserves structured targets intended to shape and control the laser-plasma interaction. A number of facilities report measured nanosecond temporal contrast values of $10^{10}$ and a range of contrasts at picosecond timescales. Improved pulse contrast can be achieved by replacing or improving laser front ends, and/or adding plasma mirrors at the end of the laser system, as discussed in Section 5.2.3.1. Better contrast is inferred but not directly measured when plasma mirrors are used. This will no longer be sufficient for increasing particle energy or improving conversion efficiency. Higher contrast must be obtained and directly measured. Detailed particle-in-cell and hydrodynamic simulations must be performed but contrast values up to at least $10^{12}$ are likely required. These simulations will need to be benchmarked against dedicated experiments. Single-shot, ultrahigh-dynamic-range pulse contrast measurements are required and developing such a technique would have significant impact.

#### 3.6.3.3 Collocation with multiple scientific capabilities

High-power lasers enable sources of secondary radiation, such as x-rays and particles, as probes that would provide extraordinarily powerful research tools. Collocating high power short-pulse lasers, long-pulse driver lasers, and other capabilities, such as x-ray free-electron laser (XFEL) make particularly powerful combinations.

---

77 J. Kim *et al.*, "Computational Modeling of Proton Acceleration with Multi-Picosecond and High Energy, Kilojoule, Lasers," Phys. Plasmas **25** (8), 083109 (2018).

#### 3.6.3.4 Higher energy, better focusing

Higher intensities are required to better understand and exploit relativistic transparency in ion acceleration mechanisms. Increasing pulse energy can build a bridge between existing and next-generation facilities. Low-energy systems can be upgraded to PW-class operation using proven commercial technology. Tighter focusing and deformable mirrors can be a relatively low cost method to increase the intensity significantly, albeit at the cost of more difficult experimental alignment. Ultimately, new laser facilities will be required.

#### 3.6.3.5 Increased experimental repetition rate

Most experiments currently operate shot-on-demand, even when the laser itself can operate at higher repetition rate. This is particularly true for solid-density target experiments. In many cases, the experimental repetition rate can be significantly increased by improving the experimental infrastructure with improved detectors, data readout, and target handling/alignment.

### 3.6.4 Summary and Recommendations

Understanding relativistic laser plasma interactions involved in particle acceleration represents a compelling scientific challenge. Advancing scientific understanding of particle acceleration promises not only great utility for many applications that can benefit national security and larger societal interests, but will facilitate advances connecting many fields, like space and planetary science, material science, and fundamental physics. Advanced laser facilities are required to produce high-energy ion beams, such as protons with energies in the 100 – 300 MeV range. These laser facilities will address a range of fundamental phenomena and important scientific questions, such as:

- Can relativistic shocks be created in the laboratory to study Fermi acceleration?
- What is the effect of ultra-strong magnetic fields on particle acceleration?

Advances in particle production can be leveraged to make effective, high-brightness, laser-based neutron sources. Accomplishing all of these science-based goals will require simultaneous advances in laser power, pulse temporal contrast and shaping, polarization control, advanced targets and diagnostics optimized for future research.

## 3.7 SRN-6 – Nuclear Photonics

Nuclear photonics is a nascent field that recently has seen a rapid increase of visibility and interest in the scientific community. The genesis of nuclear photonics can be traced to the advent of intense ultrafast lasers, which made it possible to generate ionizing radiation and induce nuclear reactions at considerably lower laser energies than those pursued for several decades in laser-driven inertial confinement fusion. The field has gathered considerable additional momentum by recognizing the potential for scaling laser-driven Compton sources to MeV energies and using their unique properties to excite and manipulate atomic nuclei. Such sources hold the promise to advance the field of nuclear spectroscopy in a manner similar to the revolution in atomic spectroscopy made possible by invention of the laser. These advances also stimulate important interactions between the electron accelerator and laser communities. Laser-driven energetic particle sources, such as protons and heavier ions, neutrons, and more exotic particles such as pions and muons, can probe nuclear physics, but also enable myriad practical applications.

*New multi-beam, high-intensity laser facilities that drive gamma ray and neutron sources will enable pulsed nuclear sources with unique properties for a wide range of science and applications.*

### 3.7.1 Nuclear Photonics Applications with Laser-Driven Photon Sources

Laser-based Compton sources, currently at the center of interest for nuclear photonics, can deliver continuously tunable, quasi-mono-energetic, well-collimated beams of x-rays at energies that readily scale to the MeV range using electrons accelerated via conventional methods (linacs) or Laser Wakefield Acceleration (LWFA), as described in Section 3.3. LWFA has a potential to deliver more compact sources with higher peak flux, whereas linacs deliver sources with higher repetition rates (MHz) and narrower energy spread. Next-generation laser-driven photon sources are expected to greatly exceed the present capabilities. They will require short-duration (few ps) laser pulses with many kilowatts of average power. When used in conjunction with low-emittance electron beams, these sources could deliver photon bandwidths as low as $\Delta E/E \approx 10^{-3}$, and efficiencies approaching unity with respect to electron participation in Compton photon production.

#### 3.7.1.1 Nuclear Physics Research with Laser Compton Sources

The high spectral brilliance of laser Compton sources makes them an attractive tool for basic research in nuclear physics [78]. Three main categories of possible studies have been identified:

78 C. A. Ur, "Gamma Beam System at ELI–NP," AIP Conf. Proc. **1645**, 237–245 (2015).

- electromagnetic dipole response of nuclei with nuclear resonance fluorescence (NRF);
- astrophysics hot topics such as heavy elements synthesis in the Universe; and
- photofission phenomena.

#### 3.7.1.2 Nuclear isomer production and research

One of the earliest proposed applications of laser-based sources would study nuclear isomers by photoexcitation of a nuclear isomeric state. Such studies are important to understanding conditions in stars, where photon baths with temperatures ranging from $10^8$ K (He inner shell) up to $3\times10^9$ K (deep O-Ne layers) are encountered. The source for isomer excitation can be bremsstrahlung driven by LWFA [79] using 1-PW to 10-PW- lasers. A narrow-energy-spread, high-energy x-ray beam from a laser-based Compton source could then be used to probe the isomeric states.

#### 3.7.1.3 Detection of special nuclear material (SNM) with low-energy-spread x-rays

Detecting SNM, particularly in the context of cargo scanning, is one of the most promising applications for nuclear photonics. Material identification via radiography benefits from narrow energy and angular spread which remove attenuated and scattered photons, reducing dose and increasing the specificity of material identification. Photofission is more effectively excited, resulting in clearer signatures with reduced radiation doses. Combined, such techniques offer a transformational approach for screening and characterizing nuclear and contraband materials. Processes used for isotope-specific detection include Nuclear Resonance Fluorescence (NRF) [80], where the narrow energy spread, continuous energy tunability, and tight collimation of Compton sources are critical [81]).

### 3.7.2 Nuclear Photonics Applications with Laser-Driven Neutron Sources

The field of nuclear photonics received a boost by the demonstration of laser-driven pulsed neutron sources that can produce as many as $10^{11}$ neutrons per pulse (Fig. 3.7.1). Laser-accelerated deuterium ions and neutrons are produced by deuteron breakup on beryllium targets with neutrons traveling predominantly in the direction of the laser pulse. Laser-driven neutron production and concepts for optimized target/moderator systems benefit from laser technology developments and the vast foundation of target/moderator design for spallation neutron sources, respectively. This promises a pathway towards high-brightness laser-driven neutron sources in the near future.

---

79 A. Berceanu, L. D'Alessi, and O. Tesileanu, "Optical Production of Nuclear Isomers," presented at the 2nd International Conference on Nuclear Photonics, Brasov, Romania, 24–20 June 2018 (Abstract P. 05).

80 F. Albert *et al.*, "Isotope-Specific Detection of Low-Density Materials with Laser-Based Monoenergetic Gamma-Rays," Opt. Lett. **35** (3), 354–356 (2010).

81 C. Geddes *et al.*, "Impact of Monoenergetic Photon Sources on Nonproliferation Applications," Idaho National Laboratory, Idaho Falls, ID, Report INL/EXT-17-41137 (March 2017).

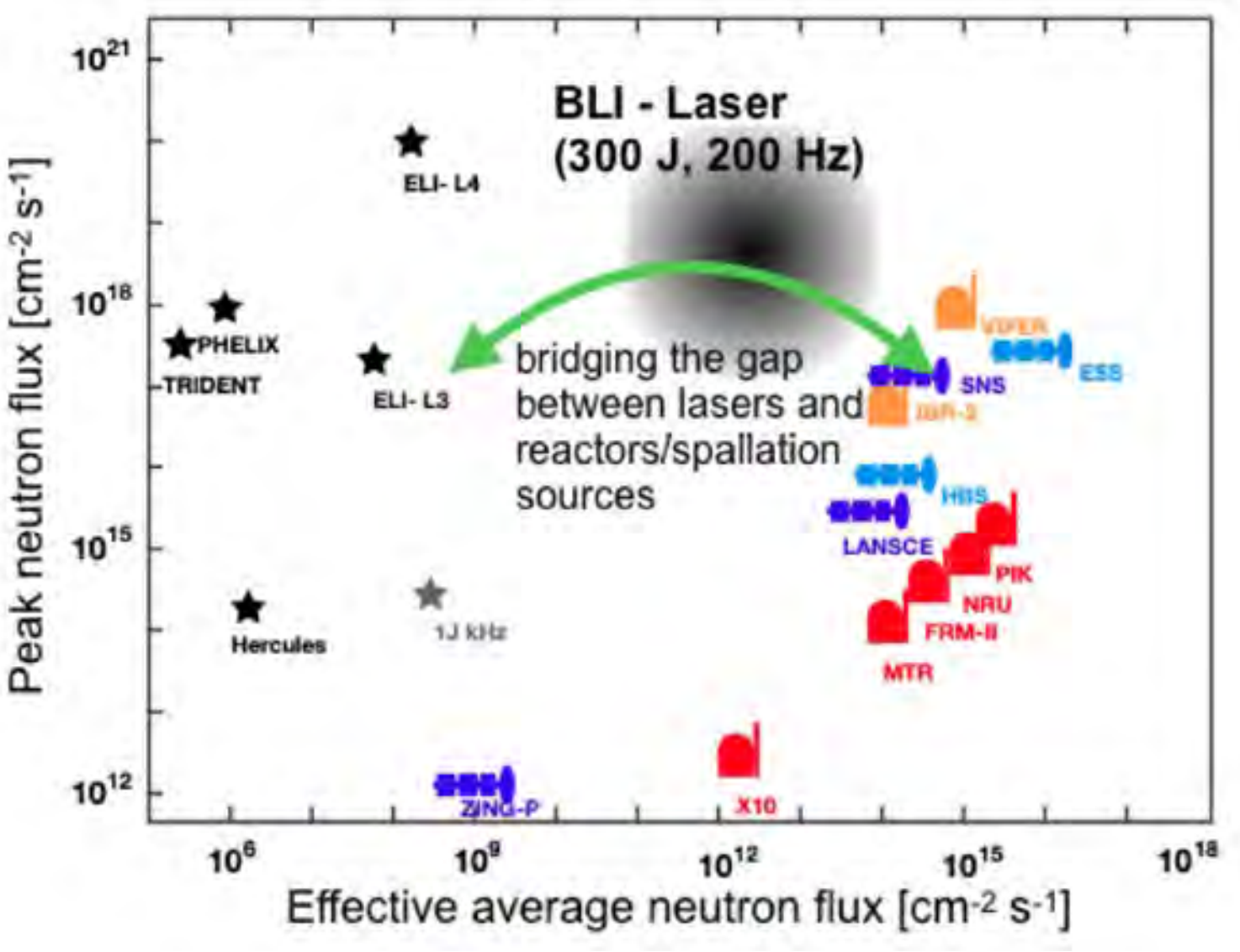

**Figure 3.7.1.** Parameter space for laser-driven neutron sources and conventional neutron sources.

Laser-based deuteron or proton accelerators offer an alternative to spallation sources, making this technology an excellent choice for potential applications in basic research, as well as in national security and in industry. While laser-driven high-peak-brightness neutron sources at present cannot yet compete with accelerator-driven neutron sources or high-flux reactors in terms of average power, they can complement these large-scale facilities as powerful and otherwise unavailable characterization tools that can be easily added to other facilities.

#### 3.7.2.1 Fast neutron radiography

The technique for short-pulse laser neutron production was demonstrated in 2012 [82]. MeV neutrons from the laser-driven source were used to radiograph a test object made with tungsten blocks of different thickness. Neutron interactions with materials are distinct from x-ray interactions; they probe complementary features of an object. Besides compactness, other advantages of laser-driven neutron source included the directionality of the energetic neutrons, which minimizes the need for shielding during the imaging and suitability for visualization of transient phenomena.

#### 3.7.2.2 Temperature measurements in astrophysical plasmas

Understanding the conditions and behavior of dense planetary cores is an important task for modern astrophysics. The goal is to produce extreme conditions in materials and study their opacity and transport properties. Ultimately, one wishes to determine the equation of state for materials relevant to the formation of planets, which present a critical input to computer models. A laser-driven source could enable such measurements using neutron resonance absorption, which could be well timed to a kJ laser to prepare compressed materials. A kJ-class, ultra-high intensity laser would be needed to provide a 100–200 MeV deuterium beam directed to a converter to produce ~$10^{13}$ n/shot and a moderator to generate thermal neutrons. Shots per minute or faster should suffice to obtain high-quality data.

---

82 M. Roth, D. Jung, K. Falk, N. Guler, O. Deppert, M. Devlin, A. Favalli, J. Fernandez, D. Gautier, M. Geissel, R. Haight, C. E. Hamilton, B. M. Hegelich, R. P. Johnson, F. Merrill, G. Schaumann, K. Schoenberg, M. Schollmeier, T. Shimada, T. Taddeucci, J. L. Tybo, F. Wagner, S. A. Wender, C. H. Wilde, and G. A. Wurden, "Bright Laser-Driven Neutron Source Based on the Relativistic Transparency of Solids," Phys. Rev. Lett. **110** (4), 044802 (2013).

#### 3.7.2.3 Detection of special nuclear material (SNM) with laser-produced neutrons

Neutrons offer additional opportunities for SNM detection and characterization using active interrogation [83]. SNM is usually detected by inducing nuclear fission and observing unique fission signatures, such as prompt and delayed fission neutrons and gamma rays. The most attractive characteristic of laser-based neutron sources in this context is their ultrashort, intense nature that produces a large number of fissions in a short-time window that could enhance the SNM detection using the characteristic delayed neutron signal [84]. In addition, short pulses of energetic laser-produced neutrons could be employed for multimodal spectroscopic transmission radiography, reducing the required time-of-flight path while reaching high energy resolution. Neutrons and photons offer complementary signatures, and the capability of laser sources to produce both is hence particularly powerful.

#### 3.7.2.4 Boron neutron capture therapy

Boron-neutron capture therapy (BNCT) is a promising medical application. Epithermal neutrons are produced and directed to a patient, where they are absorbed in boron-containing compounds. Traditionally, neutrons for BNCT have been produced by nuclear reactors, which are limited in their number and availability. The BNCT research community will benefit from a larger number of smaller, more flexible neutron sources based on intense lasers [85]. The pulsed nature of such sources, which is very different from nuclear reactors, provides an opportunity to study the differences and possible therapeutic benefits that laser-driven neutron sources may provide.

#### 3.7.2.5 Nondestructive inspection using neutron resonance spectroscopy (NRS)

Moderated epithermal or thermal neutrons can be used to identify isotopes based on their characteristic energy-dependent absorption cross sections. Neutron Resonance Transmission Analysis is a promising candidate experimental technique for spent fuel assay measurements [86].

---

83 I. Jovanovic and A. Erickson, eds. *Active Interrogation in Nuclear Security: Science, Technology and Systems*, Advanced Sciences and Technologies for Security Applications, edited by A. J. Masys (Springer International Publishing, Cham, Switzerland, 2018).

84 J. Nattress *et al.*, "High-Contrast Material Identification by Energetic Multiparticle Spectroscopic Transmission Radiography," Phys. Rev. Appl. **11** (4), 044085 (2019).

85 Y. Kasesaz, F. Rahmani, and H. Khalafi, "Feasibility Study of Using Laser-Generated Neutron Beam for BNCT." Appl. Radiat. Isot. **103**, 173–176 (2015)

86 D. L. Chichester and J. W. Sterbentz, "Neutron Resonance Transmission Analysis (NRTA): Initial Studies of a Method for Assaying Plutonium in Spent Fuel " Idaho National Laboratory, Idaho Falls, ID, Report INL/CON-A 100-J laser operating at 10 Hz with an optimized target to produce neutrons with deuteron break-up that are moderated could fulfill the requirements for this particular application.

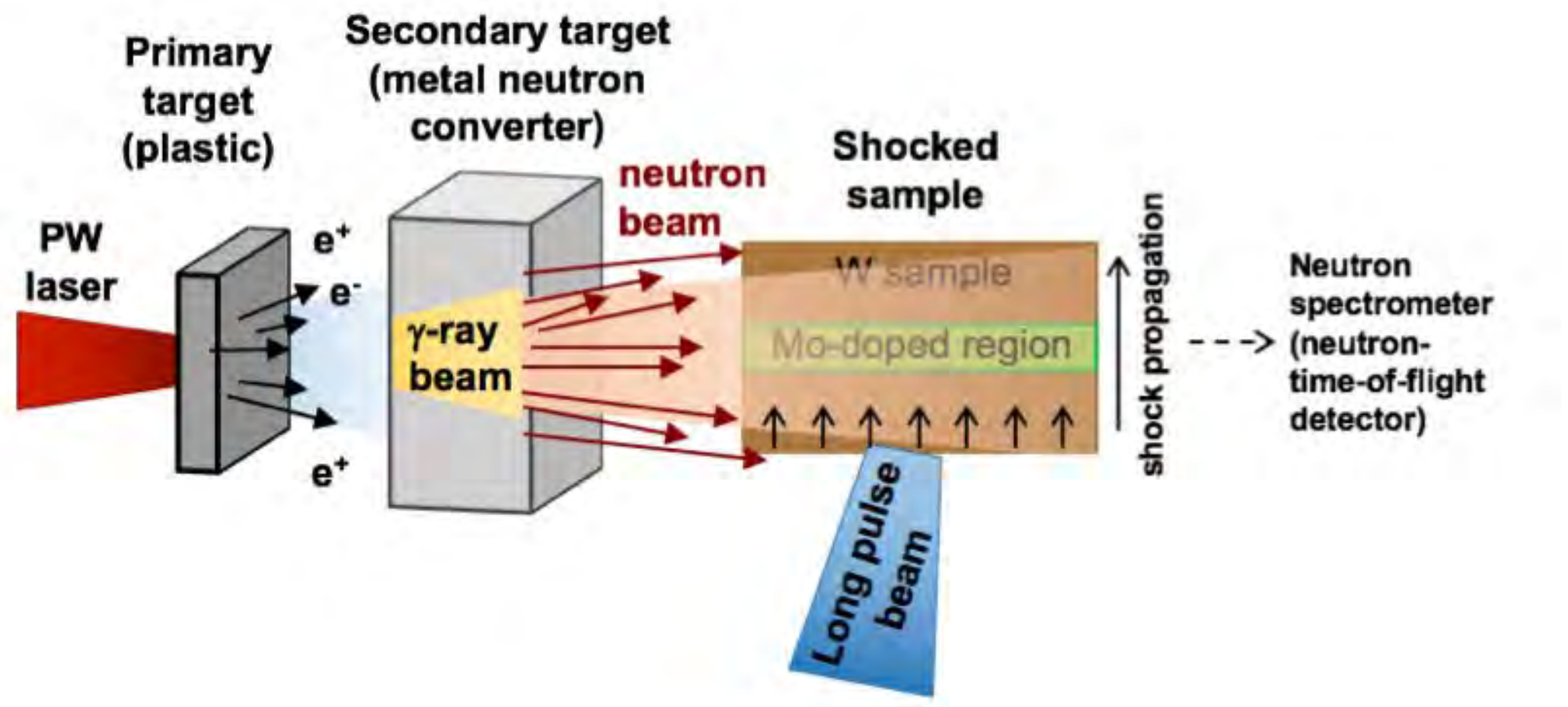

**Figure 3.7.2.** Concept for neutron resonance spectroscopy measurements of shocked materials using a laser-generated source of neutrons.

An extension of this resonance technique could determine the absorption line width in an energy spectrum and used to measure the bulk temperature of a transient sample for high energy density experiments, as illustrated in Fig 3.7.2. This approach could be applied in industry for temperature studies in samples not accessible by other thermometric methods, such as thermocouples or pyrometry.

#### 3.7.2.6 Laser-Driven Deuteron Sources

While moderation and neutron detector/beamline instrumentation are well understood, a suitable deuterium source still requires research efforts to develop targets using cryogenic deuterium jets, deuterated liquid crystals, and solids. Intense laser systems are required for research that could revolutionize production of pulsed neutrons.

### 3.7.3 Laser-Driven Sources of Exotic Particles

Recently, electrons with greater than 1-GeV energy have been generated using LWFA (Sec. 3.3). These high-energy electrons can produce high-energy x-rays through that can subsequently be used to produce sources of positrons, neutrons, and potentially more "exotic" particles such as pions and muons [87], as well as highly directional positron sources with energies greater than 100 MeV. Producing exotic particles with rest mass (>100 MeV/$c^2$) higher than

---

10-20684 (May 2011); D. L. Chichester and J. W. Sterbentz, "A Second Look at Neutron Resonance Transmission Analysis as a Spent Fuel NDA Technique," Report INL/COM-11-20783 (July 2011).

87 W Schumaker *et al.*, "Making Pions with Laser Light," New. J. Phys. **20** (7), 073008 (2018).

typically available from natural radioactive processes (decay, fusion, fission) require high-energy particles from an accelerator. These higher-energy particles have half-lives of 2 µs or less, making synchronization and high single-shot production rate crucial for potential applications and controlled measurements. Recent experiments have shown that pions can be generated through all-optical means by high-power lasers [85] (Fig. 3.7.3). To improve the statistics of these measurements it is necessary to have a much higher flux of multi-GeV electrons. Higher repetition rate (5 to 100 Hz) facilities delivering 200-TW to 1-PW pulses will be required in these applications.

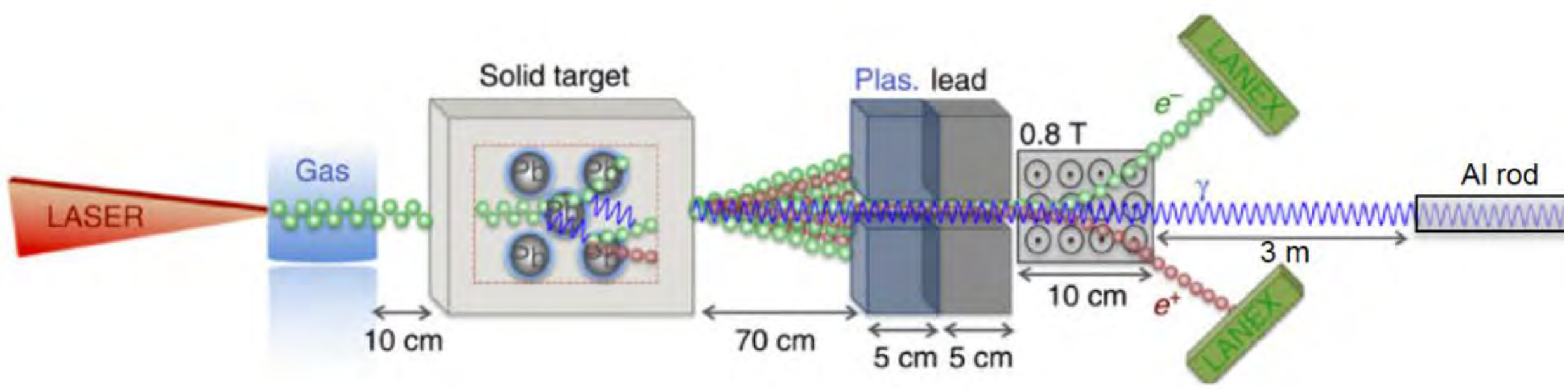

**Figure 3.7.3.** A simplified experimental geometry for measuring pion/neutron production is shown [87].

### 3.7.4 Summary and Recommendations

Nuclear photonics is a rapidly growing area of research, combining high-power lasers with nuclear techniques. Lasers are used to drive intense beams of energetic photons or particles and nuclear reactions are used to probe samples, provide quantitative, isotopic composition of large objects, verify the security of cargo, assess the integrity of vital infrastructure (bridges, buildings, turbines) and heal deep tumors in patients. This is a vital field for laser development, including:

a) Joule-class short-pulse lasers at high repetition rates (kHz to 100 kHz) for hard photon generation and fast neutron production;

b) High-energy lasers (100-300 J, 10-200 Hz) to provide single-pulse particle fluxes suitable for diagnosing transient phenomena and generating intense bursts of thermal and epithermal neutrons; and

c) High-energy, single-shot systems (shot per minute to 1 Hz) to provide highest pulsed particle yield to drive prompt nuclear reactions or for neutron spectroscopy.

# Chapter 4 – Establishing an Ultrahigh-Intensity Laser Research Ecosystem

Advancing ultrahigh-intensity scientific research in the U.S. depends on establishing a stable ecosystem that sustains existing laser facilities, upgrades them and builds new ones to stay competitive, supports new frontier science, and coordinates the research community.

## 4.1 Existing U.S. Ultrahigh-Intensity Laser Facilities – LaserNetUS

Mid-scale laser facilities play a critical role in the ecosystem of the U.S. research community. They enable breakthrough science, support development of new technology, and help educate and train a new generation of scientists that will drive innovations in the years to come.

Much of the mid-scale, high-power laser infrastructure available in the U.S. is now accessible to users through the LaserNetUS network [1] established by DOE Fusion Energy Sciences in 2018. This network serves U.S. scientists by providing access to domestic user facilities and enabling a broad range of frontier scientific research. The original LaserNetUS facilities include six university and three national laboratory–based high-intensity laser facilities, shown in Fig. 4.1.1.

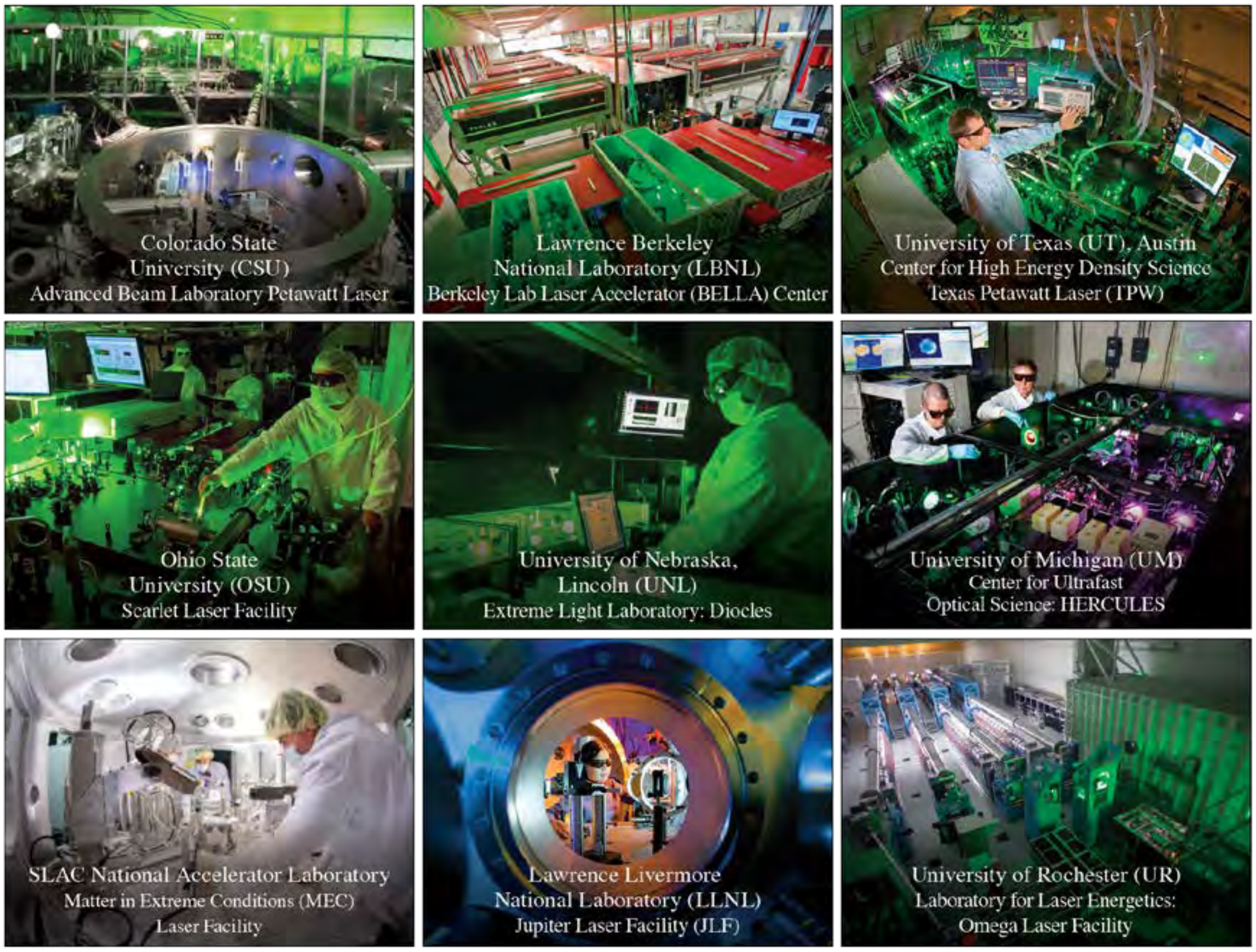

G12968J1

Figure 4.1.1. Original LaserNetUS laser facilities

1 “LaserNetUS,” https://www.lasernetus.org/.

LaserNetUS facilities are distributed geographically throughout the U.S., as shown in Fig. 4.1.2. They offer complementary laser and experimental capabilities. The ultrashort laser capabilities at these facilities are divided into two general types:

High-energy Nd:glass lasers:

- The **OMEGA EP Laser System** at the University of Rochester's Laboratory for Laser Energetics offers two kilojoule/picosecond laser beamlines with additional multi-kilojoule/nanosecond UV beamlines.
- The **Jupiter Laser Facility** at Lawrence Livermore National Laboratory (LLNL) has three operating lasers and target areas: Titan (up to 300 J depending on picosecond pulse duration), Janus (dual-beam kilojoule-class/nanosecond), and COMET (up to 10 J per picosecond). An upgrade will add a third high-energy beamline in 2020 for use in experiments in 2021.
- The **Texas Petawatt Laser** at the University of Texas at Austin delivers 150-J/150-fs laser pulses.

Ti:sapphire lasers:

- The **BELLA Center** at Lawrence Berkeley National Laboratory (LBNL) offers access to two lasers: the BELLA PW laser (up to 40 J/30 fs) and the BELLA HTW Laser with two beamlines (50-TW and 10-TW peak power) for multibeam experiments.
- The **Hercules Laser** at the University of Michigan is a 300-TW (9 J, 30 fs) laser at the Gerard Mourou Center for Ultrafast Optical Science (CUOS) that is being upgraded to 500 TW with NSF Major Research Instrumentation funding.
- The **Matter at Extreme Conditions (MEC)** instrument at the Linac Coherent Light Source (LCLS) x-ray FEL at the SLAC National Accelerator Laboratory combines the LCLS beam with high-power optical laser beams (1-J, 800-nm, 40-fs @ 5 Hz short pulses, and 4×15-J, 527-nm, shaped pulses up to 10 ns duration at a shot per 7 minutes with phase plates to shape the focal spot), and a suite of dedicated diagnostics. MEC is currently under consideration for a major upgrade by DOE Fusion Energy Sciences.
- The **Diocles Laser** at the University of Nebraska-Lincoln has a peak power and repetition rate of 100 TW and 10 Hz, respectively.
- The **Scarlet Laser** at Ohio State University produces 400-TW pulses (15 J in 40 fs) at a rate of one shot per minute and intensities at focus of $10^{21}$ W/cm$^2$.
- The **ALEPH Laser**, a 0.85 PW 800 nm system at Colorado State University, produces ultrahigh-contrast femtosecond pulses up to 10 J at 400 nm by frequency doubling 30-fs pulse at repetition rates up to 3.3 Hz in burst mode. An intensity of $6.5 \times 10^{21}$ W/cm$^2$ is obtained focusing the beam with an *f*/2 parabola. An *f*/1 parabola, available after July 2019, that is expected to deliver intensities $>1 \times 10^{22}$.

– The **Advanced Laser Light Source (ALLS)** at Université du Québec joined LaserNetUS in October 2019. ALLS includes several beamlines producing up to 220-TW pulses (4 J, 18 fs at 2.5 Hz) and intensities at focus up to $5 \times 10^{19}$ W/cm$^2$.

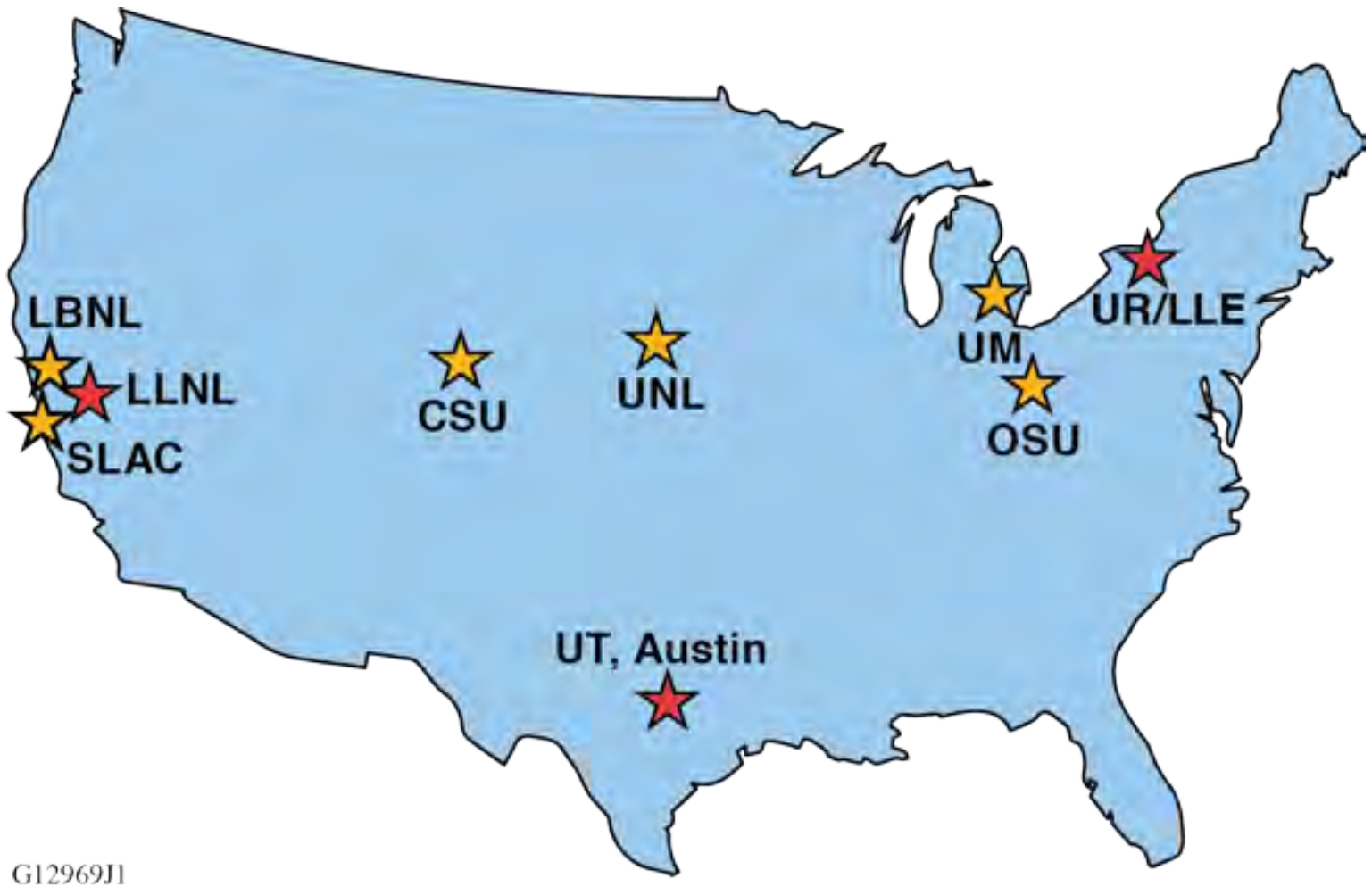

Figure 4.1.2. Original LaserNetUS high-energy Nd:glass (red) and Ti:sapphire (gold) laser facilities are geographically distributed across the U.S.

## 4.2 New Facility Capabilities Needed to Propel Frontier Science

Some of the science frontiers identified in Chap. 3 can be explored with existing laser facilities in the U.S., and another class of experiments will be possible with upgrades to them. Many of the most exciting pioneering experiments, however, will require laser facility capabilities beyond those currently available in the U.S. or even internationally. Developing new facilities that leapfrog the U.S. into a new position of leadership would certainly enable new discovery science.

LaserNetUS must develop over time to stay competitive with international facilities and serve as a key element in driving national research forward in high-field and high-energy-density plasma sciences. Near-term upgrades to existing facilities serve two purposes: (1) meet the ever-increasing demands of frontier science and (2) provide platforms for developing technology required in new facilities.

### 4.2.1 Near-term improvements to existing facilities

A first stage of future experiments to pursue new science opportunities could be realized given the following three general types of near-term (<five years) upgrades to existing laser facilities:

1. Deliver multiple beams simultaneously: Delivering multiple laser pulses at various wavelengths, intensities, and pulse parameters that are accurately synchronized into the same interaction volume is increasingly important for many "pump-and-probe" experiments. These include cases requiring secondary radiation produced by high-intensity laser pulses, as well as high-energy nanosecond pulses for compressing and/or preheating samples.

2. Precision laser performance: Many future experiments depend on tailoring the spatial (focal-spot shape), temporal (pulse shape and contrast), and polarization properties of laser pulses, as well as improved energy and pointing stability, to achieve high-precision studies of the extreme science described in Sec. 5.2.

3. Experimental systems for increased repetition rates: Delivering high-energy ultrafast laser pulses at repetition rates starting at one shot per minute and even higher to several Hz will enable acquiring orders-of-magnitude more data to fully characterize and understand complex phenomena, both directly with the laser and with high-flux secondary radiation sources. New experimental systems will need to be developed for this new high-rate experimental paradigm, including target diagnostics, target production and positioning, and debris mitigation to protect expensive final optics.

#### 4.2.2 Significant capability upgrades requiring larger and longer-term investments

Significant upgrades to existing facilities will require larger and longer-term (> five years) investments:

1. Increasing the repetition rate of petawatt-class lasers: Significant opportunities that require high-peak-power (>100-TW) ultrafast laser pulses are presently limited to Hz-class repetition rates. They include plasma particle accelerators, bright high-average-power x-ray and gamma-ray sources, high-average-power soft x-ray lasers, and high-average-flux ion and neutron sources. Some applications require average powers and repetition rates that exceed the present state of the art by orders of magnitude. Higher repetition rates at kHz and beyond will enable feedback control, leading to greatly increased beam stability and improved experimental accuracy.

   The highest-repetition-rate PW-class laser routinely used for applications has been limited to 1 Hz [2], even though petawatt-class Ti:sapphire lasers have been demonstrated up to 10 Hz [3]. Diode pumping and improved thermal management described in Sec. 5.1 would lead to multi-kW average-power lasers needed for plasma accelerators. A proposed k-BELLA laser (3-J, 30- to 100-fs pulses at 1 kHz) is the first step in a technology road map for laser wakefield acceleration (LWFA) [4]. Longer term development leading to megawatt-class average powers is needed that would rely on emerging architectures, such as coherently combining many fiber lasers, bulk thulium-doped crystals, or novel laser ceramic hosts.

2 K. Nakamura *et al.*, "Diagnostics, Control and Performance Parameters for the BELLA High Repetition Rate Petawatt Class Laser," IEEE J. Quantum Electron. **53** (4), 1200121 (2017).

3 Y. Wang *et al.*, "0.85 PW Laser Operation at 3.3 Hz and High-Contrast Ultrahigh-Intensity $\lambda$ = 400 nm Second-Harmonic Beamline," Opt. Lett. **42** (19), 3828–3831 (2017); E. Sistrunk *et al.*, "All Diode-Pumped, High-Repetition-Rate Advanced Petawatt Laser System (HAPLS)," in *Conference on Lasers and Electro-Optics*, OSA Technical Digest (online) (Optical Society of America, Washington, DC, 2017), Paper STh1L.2.

4 *Report of Workshop on Laser Technology for k-BELLA and Beyond*, Lawrence Berkeley National Laboratory, Berkeley, CA (9–11 May 2017).

2. Extending wavelengths to 10 $\mu$m and beyond: The ponderomotive force that accelerates electrons and ions from laser-ionized plasmas scales favorably with wavelength. High-power, mid-infrared (mid-IR) lasers could generate millimeter-scale plasma bubbles, self-channel in the atmosphere, generate light spanning the spectrum from x rays to THz waves, and exploit new regimes in nonlinear optics. At present, most mid-IR, high-peak-power sources rely on optical parametric amplifiers (OPAs) pumped by 1- to 2-$\mu$m lasers. These parametric amplifiers become prohibitively inefficient in the long-wave IR range (8 to 14 $\mu$m). $CO_2$ lasers operating at 10 $\mu$m are currently the most viable candidates for reaching very high relativistic intensity (>$10^{16}$ W/cm$^2$) and high-energy pulses in the long-wavelength portion of the IR spectrum given their ability to store a great deal of energy within the active medium. Significant progress has been made in the U.S. by amplifying few-picosecond pulses to greater than 10-TW peak powers [5].

   Switching from the traditional electrical-discharge pumping to optical pumping of high-pressure centimeter-scale $CO_2$ cells could lead to compact systems that support 10×– shorter pulse lengths while simultaneously increasing the repetition rate to 10 to 100 Hz. A long-term, ten-year goal is to develop a ≥10-TW (3 to 5 J, 300 to 500 fs) optically pumped $CO_2$ laser system running at 10 to 100 Hz.

### 4.2.3 New Laser Facilities with Parameters Well Beyond Any Current Lasers

Many of the most exciting scientific research identified in Chap. 3 will require new laser facilities in the US with parameters well beyond any lasers currently operating**.** Generating scientifically interesting conditions in plasmas or materials requires a certain amount of energy per unit volume. Higher laser pulse energies make it possible to produce these conditions at larger volumes and with more spatial homogeneity, higher temperatures, and pressures. Higher peak fluxes are also possible from high-energy photon or secondary-particle sources.

The U.S. still leads in the development of technology needed for next-generation laser facilities: both high-intensity lasers deployed at ELI-Beamlines in the Czech Republic were built in the U.S. The L3 [also called High-Repetition-Rate Advanced Petawatt Laser System (HAPLS)] laser developed by LLNL delivers petawatt pulses (30 J in 30 fs at 800-nm wavelength) at repetition rates up to 10 Hz. It uses helium gas cooling to remove heat from the pump and Ti:sapphire gain elements. The L4 laser developed by National Energetics is designed for two different modes of operation: one delivers 10-PW (1500 J in 150 fs) pulses at 1-$\mu$m wavelength; another can simultaneously produce high-energy shaped nanosecond pulses (up to 2 kJ) and 1-PW pulses. It uses liquid-cooled, split-disk Nd:glass amplifiers to operate at up to one shot per minute. ELI Nuclear Physics recently installed two 10-PW Ti:sapphire lasers developed by Thales (France) that are being commissioned.

---

5 D. Haberberger, S. Tochitsky, and C. Joshi, "Fifteen Terawatt Picosecond $CO_2$ Laser System," Opt. Express **18** (17), 17,865–17,875 (2010); M. N. Polyanskiy, I. V. Pogorelsky, and V. Yakimenko, "Picosecond Pulse Amplification in Isotopic $CO_2$ Active Medium," Opt. Express **19** (8), 7717–7725 (2011).

Significant opportunities for scientific exploration exist when new capabilities are collocated with existing scientific infrastructure. For instance, the Linac Coherent Light Source (LCLS) x-ray free-electron laser (FEL) at the SLAC National Accelerator Laboratory has a rich history of ground-breaking discoveries enabled by its coherent, femtosecond x-ray beamline. Likewise, the Omega Laser Facility at the University of Rochester's Laboratory for Laser Energetics (LLE) facilitates cutting-edge high-energy-density research. Establishing two new and complementary U.S. facilities would leapfrog the laser capabilities being developed and deployed now in Europe and Asia and provide unique capabilities that would reestablish U.S. leadership in high-field physics.

#### 4.2.3.1 High-energy and high-intensity lasers collocated at LCLS

The LCLS MEC instrument currently uses the XFEL beam to probe materials under dynamic compression and heating produced with relatively low-energy, ultrashort-pulse and nanosecond compression lasers. A proposed DOE Fusion Energy Sciences MEC laser and facilities upgrade, shown in Fig. 4.2.1, would add a 10-Hz, multi-PW laser along with a kJ, UV compression laser system with temporally shaped nanosecond pulses. Including multiple kJ compression lasers is also being considered. These new lasers would significantly increase the range of density, pressure, and temperature regimes of materials under study. Experiments combining the PW laser with a relativistic electron beam would also enable studies of nonlinear Compton scattering, as well as strong-field QED.

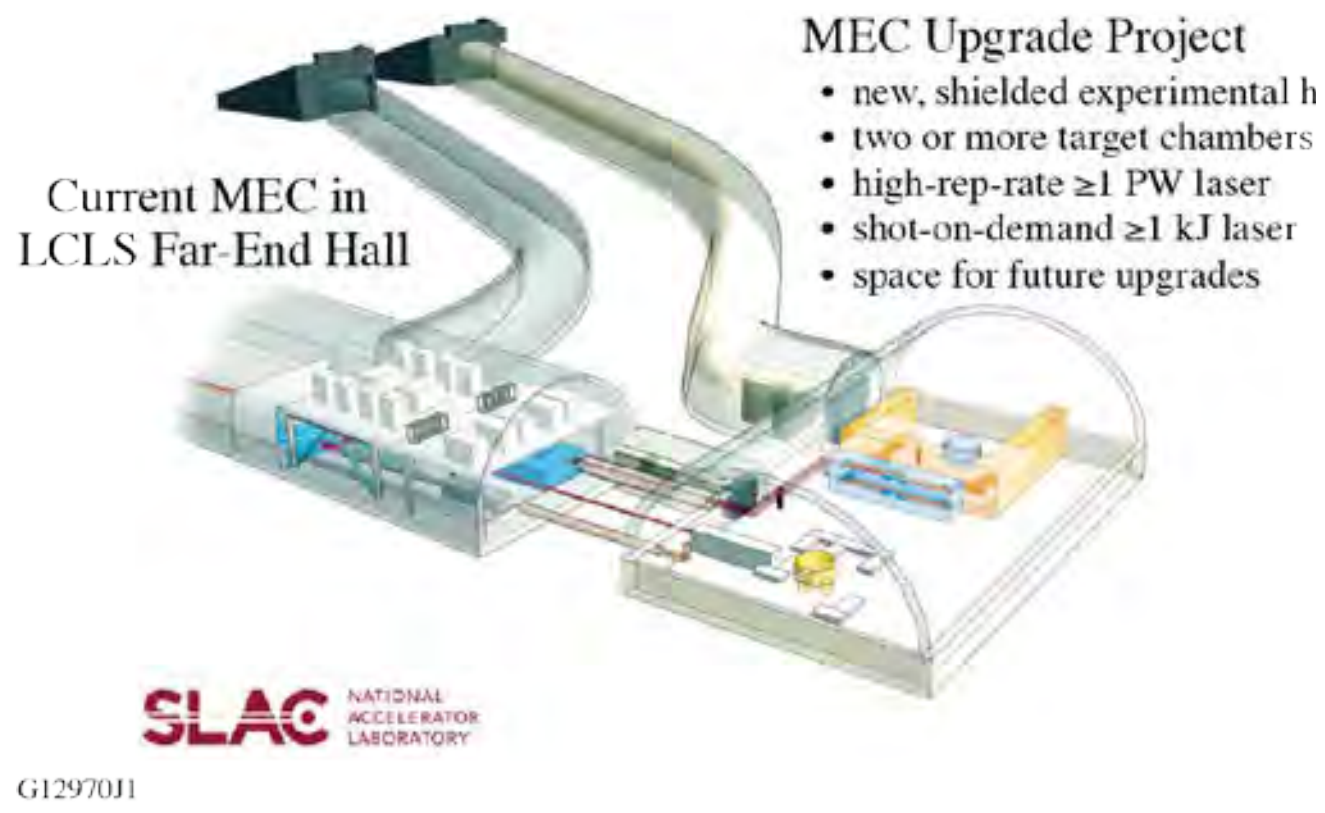

Figure 4.2.1: Notional layout of a PW and kJ nanosecond drive laser upgrade in the MEC end station section of the LCLS x-ray FEL.

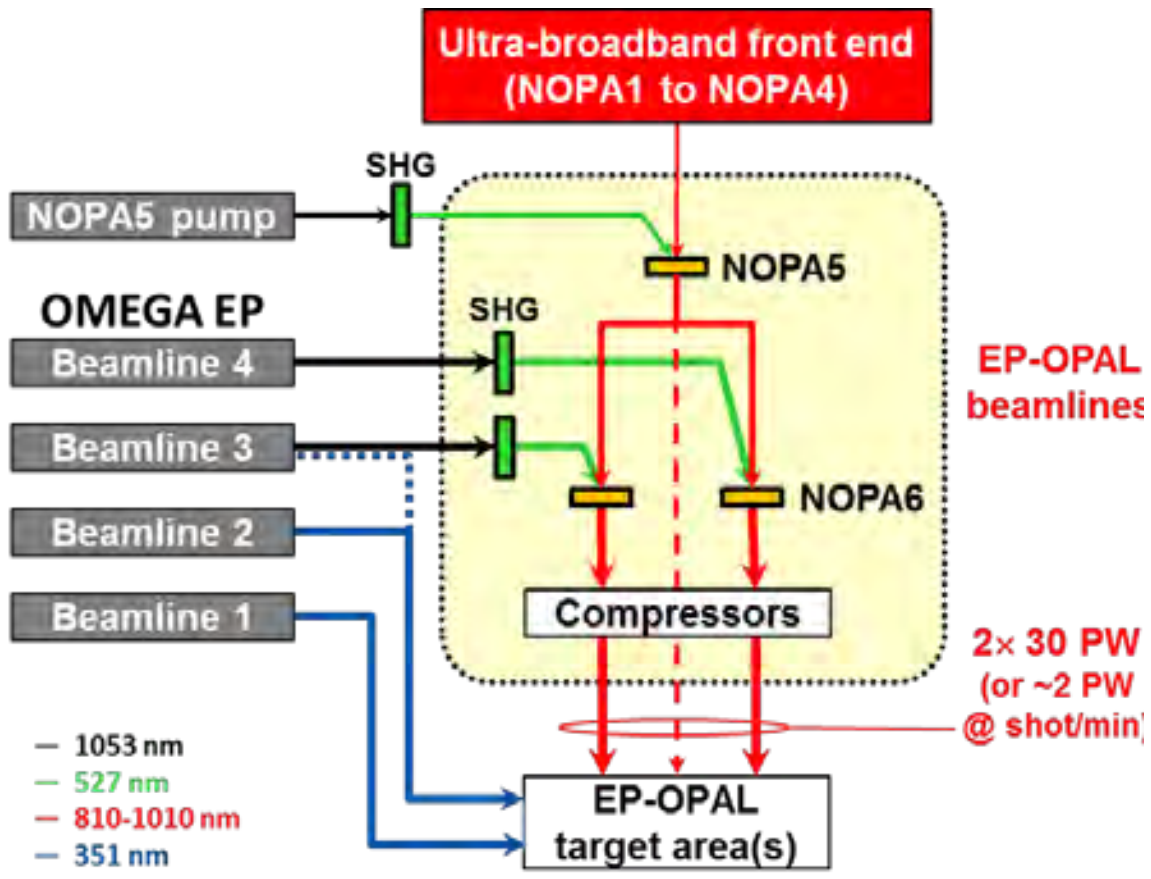

Figure 4.2.2: Schematic of a 2×30 PW laser using beamlines from the OMEGA EP laser to drive femtosecond, high-energy OPCPA amplifiers.

#### 4.2.3.2 Ultrahigh-field laser facility with multiple multi-10 PW beamlines

An ultrahigh-field laser facility with at least two beamlines delivering peak powers greater than 10 PW and focal intensities greater than $10^{23}$ W/cm$^2$ would establish a world-leading capability. One proposed approach, shown in Figure 4.2.2, would use two OMEGA EP beamlines reconfigured to pump independent 30-PW optical parametric amplifier lines (OPALs) delivering 600 J in 20-fs pulses. The two EP-OPAL beamlines combined with the remaining OMEGA EP beamlines would enable a wide array of compelling science, including high-energy-density sciences, quantum science with extreme fields, relativistic particle beams, and advanced light sources.

### 4.2.4 Long-Term (10+ years) Facilities Requiring Further Technology Development

Chapter 3 identifies science opportunities that would demand extreme focused intensities as high as $10^{26}$ W/cm$^2$, such as producing electron/positron pair plasmas, studying high-field physics, and generating bright relativistic ion, neutron, and electron sources for studying nuclear physics and nonlinear QED. Setting a goal to build an exawatt-class laser producing focal intensities up to $10^{26}$ W/cm$^2$ would ensure U.S. scientific leadership in high-field physics for a generation.

Figure 4.2.3 shows a high-level concept for such an exawatt laser facility with an array of coherently combined 100-PW beams with low effective *f*-number focusing. This system architecture assumes that coherently combining fewer units with the largest practical output beam size and energy is less challenging than combining a larger number of smaller units.

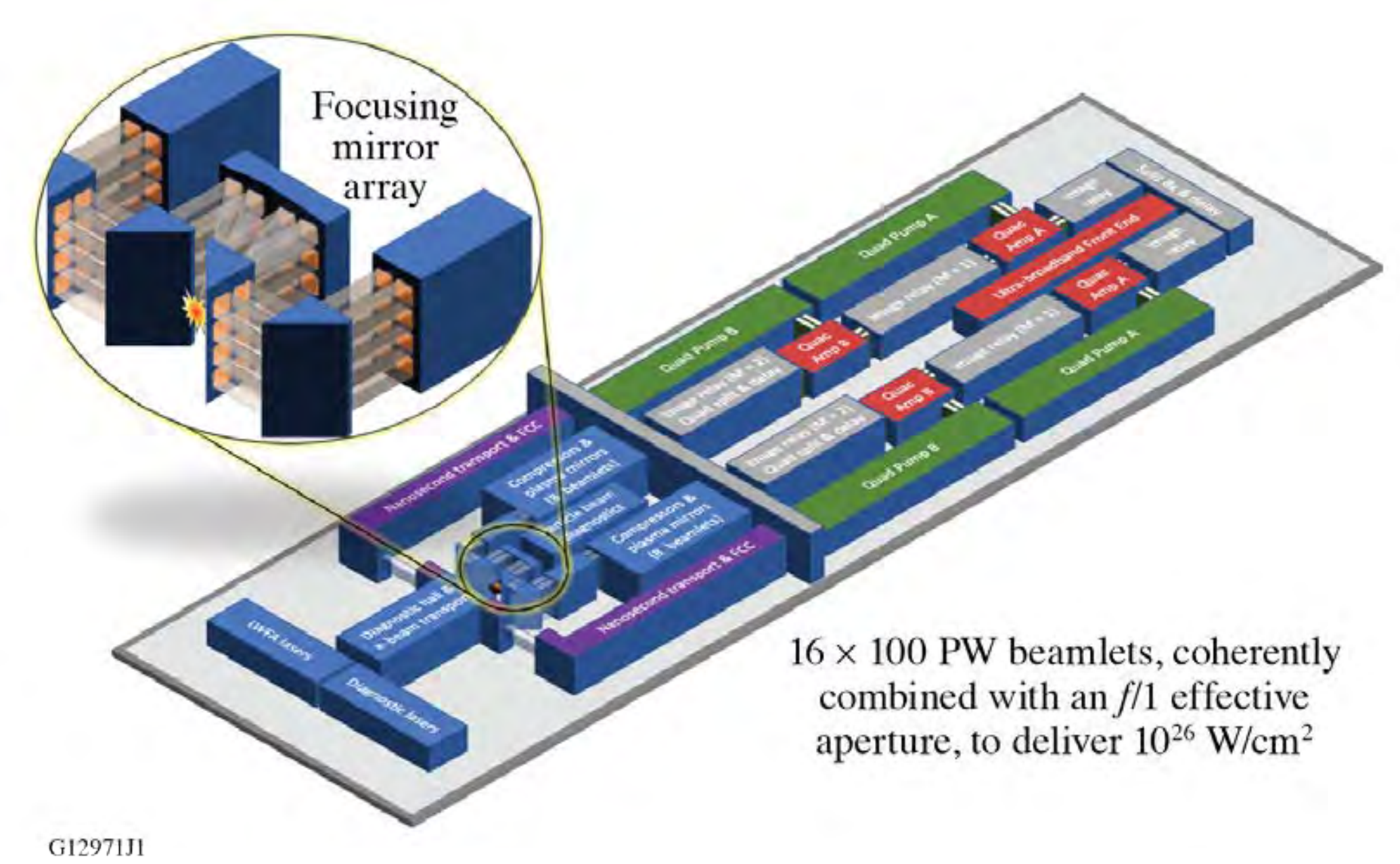

Figure 4.2.3. Concept for a future facility including a laser system capable of focused intensities up to $10^{26}$ W/cm$^2$, high-energy particle diagnostics, and auxiliary experimental and laser systems.

A common front end seeds parallel beamlines with final amplifiers and compressors. The front end needs sufficiently broad bandwidth, ultrahigh temporal contrast, and high energy before splitting. An OPCPA version would target 100-PW, 20-fs, 2-kJ beamlet units, while a Nd:glass–based chirped-pulse–amplification (CPA) version would target 100-PW, 100-fs, 10-kJ units.

A vision for a future system capable of $10^{26}$ W/cm$^2$ is required to identify and prioritize technology research and development needs that must be fulfilled. Based on current near-infrared laser technologies such as glass lasers, or glass-laser–pumped OPCPA (or Ti:sapphire), the most plausible strategy for such a laser would be to coherently combine multi-tens of PW units with near-diffraction-limited focusing to an approximately 1-$\mu$m$^2$ spot. Realizing an exawatt laser facility requires a focused program to develop technologies, including:

- coherent combination of 20- to 100-fs, 100-PW pulses for an effective *f*/1 focus,
- large-aperture, low-dispersion, broadband gratings (for a 2-m$^2$ beam) with high uniformity and low scattering,
- high-precision pointing stability and control (sub-$\mu$rad) and wavefront control,
- repetition-rated lasers for pumping broadband kilojoule amplifiers (OPCPA or Ti:sapphire), or chirped-pulse amplification,
- ultrahigh temporal contrast (>$10^{14}$),
- plasma optics for contrast improvement and focusing-configuration flexibility,
- on-target intensity measurements for *in-situ* confirmation,
- femtosecond-scale timing control and fiducial distribution across facility elements, and
- diagnostics for high-energy particles (multi-GeV electrons, GeV ions) and photons.

Another long-term technology goal requiring technological development is very high-repetition-rate, high-energy (e.g. >10 J and >50 kHz), femtosecond lasers. Near-term development of high repetition rate lasers delivering pulse energies of a few joules at a few kHz will enable high-precision laser wakefield acceleration experiments and light sources. Much higher average powers will be necessary to drive collider stages and other applications that will require research and development in several areas, including: gain materials, thermal management, and high damage threshold multilayer coating with high durability. A push towards high energy and high repetition rates would maintain U.S. leadership in advanced laser development.

## 4.3 Community Coordination

The NAS report *Opportunities in Intense Ultrafast Lasers: Reaching for the Brightest Light* identified improving coordination across the U.S. high-intensity laser community—universities, national laboratories, and industry—as an important element of advancing important science and applications, as well laser facilities and the underlying technology. Conclusion 1 of the report noted, "*Coordination between industry and government is limited and often inadequate*" and Conclusion 7 highlighted "*University/Laboratory/ Industry cooperation is necessary to retain and renew the talent base. Cooperation among all sectors ... has proved essential and the current situation could be improved to develop a robust national talent pool and a strong technology base for this fast growing area*." The NAS report recommended DOE lead development of a **comprehensive interagency national strategy for high-intensity lasers**. It also called for programs to develop and operate mid- and large-scale projects, as well as establishing a technology development program with technology transfer across the community. This section addresses workshop outcomes related to these recommendations and conclusions.

Adopting **standards and sharing best practices** promises a way to reduce both capital and operating costs while improving performance and maximizing compatibility, interoperability, safety, repeatability, and quality. One of the greatest benefits would be **enhanced "network effects."** Standards and coordination increase compatibility and interoperability, allowing information to be shared within a larger network and attracting more users of new technologies, further enhancing network effects. Standardization can facilitate commoditization of formerly custom products and processes, which can lead to broader markets where only niche markets might otherwise exist.

Chapter 5 identifies technical research needs (TRNs) that must be addressed to meet the scientific research needs presented in Chap. 3. These TRNs include five critical technologies along with an overarching need to develop a U.S. supply chain for intense lasers. These technology challenges represent general needs that might be solved individually for every project, but taking this approach would essentially "reinvent the wheel" multiple times. **Standardizing technology** can avoid competition of rival and incompatible approaches that can slow or even kill the growth of a promising technology because of market fragmentation. **Modularization** within a standard can lead to increased flexibility and rapid introduction of new features and products to better meet specific needs. Standardized components, subsystems, testing methods and instrumentation, and data formats can optimize logistics and reduce the total cost of ownership.

*De facto* standards can arise organically, but **coordinating research and development** to satisfy the broader needs of the community and aligning it with a national strategy would be more efficient. For example, high-energy, repetition-rated lasers are needed for several purposes: directly amplifying short pulses using chirped-pulse amplification, and pumping broadband laser and parametric amplifiers, as well as compressing and heating experimental targets. **Developing "laser classes" to address these applications offers significant advantages**, much as navies build classes of ships or automobile and aircraft manufacturers develop lines of products with similar designs that reuse proven components and subsystems like engines, drive trains, controls, detectors, etc. Nonrecurring engineering costs can be spread across multiple units. Proven designs can be easily adapted and extended to new applications while minimizing cost and development time.

**Adopting standards** depends on setting performance specifications and timelines that can meet broad needs. **Road maps** can identify facilities required to meet established scientific research needs, as well as needed technologies that can benefit by taking a standards-based approach to research and development. Developing and implementing technical standards requires consensus and compromise among parties involved, including both users and suppliers. All parties can realize mutual gains by making mutually consistent decisions. A community-based organization, like LaserNetUS, is well suited to facilitate this process if it includes industry and the initiative enjoys broad community and cross-agency federal support.

**Joint research centers located at universities**, such as the Fraunhofer Institutes in Germany that are supported by industry and government, **have proven extremely effective at developing breakthrough technologies, spawning new industries, and training a highly skilled workforce**. For example, the Fraunhofer Institute for Applied Optics and Precision Engineering (Fraunhofer IOF) is located close by Friedrich-Schiller-Universität (FSU) in Jena, Germany. Together with the FSU Institute of Applied Physics, Fraunhofer IOF conducts basic and applied research to control light from its generation and manipulation to applications. Technology transfer from one of these collaborations to a spin-off company, Active Fiber Systems (AFS) GmbH, led to commercial production of reliable laser systems suitable for scientific and industrial applications. AFS GmbH started with high-end, customized solutions for cutting-edge facilities like the high-repetition-rate lasers (HR1 and HR2) at ELI-ALPS. They refined this expertise to produce industrial-grade lasers for materials processing and high harmonic generation (HHG) and lasers for nonlinear microscopy. FSU students and graduates are highly sought after by research institutions and companies across Germany and the world.

The U.S. can replicate this type of community by partnering with existing companies and launching new ones to pursue comparable results in the field of ultraintense lasers. Some successful examples of technology transfer and synergy between universities and their spin-off in this field already exist in the US: the University of Texas at Austin and National Energetics, Colorado State University and XUV Lasers, and the University of Colorado and KM Labs.

**Laser facility operations** represent another opportunity to coordinate the U.S. high-intensity laser community. The International Laser Operations Workshop (ILOW) provides a forum for organizations that operate large-scale laser facilities to discuss issues and exchange information related to laser facility operations, governance, maintenance, diagnostics, safety and security, and technology. The biennial workshop presents a significant opportunity for laser operations staff to establish professional contacts and share best practices, as well as lessons learned, to strengthen the international community. These exchanges serve to build cooperation while also spurring healthy competition among the participants. ILOW was initiated by major laser fusion facilities, but it has recently expanded to include ultra-intense laser facilities, including the Extreme Light Infrastructure. Integrating facilities from LaserNetUS, LaserLab Europe, and the Asian Intense Laser Network would be a natural next step.

# Chapter 5 – Technology Research Needs (TRNs) and Priorities

A significant focus of the BLI workshop was technology research and development needs to realize the capabilities that will enable the exciting science discussed in Chap. 3 and facility capabilities in Chap. 4. This chapter addresses five key areas of technology development and one overarching area requiring attention. The technology needs span new components, materials, system architectures, diagnostics and control schemes. Two areas promise significant opportunities to advance ultrahigh-intensity science: improved thermal management to enable high-average-power and high-repetition-rate lasers (Sec. 5.1); and cutting-edge laser technology for high-precision extreme science (Sec. 5.2). Other important needs include: new pulse compression technologies (Sec. 5.3); advances in large-aperture and high-damage-threshold optics for femtosecond pulses, and basic optical material development, such as new broad-bandwidth gain materials (Sec. 5.4); and advanced experimental systems (Sec. 5.5). After introducing each technology need, each section below maps it to BLI scientific research needs and required facility capabilities, outlines related technology R&D efforts, and identifies priorities.

Success in developing these technologies depends on active participation by the lasers and optics industry. Revitalizing commercial capabilities and developing strong supply chains in the U.S. (Sec. 5.6) represents a strategic opportunity to reverse a trend where leadership has been shifting from the U.S. to Europe and Asia.

## 5.1 Thermal Management to Enable High-Average-Power and High-Repetition-Rate Lasers

### 5.1.1 Introduction

Increasing the repetition rate of high-energy and high-intensity lasers enables higher experimental productivity and promises improved performance by using feedback from the experiment to the laser. High repetition rate maximizes the amount of data per experiment campaign and/or the number of experiments per week/month. Large volumes of data can be collected to enable statistical analysis of experimental results, which is essential for experiments in which signals of interest are small relative to background, signals that occur rarely, or when experimental conditions vary from shot to shot. High repetition rate also enables scanning laser parameters to find experimental signals or vary experimental conditions. For experiments in which the laser produces secondary particles, high repetition rate enables a high flux of these particles. Even for experiments in which the rate of experiments is limited by other factors, such as target delivery, data acquisition, or debris/radiation generation, operating lasers at high repetition rate enables feedback control and the application of modern data processing and optimization methods through machine learning to yield high stability and repeatability with consistent performance parameters at whatever rate is needed for experiments.

Thermal management is a critical issue for high-energy and high-average-power laser systems. Gas (e.g., $CO_2$) and liquid (e.g., dye lasers) lasers can use flow architectures to "refresh" the gain medium in the active region for every laser pulse, or on a continuous basis for cw laser systems, to remove heat, but neither of these type lasers are generally well suited for achieving the highest peak powers because of narrow gain bandwidths or poor energy storage. With some exceptions, olid-state lasers are most appropriate but they require careful attention to thermal management.

*Advanced thermal management technologies will enable high-average-power and high-repetition-rate lasers needed for high-precision extreme laser science and applications.*

The source of heat in solid-state lasers arises from pumping the gain medium. Ideally, the total heat load results only from the quantum defect of the laser gain medium—the difference between the energy of the pump photons and the energy of the extracted photons. In reality, other heat sources also exist, including energy transfer up-conversion, excited-state absorption, and other nonradiative loss mechanisms induced by co-dopants, impurities, and other defects, as well as absorption of amplified spontaneous emission (ASE), particularly in edge cladding material. Both steady-state and transient (sometimes called "prompt") thermal effects can induce thermal gradients in optical materials that cause stress-induced depolarization (most problematic in isotropic materials like glass and YAG) and spatially dependent changes in the optical path length that degrade beam quality. Ultimately thermal loads and restricted geometrical constraints may induce so much stress, that it can fracture the solid-state laser gain medium.

Solid-state lasers fall into three main types based on their pumping and thermal management geometries: rods, disks, and multi-slabs. **Figure 5.1.1** depicts these three categories and the strengths and weaknesses of the thermal techniques they employ.

**Rod amplifiers** constitute the majority of energetic industrial lasers since they are simple to pump and capable of storing many joules of energy, and the laser beam does not pass through the coolant. Zig-zag slab lasers (not shown) similarly remove heat in a direction transverse to beam propagation but only in one direction, which allows scaling the beam size in the other transverse dimension to increase pulse energy while maintaining favorable thermal characteristics. This usually requires complicated optical systems to convert to highly asymmetric beams and back to symmetric beam profiles for practical applications.

Increasing the cooling surface area for a given gain volume of the rod-shaped laser medium improves heat extraction. Large surface area to volume ratios support efficient heat transfer and maintain small temperature gradients, resulting in improved repetition rate and average power. Simultaneously increasing the length of the rod and decreasing the diameter greatly increases this ratio.

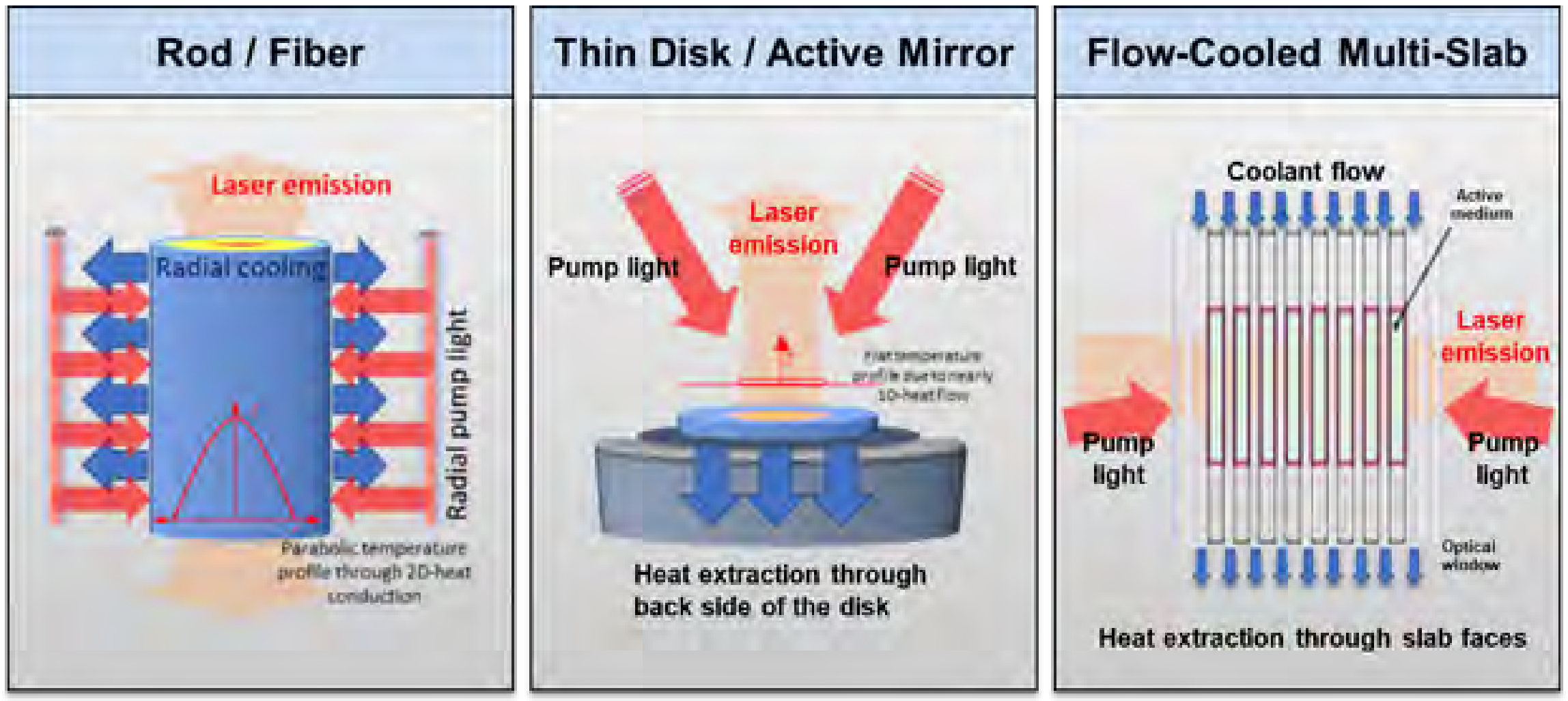

**Figure 5.1.1.** (a) Rod amplifiers have excellent pumping and energy storage, but radial heat flow produces significant transverse temperature gradients that cause optical distortions. (b) Thin-disk lasers minimize temperature gradients that are primarily longitudinal with respect to the laser propagation direction, but limit gain media volume for energy storage. (c) Multi-slab lasers provide large surfaces and large volumes of active material for energy storage, but the laser beam traverses the coolant.

**Fiber amplifiers** represent the limiting case of the rod geometry. As such, fibers have been used in many of the highest average-power laser systems. The relatively small aperture of a single fiber limits pulse energies, which in ultrashort pulse CPA architectures can only reach the µJ to mJ range. These energy scaling limitations may be overcome using coherent signal combining in the spatial, temporal, and/or spectral domains to reach sufficiently high combined pulse energies at very high average powers for multi-kHz repetition rate sources that are required for the applications described in sections 3.3., 3.5 and 3.7. Fiber lasers boast high efficiency and the potential for monolithic integration to develop compact and robust high power systems.

**Thin-disk lasers** use “active mirror” gain elements that conductively extract heat through one side of the disk. This enables efficient heat extraction in the direction of laser beam propagation but it limits the volume of the gain medium, which is proportional to energy storage. Thin-disk lasers can deliver high average powers, but the highest energy from thin-disk laser systems is currently limited to ~1 J. Designs with excellent thermal conductivity, like Ti:sapphire and/or cryogenically cooled Yb:YAG, show promise in scaling to thicker disks and higher energy storage up to a few joules while scaling to average power beyond ~1 kW.

**Thick-disk lasers** overcome the energy-storage limitation of thin disks by trading off a thicker gain medium in an active mirror geometry. Cryogenically cooled Yb:YAG thick-disk amplifiers generating 1.4 J at repetition rates of 0.5 kHz have been demonstrated [1].

**Multi-slab amplifiers** with face cooling promise a way to realize high average power (>1 kW) and high energy (~10 to 1000 J) by splitting the laser gain volume into multiple thin slabs and extracting heat with coolant flowing across their surfaces. This vastly increases the surface area available for cooling and reduces the path length for heat diffusion, but the laser beam must propagate through the coolant. Traversing the coolant can degrade beam quality because of spatial variations in coolant density and temperature. In principal, this approach can be scaled to a very large (20- to 40-cm) aperture, enabling amplification of up to multi-kJ energies.

**A** number of laser systems use **gas cooling** [2]. Typically, helium gas is used because it is chemically inert and has the lowest thermo-optic coefficient, $dn/dT$, resulting in the smallest impact to beam quality. This approach has been used to construct high-energy pump lasers for Ti:sapphire amplifiers with repetition rates of 3 to 10 Hz and energy ~100 J. **Liquid cooling** implemented in large-aperture (up to 30 cm), flash-lamp–pumped Nd:glass disk amplifiers produce pulse energies up to 2 kJ in a Nd:glass CPA system at a repetition rate approaching one shot per minute [3].

**Coherently combined laser arrays** amplify signals in multiple separate parallel paths and that are subsequently recombined into a single signal. This approach leverages high-average power laser technologies with limited maximum energies, such fibers and thin disks, and offers a pathway to very high average powers using distributed heat removal from each channel.

**Optical parametric amplification** (OPA) at high average power is possible in principle since the residual energy (quantum defect) from the pump-to-signal photon conversion process takes the form of an idler photon that propagates out of the amplifier; however, pump, signal or idler absorption can lead to thermally induced phase mismatches that degrade parametric amplification in this three-wave mixing process. OPA thermal management represents a new frontier since it represents a promising approach for future ultra-broadband and tunable sources.

---

1 C. Baumgarten *et al.*, “1 J, 0.5 kHz Repetition Rate Picosecond Laser,” Opt. Lett. **41** (14), 3339–3342 (2016).

2 A. Bayramian *et al.*, “The Mercury Project: A High Average Power, Gas-Cooled Laser for Inertial Fusion Energy Development,” Fusion Sci. Technol. **52** (3), 383–387 (2007); E. Sistrunk *et al.*, “All Diode-Pumped, High-Repetition-Rate Advanced Petawatt Laser System (HAPLS),” in *Conference on Lasers and Electro-Optics,* OSA Technical Digest (online) (Optical Society of America, Washington, DC, 2017), Paper STh1L.2; M. Siebold, F. Roeser, M. Loeser, D. Albach, and U. Schramm, “PEnELOPE: a High Peak-Power Diode-Pumped Laser System for Laser-Plasma Experiments,” Proc. SPIE **8780**, 878005 (2013).

3 G. Chériaux *et al.*, “kJ-10 PW Class Laser System at 1 Shot A Minute,” Proc. SPIE **10898**, 1089806 (2019).

### 5.1.2 Mapping to BLI Scientific Research Needs and Facility Capabilities

Substantial progress in thermal management of high-energy lasers has been made recently for both CPA and pump lasers, but additional research and development is required to realize the high-repetition-rate systems needed to address the science described in Chap. 3 and the new facility capabilities identified in Chap. 4. Providing highly focusable pulses with up to kilojoule energies and peak powers up to and beyond 10 PW at repetition rates up to 10 Hz and beyond poses a significant technical challenge that requires concentrated and sustained research and development.

All Scientific Research Needs (SRNs) described in Chap. 3 will benefit greatly with higher repetition rates to yield more data. Repetition rate and pulse energy determine laser average power, which needs to scale from current systems that operate at several hundred watts to multiple kilowatts and ultimately to megawatt-class systems. At each step, repetition rates and energies trade-off for each SRN. Those addressing extreme science (Secs. 3.1, 3.2, and 3.4) initially will require kilojoule-class energies and be satisfied with shot per minute or multi-Hz repetition rates, while others directed toward applications (Secs 3.3, 3.5–3.7) demand multi-kHz repetition rates at lower energies.

### 5.1.3 Related BLI Technology R&D Areas

**5.1.3.1 New broadband laser materials** with smaller quantum defects would reduce thermal loading. Materials with high thermal conductivity and thermal shock parameters would maintain excellent beam quality. **Yb-doped gain materials** boast intrinsically low quantum defects for lasing wavelengths near 1 $\mu$m, and large storage times that allow high-energy operation with diode pump lasers; however, many of these hosts have bandwidths that support only picosecond pulses, though some materials, like Yb:glass or Yb:CaF can amplify 100-fs pulses. Yb-based laser materials present many advantages, like broad gain bandwidth and they can be diode pumped, but they can be challenging for efficient high-energy pulse amplifiers due to their high saturation fluence.

**Cubic sesquioxides** are examples of potential host materials for high-power ultrafast lasers, but growing these crystals is very challenging because of their extremely high melting points. Ytterbium-doped sesquioxides, such as $Lu_2O_3$, $Sc_2O_3$, and $Y_2O_3$, have wide emission bands and high thermal conductivity suitable for short-pulse amplification at high average power [4]. Substitution of $Yb^{3+}$ ions in sesquioxides does not require any charge compensation, which makes them easier to fabricate, especially at high doping levels. Developing transparent laser ceramics composed of polycrystalline sesquioxide or other materials that can be scaled to large apertures for high-energy CPA would represent a significant advance.

**Tm-doped laser materials** provide a favorable alternative to Yb-doped crystals for laser operation in the 2-μm wavelength region that can support pulses as short as 100 fs. An example is Tm:YLF, where calculations suggest diode-pumped operation up to 300-kW average power [5]. Like Yb-doped crystals, Tm-doped materials typically operate as quasi-three-level lasers with high saturation fluence and low stimulated emission cross section unless cryogenically cooled. $Tm^{3+}$ excited states in several crystalline media have lifetimes nearly an order of magnitude longer than Yb. Pump energy stored in the laser medium can be efficiently extracted at well below the saturation fluence if operated in "multi-pulse extraction" mode where extraction takes place many times per gain lifetime, which is analogous to continuous-wave operation.

---

4 A. Pirri *et al.*, "An Overview on Yb-Doped Transparent Polycrystalline Sesquioxide Laser Ceramics," IEEE J. Sel. Top. Quantum Electron. **24** (5), 1602108 (2018).

**5.1.3.2 Resonant diode pumping** into the upper laser level is a proven technique to increase quantum efficiency and reduce thermal loading. Wavelength-stabilized diode arrays for pumping broadband laser materials would result in higher-power operation than current diode pumping technology.

**5.1.3.3 Passive and adaptive wavefront control** to deal with mm- or sub-mm-scale spatial wavefront distortions from coolant temperature and density fluctuations would improve beam quality when using aggressive cooling scenarios. Passive techniques include minimizing the fluctuations by optimizing coolant flow properties and design, while active control schemes need bandwidths sufficiently high address relevant time scales as short as milliseconds.

**5.1.3.4 Operation at reduced or cryogenic temperatures** improves laser material properties, such as thermal conductivity and thermo-optic coefficient, d*n*/d*T*. New thermal management approaches to extract high heat fluxes (>100 W/cm$^2$), such as evaporative spray cooling and microchannel heat sinks, would make compact multi-kW amplifiers possible. A promising approach to high-intensity, high-energy ultrashort lasers operating at the multi-joule level at much higher repetition rate on Ti:Sa pumped by diode-pumped, cryo-cooled Yb:YAG pump lasers. Multiple, multi-pass cryogenically cooled Ti:sapphire active mirror amplifiers could reach ~4 J energies when pumped by the frequency doubled output of this pump laser generating 1.5 J pulses at 500 Hz (0.75 kW) [1] that can likely be scaled to ~3 kW average power.

**5.1.3.5 Solid-state heat spreaders** used in "sandwich" structures promise another way to increase heat extraction from laser elements. A high-thermal-conductivity, transparent material serves to provide a low-resistance thermal pathway to maintain tight tolerances on temperature changes and temperature uniformity. These structures typically use ultrapure diamond or sapphire to minimize absorption. Heat spreaders are particularly attractive for improving thermal performance of broadband nonlinear optics where material choices are limited. Research and development is needed to develop methods to bond and/or coat laser gain materials and nonlinear crystals with thermally conductive materials, like diamond, SiC, YAG, and sapphire.

---

5 T.M.Spinka, *et al.*, "Laser Technologies for PW-Class Peak Power at Multi-kW Average Power," CLEO 2019, San Jose, CA, USA, May 2019; C. W. Siders *et al.*, "Wavelength Scaling of Laser Wakefield Acceleration for the EuPRAXIA Design Point," Instruments **3** (3), 44 (2019).

### 5.1.4 Research and Development Priorities

Research to improve thermal management in the following key areas would pay off:

1. Develop diode-pumped liquid- and gas-cooled amplifiers operating at 5- to 10-kW average power;
2. Scale liquid-cooled, flash-lamp–pumped disk amplifiers to the 40-cm aperture to increase the repetition rate of current multi-kJ lasers for 10- to 100-PW lasers;
3. Develop and characterize new gain broadband materials, such as Yb- and Nd-doped crystalline and ceramic materials, like cubic sesquioxides and CaF;
4. Develop compact and efficient cryogenic cooling systems and components;
5. Develop solid-state heat spreader technologies for important laser and nonlinear optical materials.

## 5.2 Laser Technology for High-Precision Extreme Science

### 5.2.1 Introduction

New laser facilities in Europe and Asia will push the frontiers of high-intensity science with world-leading capabilities: laser pulses delivering up to 10-PW peak power at repetition rates up to one shot per minute, and PW lasers at 10 Hz. These facilities will be the first to access extreme conditions that will doubtlessly uncover new science that will, in turn, require more exploration with careful and in-depth studies. Higher laser repetition rates, as discussed in Sec. 5.1, can yield more data to advance these studies, but highly precise and repeatable laser performance will be just as important, if not more so.

> *High-precision extreme laser science depends on highly precise and repeatable lasers that produce large data sets with data rates well suited to modern big-data and machine-learning approaches.*

All four TRN panels recognized that making technical advances to improve laser repetition rates and average powers could enable a new era of high-precision science. Large data sets taken across many shots that precisely capture both experimental conditions and results can facilitate new ways to study frontier science not possible with existing single-shot lasers. Statistical and big-data analytical approaches have been developed and employed in other fields, like high-energy physics, to study rare events, validate predictions of quantum statistics, characterize background, and identify weak signals to uncover hidden patterns and unknown correlations. Machine-learning algorithms using feedback from laser and/or experimental measurements could be applied to deliver stable and repeatable performance or to systematically survey broad ranges of parameter space.

### 5.2.2 Mapping to BLI Scientific Research Needs and Facility Capabilities

Laser technologies enabling high-precision extreme science apply to all SRN areas described in Chap. 3 and the new facility capabilities identified in Chap. 4. Ultrahigh temporal contrast is required for laser interactions with solid targets used for ion acceleration and its applications (Secs. 3.1, 3.2, 3.5–3.7). High-energy, coherent beam combination applies to research activities needing the absolute highest intensities (Sec. 3.4), as well as high repetition rates required for accelerator applications (Sec. 3.3). Focal-spot intensity measurements and control along with femtosecond timing control apply to every SRN field.

### 5.2.3 Related BLI Technology R&D Areas

**5.2.3.1 Ultrahigh temporal contrast –** The temporal contrast required for many future ultraintense lasers exceeds current capabilities. Many of the science research needs discussed in Chap. 3 require high contrast to avoid creating preplasmas or even destroying targets. Examples are ion acceleration and the interaction with nano- and micro-structured targets that prepulses can destroy or significantly alter prior to the arrival of the ultra-intense pulse. Achieving high temporal contrast is particularly challenging for fiber laser systems, but some applications, like LFWA in plasma channels, may only require substantially lower contrast values.

Achieving high temporal contrast requires careful system design and component choice. Front-end laser systems are available with ultrahigh contrast ($>10^{14}$), typically by using either double-CPA with pulse-cleaning techniques [6] or picosecond-pumped optical parametric amplifiers [7]. Preserving the contrast depends on the quality of subsequent optical systems, such as the pulse stretcher [8] and OPCPA pump lasers [9]. Nonlinear frequency conversion, such as second-harmonic generation, can significantly improve output contrast at full scale that requires technical developments to produce high-aspect-ratio crystals.

**Plasma mirrors** (Fig. 5.2.1) are now widely used to remove prepulses from joule-class lasers. Prepulses pass through these devices that start reflecting once plasma is generated at the surface of the optic when the ionization threshold intensity is exceeded. Significantly, plasma responses to intense light are ultrafast, spectrally broadband, and highly nonlinear. A single plasma mirror typically increases picosecond contrast by several of orders of magnitude [10], and several plasma mirrors can operate in series to achieve higher temporal contrast. Plasma mirror technology needs to be scaled for two classes of next-generation lasers: (1) kilojoule–femtosecond laser systems where refocusing can also provide needed experimental flexibility (tighter or looser focuses from a focusing element); and (2) high-repetition-rate (50- to 100-kHz), high-average-power lasers, where fresh plasma mirrors must be continuously supplied for sustaining long-term operation. Both classes of lasers require plasma mirrors with high reflectivity (>90%) and wavefront quality to maintain focusability.

---

6 M. P. Kalashnikov *et al.*, “Double Chirped-Pulse-Amplification Laser: A Way to Clean Pulses Temporally,” Opt. Lett. **30** (8), 923–925 (2005).

7 C. Dorrer, A. V. Okishev, and J. D. Zuegel, “High-Contrast Optical-Parametric Amplifier as a Front End of High-Power Laser Systems,” Opt. Lett. **32** (15), 2143–2145 (2007).

8 V. Bagnoud and F. Salin, “Influence of Optical Quality on Chirped-Pulse Amplification: Characterization of a 150-nm-Bandwidth Stretcher,” J. Opt. Soc. Am. B **16** (1), 188–193 (1999).

9 C. Dorrer, “Analysis of Pump-Induced Temporal Contrast Degradation in Optical Parametric Chirped-Pulse Amplification,” J. Opt. Soc. Am. B **24** (12), 3048–3057 (2007).

10 A. Lévy *et al.*, “Double Plasma Mirror for Ultrahigh Temporal Contrast Ultraintense Laser Pulses,” Opt. Lett. **32** (3), 310–312 (2007).

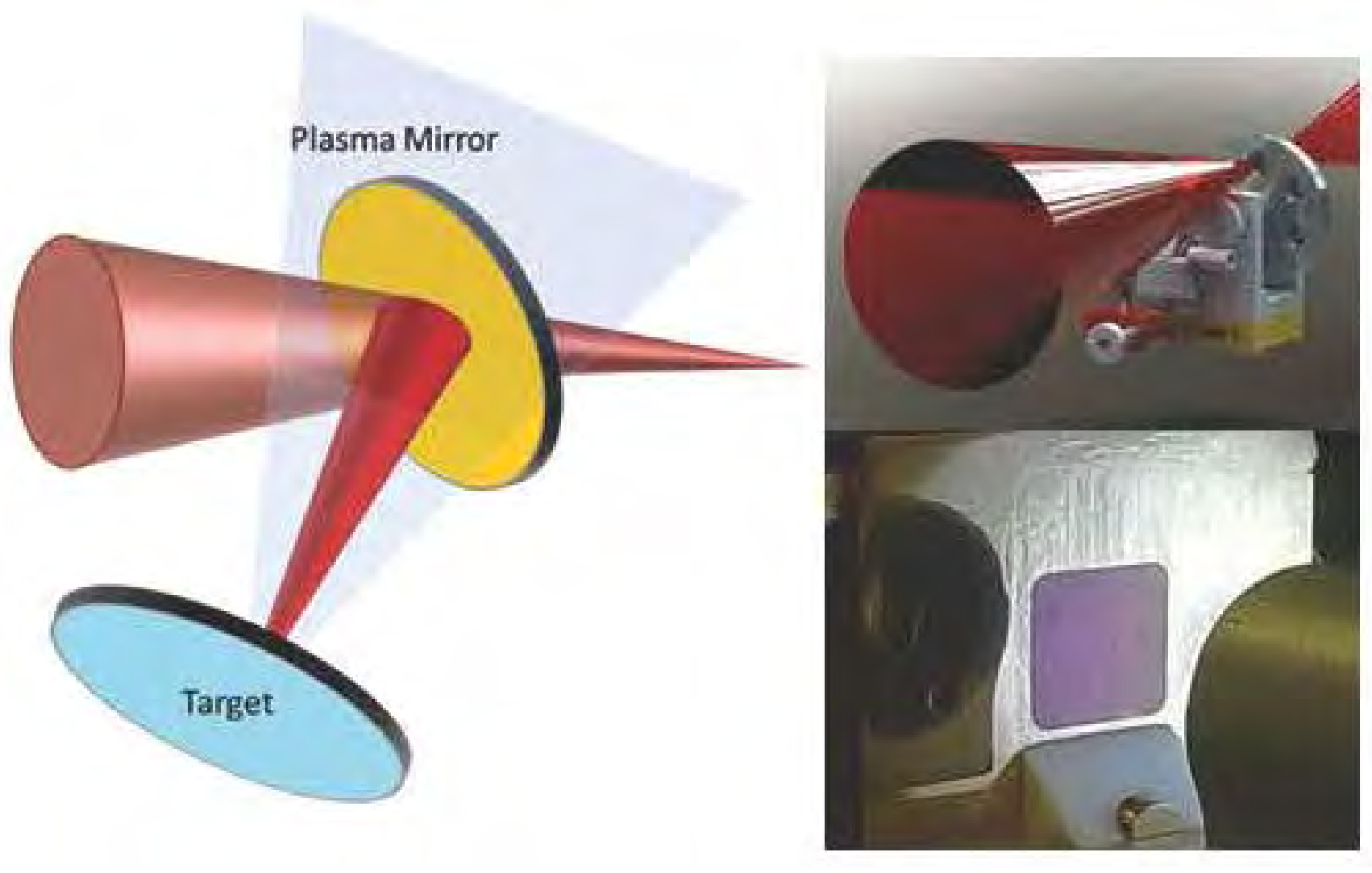

(a) (b)

**Figure 5.2.1.** (a) General configuration of a plasma mirror. A focusing beam passes through an anti-reflection coated optic until the laser intensity reaches a threshold for plasma formation on its surface. The plasma layer becomes highly reflective and directs the rest of the pulse to focus. (b) Free-standing, ultra-thin films formed dynamically and *in-situ* have served both as plasma mirrors and ion acceleration targets. [11]

Delivering intense laser pulses with ultrahigh temporal contrast (>$10^{14}$) requires **extremely high dynamic-range measurements**. The next generation of ultraintense lasers will require measurements at least 1000× more sensitive than currently available. New techniques must be developed to distinguish between temporal contrast degradation in the near or far field caused, for example, by the surface roughness of stretcher optics or parametric fluorescence. Single-shot, third-order cross-correlators have been demonstrated [12], but reliable measurements depend on high beam quality. (Secs. 5.2.3 and 5.2.4).

11 P. L. Poole *et al.*, “Experiment and Simulation of Novel Liquid Crystal Plasma Mirrors for High Contrast, Intense Laser Pulses,” Sci. Rep. **6** (1), 32041 (2016).

12 C. Dorrer, J. Bromage, and J. D. Zuegel, “High-Dynamic-Range Single-Shot Cross-Correlator Based on an Optical Pulse Replicator,” Opt. Express **16** (18), 13,534–13,544 (2008).

**5.2.3.2 High-energy, coherent beam combination –** Current laser technology poses practical limits to scaling single-aperture, ultra-intense lasers to higher energies. Large-aperture laser and OPCPA crystals can amplify chirped pulses to kilojoule energies, but grating technology limits compressed pulse energies, as discussed in Sec. 5.3. Long considered a *holy grail* for overcoming the limits for a single laser aperture, significant progress has been made in combining cw lasers, where the most impressive published result in terms of average power is the coherent combination of seven 15-kW cw Nd:YAG lasers to produce a 100-kW system [13]. Coherently combining ultrashort pulses leverages ongoing efforts in cw laser combining; however, it is in the early stages of development. So far, only demonstrations combining of up to 16 spatial channels with up to ~1 to 2 kW of average power, and up to 81 pulses time-domain combined, with the total energy of ~10mJ [14,15].

**Coherently combined laser arrays** can incorporate three types of coherent combining: time-domain, spatial-domain and spectral-domain, as illustrated in Fig. 5.2.2 for an all-fiber based architecture. **Spatial-domain combining** [14] promises multi-kW average power and Joule pulse energy scalability. **Time-domain combining** [16] promises approximately 100-fold higher energies up to 10s of mJ from individual fiber amplifiers compared to conventional single-pulse fiber CPA [15] by amplifying trains of pulses that are recombined. **Spectral-domain combining** can extend the overall amplification bandwidth by combining groups of channels operating at different wavelengths to achieve very short pulse durations (down to ~30 fs) [17]. Solid-state versions offer another approach where several different crystal types can be chosen to provide overlapping spectral coverage.

Significant further R&D is required to continue development of this fiber-based architecture towards multi-kW average powers and Joule-level output energies, with peak powers in the 10 to 50-TW range, as required for laser plasma acceleration applications in Sec. 3.3, and some applications in Sec. 3.7. This R&D includes developing high-power beam combiners, and monolithic fiber integration technology (specialty large core fibers, fusion-integrated pump combiners, etc.), and leverage pre-pulse contrast enhancement techniques. Attosecond and HHG applications (section 3.5) do not require such large pulse energies, and multi-kW fiber laser R&D for these applications needs to focus on achieving few-optical cycle generation with pulse energies in the ~10-30 mJ range.

---

13 S. J. McNaught *et al.*, “100 kW Coherently Combined Slab MOPAs,” in *Conference on Lasers and Electro-Optics/International Quantum Electronics Conference*, OSA Technical Digest (CD) (Optical Society of America, Baltimore, MD, 2009), Paper CThA1.

14 H. Stark *et al.*, “23 mJ High-Power Fiber CPA System Using Electro-Optically Controlled Divided-Pulse Amplification,” Opt. Lett. **44** (22), 5529–5532 (2019).

15 H. Pei, J. Ruppe, S. Chen, M. Sheikhsofla, J. Nees, Y. Yang, R. B. Wilcox, W. Leemans, and A. Galvanauskas, “10 mJ Energy Extraction from Yb-doped 85 µm Core CCC Fiber Using Coherent Pulse Stacking Amplification of fs Pulses,” in *Laser Congress 2017 (ASSL, LAC)*, OSA Technical Digest (online) (Optical Society of America, Washington, DC, 2017), Paper AW4A.4.

16 T. Zhou *et al.*, “Coherent Pulse Stacking Amplification Using Low-Finesse Gires-Tournois Interferometers,” Opt. Express **23** (6), 7442–7462 (2015).

17 W. Chang, T. Zhou, L. A. Siiman, and A. Galvanauskas, “Femtosecond Pulse Spectral Synthesis in Coherently-Spectrally Combined Multi-Channel Fiber Chirped Pulse Amplifiers,” Opt. Express **21** (3), 3897–3910 (2013).

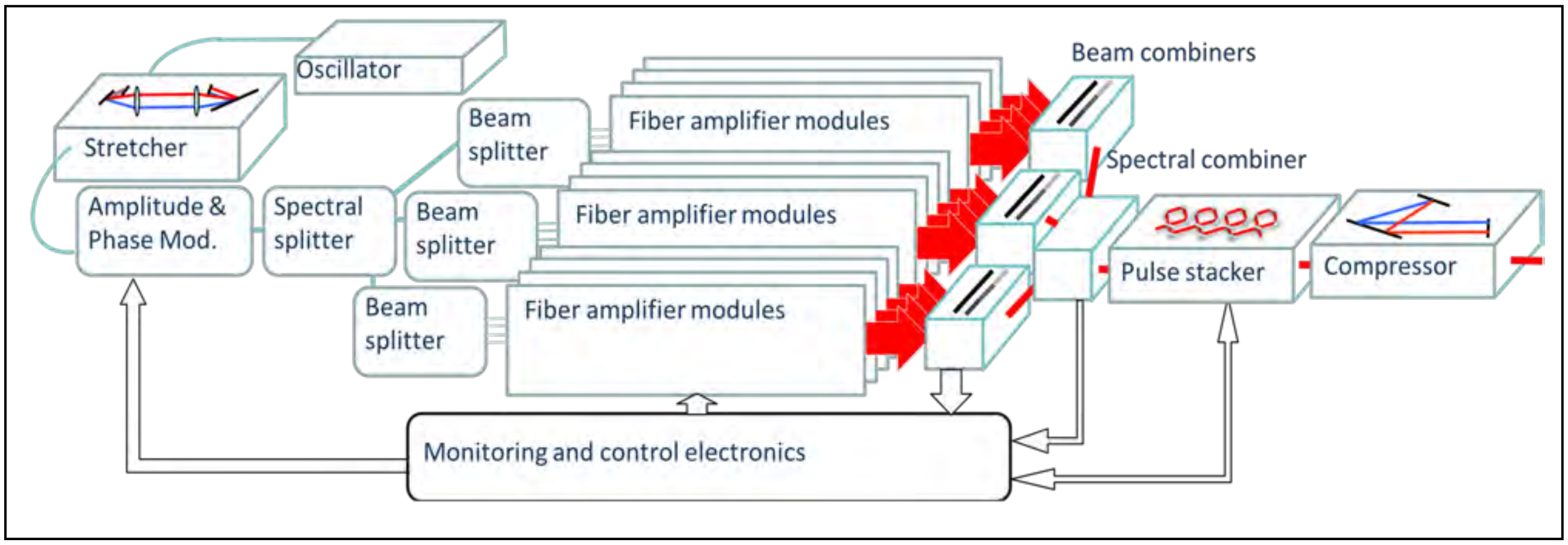

**Figure 5.2.2** Architecture of coherently combined fiber laser arrays

Significant further R&D is required to continue development of this fiber-based architecture towards multi-kW average powers and Joule-level output energies, with peak powers in the 10 to 50 TW range, as required for laser plasma acceleration applications in Sec. 3.3, and some applications in Sec. 3.7. This R&D includes developing high-power beam combiners, and monolithic fiber integration technology (specialty large core fibers, fusion-integrated pump combiners, etc.), and leverage pre-pulse contrast enhancement techniques. Attosecond and HHG applications (Sec. 3.5) do not require such large pulse energies, and multi-kW fiber laser R&D for these applications needs to focus on achieving few-optical cycle generation with pulse energies in the ~10-30 mJ range.

Combining bulk ultrashort pulse lasers is difficult because of the additional degrees of freedom. The best results so far have been modest in terms of average-power performance [18]. Several approaches have been pursued; the technique conceived above for the future ultra-intense facility is referred to as “tiled aperture combining” [19]. This concept is being actively pursued for the SEL project [20] in China. A variation that uses a “dipole” focusing scheme has been proposed for an ultra-intense facility [21] in Russia.

Lossless coherent combination requires perfect beams [22]. The challenges that must be overcome to ensure high-efficiency combining of the beams include obtaining beamlet wavefront

18 A. Klenke *et al.*, “Coherently Combined 16-Channel Multicore Fiber Laser System,” Opt. Lett. **43** (7), 1519–1522 (2018).
19 V. E. Leshchenko, “Coherent Combining Efficiency in Tiled and Filled Aperture Approaches,” Opt. Express **23** (12), 15,944–15,970 (2015).
20 C. Peng *et al.*, “Phasing Methods of Tiled-Aperture Coherent Beam Combining for High Peak Power Lasers,” EPJ Web Conf. **205**, 01001 (2019).
21 A. V. Korzhimanov *et al.*, “Horizons of Petawatt Laser Technology,” Phys.-Usp. **54** (1), 9–28 (2011).)
22 G. D. Goodno, C. C. Shih, and J. E. Rothenberg, “Coherent Combining with Imperfect Beams,” in *Advances in Optical Materials*, OSA Technical Digest (CD) (Optical Society of America, Istanbul, 2011), Paper AMC4.

uniformity to tenths of a wave; precision timing and pointing control to fractions of the pulse width and *coherently combined* focal spot, respectively; and ensuring sufficient passive stability, active control, and full-energy characterization. Technical research and development extending this technique to bulk laser systems could lead to multi-kilojoule, femtosecond exawatt-class lasers of the type shown conceptually in Fig. 4.2.3.

**5.2.3.3 *In-situ* focal-spot intensity measurements –** Peak intensity is arguably the most fundamental parameter of any ultra-intense laser experiment. Standard practices to evaluate focal intensity are indirect based on separate measurements of pulse energy, duration, and focal-spot size, assuming that most of the energy is contained within the ultrashort pulse in the laser focus. Recent reports [23,24] show many reasons why the actual intensity may be much lower than this approach estimates. *In-situ* and direct measurements may be required to accurately gauge intensity effects beyond $10^{23}$ W/cm$^2$. Detecting high-charge states [25,26] resulting from ultrahigh-field ionization can measure *in-situ* peak intensity, but Coulomb explosion effects complicate this technique for pulse energies and intensities greater than 20 J and $10^{22}$ W/cm$^2$, respectively.

Charged particles driven by laser fields radiate, and this Thomson-scattering radiation depends on the fields experienced by the charged particle. For intensities between $10^{18}$ and $10^{22}$ W/cm$^2$, Thomson scattering from electrons can be a useful tool for measuring the peak intensity [27]. Doppler shifts of this radiation coming from inner-shell electrons can be matched to the peak intensity [28]. Beyond $10^{24}$ W/cm$^2$, radiation reaction dominates for electrons and Thomson scattering from electrons is too complicated and directional. In this ultrahigh-intensity region, ion motion starts to be relativistic. Protons (coming from ionized hydrogen) are relativistically driven by the laser field and consequently radiate with a convenient Doppler shift that can be measured close to the plane perpendicular to the laser propagation, whereas electrons do radiate close to the forward direction [27].

**5.2.3.4 Focal-spot control –** Interactions of ultra-intense lasers with high-energy electron beams will require pointing stability defined by the electron beam width (a few microns); while high-efficiency, coherent beam combination will require stability better than a few percent of the laser beam divergence [19]. Optomechanical designs with high passive stability will be necessary but likely insufficient. **Active pointing stabilization and wavefront control systems** must be developed with sufficient control bandwidth to account for acoustic vibrations and slow drifts. Pilot beams, control hardware, and measurement techniques consistent with the overall laser facility design and concept of operations can leverage advances developed for large segmented-primary telescopes.

---

23 G. Pariente *et al.*, "Space–Time Characterization of Ultra-Intense Femtosecond Laser Beams," Nat. Photonics **10** (8), 547–553 (2016).

24 A. Jeandet *et al.*, "Spatio-Temporal Structure of a Petawatt Femtosecond Laser Beam," J. Phys. Photonics **1** (3), 035001 (2019).

25 E. A. Chowdhury, C. P. J. Barty, and B. C. Walker, " 'Nonrelativistic' Ionization of the *L*-Shell States in Argon by a 'Relativistic' $10^{19}$ W/cm$^2$ Laser Field," Phys. Rev. A **63** (4), 042712 (2001).

26 A. Link *et al.*, "Development of an *In Situ* Peak Intensity Measurement Method for Ultraintense Single Shot Laser-Plasma Experiments at the Sandia *Z* Petawatt Facility," Rev. Sci. Instrum. **77** (10), 10E723 (2006).

27 C. He *et al.*, "Relativistic Thomson Scattering: A Tool for Pulse Diagnostics and Exploring Inner-Shell Dynamics," Bull. Am. Phys. Soc. **64** (4), Q03.00004 (2019).

28 O. Har-Shemesh and A. Di Piazza, "Peak Intensity Measurement of Relativistic Lasers via Nonlinear Thomson Scattering," Opt. Lett. **37** (8), 1352–1354 (2012).

Providing more control of **spatial intensity and polarization distributions** in ultrafast laser beams can lead to improved laser and experimental performance. For example, a novel concept based on recently demonstrated **"flying focus" technology** [29] suggests a new paradigm in laser-plasma acceleration that could advance the dream of a TeV linear accelerator using a single-stage system without guiding structures. An achromatic flying focus [30] allows the focal spot of an ultrashort laser pulse to propagate through meters of plasma at a velocity matched to the accelerated electrons. The concept decouples the plasma conditions from the acceleration length, removes the need for a plasma channel, and provides a novel injection scheme for monoenergetic electron beams that could lead to single-stage, TeV-class linear acceleration.

Similarly, **arbitrary focal-intensity distributions** needed for applications require new laser beam mode conversion techniques. Super-Gaussian laser output beam profiles are preferred for efficient laser operation but they result in highly structured focal spots. For example, laser wakefield acceleration in capillaries requires center-peaked, Gaussian-like focal spots with truncated wings to avoid damage to the capillary walls, while super-Gaussian focal spots with square temporal profiles and femtosecond rise times are needed to accelerate ions from thin planar targets and to study other relativistic laser–plasma interactions.

To implement **focusing schemes with flexible parameter ranges** that are adaptable to various experimental goals, beam-control technologies must advance beyond currently available approaches, such as static beam shapers and phase correctors along with deformable mirrors with low spatial resolution and temporal bandwidth. **Programmable optical devices linked with high-speed, closed-loop feedback control** must be developed to meet the opportunities enabled by and high-precision needs of future high-repetition laser systems. Pulse repetition rates greater than ~100 Hz require faster image detection systems and aggressive processing requirements to implement hardened machine vision for feedback control of laser parameters, as well as damage/defect detection for machine safety.

---

29 D. H. Froula *et al.*, "Spatiotemporal Control of Laser Intensity," Nat. Photonics **12** (5), 262–265 (2018).
30 J. P. Palastro *et al.*, "Dephasingless Laser Wakefield Acceleration," submitted to Nature Photonics.

**5.2.3.5 Femtosecond timing control –** Co-timing femtosecond lasers and diagnostics for ultra-intense laser experiments requires **high-precision timing signals** distributed throughout the facility where time-sensitive measurements and controls are located. Timing "fiducial" signals are often implemented as optical waveforms distributed by fiber-optic systems. Synchronizing femtosecond pulses is now possible at the few-fs level with available commercial products [e.g., Cycle GmbH] that implement bi-directional stabilized optical links for fs-scale timing distribution where needed. These systems were developed for particle accelerators and free-electron laser facilities. Developing low-loss fiber-optic cables with significantly reduced temperature-dependent group delay and/or with passive acoustic damping may improve the stability and accuracy of these timing links. Standardizing on modular systems and integrating them into future facilities of all scales would deliver required performance at lower costs and enable the use of lasers and other ultrafast devices across a network of U.S. and worldwide facilities.

### 5.2.4 Research and Development Priorities

Key development efforts to mature the technologies required for high-precision extreme science include

1. Scale plasma-mirror technology for kilojoule–femtosecond laser systems where refocusing can provide needed experimental flexibility;
2. Develop plasma-mirror technology for high-repetition-rate (50- to 100-kHz), high-average-power lasers;
3. Develop instruments capable of extremely high dynamic-range ($>10^{15}$) and single-shot measurements;
4. Develop *in-situ* and direct focal-spot measurements capable of intensities up to and beyond $10^{23}$ W/cm$^2$;
5. Develop active pointing stabilization and wavefront control systems to enable coherent beam combination and diffraction-limited focal spots that can reliably interact with tightly focused particle beams;
6. Develop multi-kW average power and 10- to 50-TW peak power coherently combined fiber laser arrays for applications.
7. Develop techniques to control and manipulate the focal intensity and polarization state of high-energy, femtosecond lasers, including programmatic control to provide flexible parameter ranges; and
8. Develop precision timing systems with reference-signal distribution throughout facilities to standard receivers that can be integrated into laser/experimental sub-systems.

## 5.3 High-Energy Pulse-Compression Technology

### 5.3.1 Introduction

A critical technology bottleneck in any CPA laser design is the method of optical compression to reverse the chirp placed on the pulse for amplification. As pulse energy along with peak and average powers continues to increase in the next generation of frontier laser facilities, pulse compression technology must be a central R&D theme for new with a focus particularly centered on large-aperture diffraction gratings. Advanced pulse compression schemes, like plasma or thin-film compression, also represent breakthrough opportunities.

*Pulse compression technology requires focused R&D to realize the next generation of frontier laser facilities operating at high peak and average powers, including advances to diffraction gratings, plasma-based and other novel approaches.*

Practically all high-energy, high-intensity lasers considered in this workshop are limited by damage of the final compressor grating. To date, this limit has been addressed simply by increasing grating size to nearly the meter scale, but steep scaling of cost and laser system complexity have slowed progress. Current state-of-the-art grating technology use either metallic coatings or multilayer dielectric coatings (MLD) with apertures roughly up to 1.5 $m^2$. Damage thresholds range from 0.3 J/$cm^2$ for metallic-coated gratings to approximately 1 J/$cm^2$ for MLD gratings operating with sub-20-fs to a few hundred femtosecond pulses, respectively. The challenge for future multi-PW laser systems will be to increase the damage threshold of diffraction gratings and increase the size of gratings that can be manufactured. Significant improvements in damage threshold, aperture, multi-optic phasing (tiled gratings), and average power handling will be necessary for the next generation of high-intensity lasers.

### 5.3.2 Mapping to BLI Scientific Research Needs and Facility Capabilities

Pulse compression currently limits all ultra-intense lasers required for all science cases in Chap. 3 and both high-energy, single-shot and high-average-power/high-repetition-rate laser systems. Increasing laser-induced damage thresholds would benefit all applications. Increasing grating size benefits the highest energy applications (Secs. 3.1–3.3), while thermal management applies to applications (Secs 3.3, 3.5–3.7) that demand multi-kHz repetition rates at lower energies. New diffraction-grating technology for applications that scale favorably with longer wavelengths represents another area demanding R&D. Advancing plasma-based amplification/pulse compression and thin-film compressor technology offers novel approaches to producing high-energy/high-intensity pulses.

### 5.3.3 Related BLI Technology R&D Areas

**5.3.3.1 Laser-induced–damage threshold (LIDT) –** A potential path to increasing the energy and peak power of next-generation CPA laser systems is to increase the operating fluence of the compressed beam. Currently there are generally three kinds of diffraction gratings summarized in Fig. 5.3.1: metal (such as gold), multilayer dielectric (MLD), and hybrid metal-dielectric.

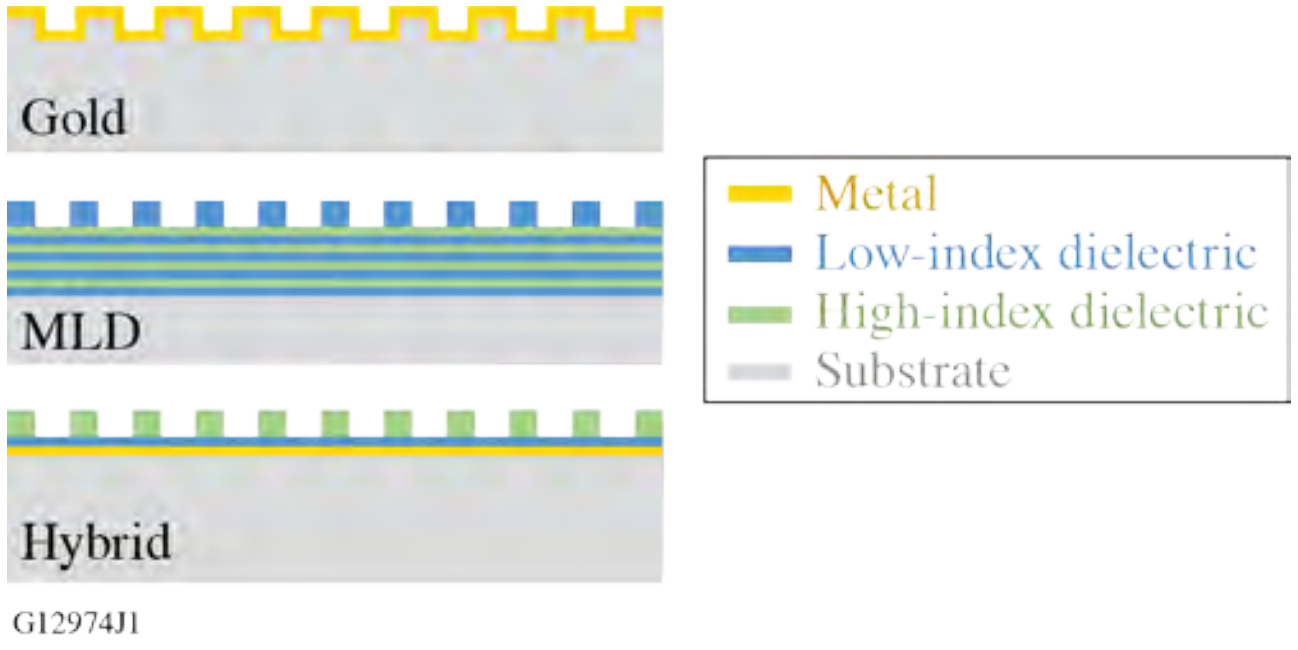

**Figure 5.3.1.** Three types of reflective diffraction gratings used for ultrashort pulse compression: gold, multi-layer dielectric (MLD), and hybrid metal-dielectric.

**Metallic pulse compression gratings** typically employ gold coatings on holographic structures etched into photoresist on a substrate or directly into the substrate. They support broad bandwidths with high diffraction efficiency, but incur significant absorption losses that lower damage thresholds and capacity for average power handing.

**MLD gratings** are advantageous due to their high efficiency, resulting in less overall loss, and orders-of-magnitude-lower absorption than gold gratings, but generally support narrower bandwidths. MLD gratings for typical petawatt-class laser systems require high-quality multilayer coatings at meter-scale apertures. The **availability of MLD coatings is at risk of being completely lost** as there is currently only one domestic and one foreign coating vendor qualified to produce the required densified e-beam multilayer coatings at large dimensions. Without investment and a sustained market, the domestic capability to fabricate large MLD diffraction gratings may be lost.

**Increasing LIDT must be a very high priority for near-future grating R&D.** Developing improved cleaning processes and designs to engineer electric-fields in the grating structure could increase LIDT of gratings by factors of two to four for picosecond laser systems. For femtosecond laser systems, the LIDT for gold gratings likely cannot be increased much beyond 0.3 J/cm$^2$, but new grating designs, including hybrid metal-dielectric gratings, are promising. These designs apply a few dielectric layers including a grating structure over a metal mirror. Hybrid metal-dielectric gratings promise the best properties of metal and MLD gratings—broader bandwidths, higher diffraction efficiencies and improved damage thresholds—but it is not a mature technology with manufacturing process challenges, like dielectric adhesion to metals. Further development is required to enable reliable and affordable manufacturing using better process control.

Areas of investment that would provide incremental improvements include incorporating new high-index, high-band-gap materials into gratings. Areas with higher risk and higher payoff include optimizing designs to reduce electric-field enhancements in the grooves, investigating and eliminating defects, studying optimal coating techniques and process development, and exploring new designs for *p*-polarized gratings that show promise for higher LIDT values with high diffraction efficiency over broad bandwidths. More detailed understanding of the laser damage mechanisms is needed with enhanced monitoring, and predictive diagnostics for the onset of laser damage.

The rate at which LIDT can be improved for any grating type is limited by the poor feedback loop involving testing and process development. Past experience in areas like high-LIDT thin-film coatings teaches that **integrating process development and testing seamlessly leads to the best and fastest progress**.

**5.3.3.2 Grating size – Larger gratings are required for achieving the highest possible peak powers**. Doubling linear grating dimensions results in a four-fold increase in surface area, and represents a reasonable goal given the current state of the art. Larger grating sizes will be limited by manufacturing technology and substrate costs can grow very rapidly, as capital equipment for processes and handling costs can scale nonlinearly with size.

Scaling gratings is best achieved using holographic patterning technology, such as **Scanning Beam Interference Lithography (SBIL)** pioneered in the U.S. by Plymouth Grating Laboratory (PGL). Analogous technology is being developed in China, but it is not yet mature. Scaling SBIL to larger sizes is technically feasible but it requires capital investments for new holographic patterning and etching tools, as well as infrastructure for preparing, cleaning, and handling larger optics.

Another direction that could enable multi-10 PW systems is **coherent phasing of gratings or multiple compressed beams**. The OMEGA EP Laser System pioneered “tiled gratings,” where three coherently phased gratings produce an effective aperture needed to compress kJ-picosecond pulses. The L4 system design for ELI-Beamlines employs a similar approach where a fold mirror effectively doubles the width of a single grating. Further research on applying these techniques and coherent beam combination for high-energy femtosecond pulses (Sec. 5.2.3.2) would significantly advance the field.

**5.3.3.3 New Compressor Geometries –** New compressor geometries might also enable operation at high beam fluence in future facilities. For example, the 10-PW L4 laser at ELI-Beamlines will be the first multi-PW-class laser implementing a low-dispersion MLD compressor architecture that has the potential to increase total compressor energy output given a fixed grating size [3]. Until recently there have been no high-efficiency grating designs for wavelengths near 800 nm when mid-scale MLD gratings supporting >20-fs pulse durations were demonstrated [31]. MLD gratings for mid-IR lasers with ~100-fs pulse durations were fabricated [32], and active

31 E. Sistrunk *et al.*, “Laser Technology Development for High Peak Power Lasers Achieving Kilowatt Average Power and Beyond,” Proc. SPIE **11034**, 1–9 (2019).

32 B. Rus *et al.*, “ELI-Beamlines: Progress in Development of Next Generation Short-Pulse Laser Systems,” Proc. SPIE **10241**, 1–9 (2017).

cooling [33] has the potential to support petawatt peak power lasers with average powers of 300 kW. Polarization-independent gratings for spectral beam combining at MW-level average powers have been developed that could be adapted for high-peak-power systems to further extend average power capabilities.

**5.3.3.4 Diffraction grating thermal management –** Thermal management is the dominant issue for high-average-power gratings. There are three complimentary approaches to avoid deleterious effects from heating in gratings as average power increases: (1) reduce or eliminate absorption in the gratings, (2) minimize the effects caused by heating, and (3) remove the heat and associated thermal gradients. R&D in all three areas will be necessary for multi-kW systems that are needed for applications like laser-based electron wakefield acceleration. Combining low-absorption gratings with low expansion substrates that are actively cooled will require coordinated development of appropriate materials, coatings and active cooling systems. Developing actively cooled MLD-based gratings promises average-power levels greater than 100 kW.

**5.3.3.5 New wavelengths –** Most high-average-power femtosecond laser systems are based on Ti:sapphire and operate around 800 nm, but OPCPA and novel broadband laser materials for the mid- and deep-infrared portion of the spectrum with wavelengths up to 10 $\mu$m are opening the door to alternative wavelength ranges and increased bandwidths (shorter pulses). Grating technology to support these new wavelengths needs to be fully developed.

**5.3.3.6 Plasma-based amplification and pulse compression –** Plasma photonics (Sec. 3.1) offers new approaches to manipulating laser light that promise breakthroughs to overcome conventional bottlenecks to achieving ultrahigh intensities, like optical damage of laser components. Two schemes include laser-plasma amplification and relativistic plasma mirrors. Research and development of these plasma-based technologies represent high-risk/high-reward paths to high-energy femtosecond pulses.

Studies are underway to evaluate **Raman scattering as an efficient ultrashort-pulse laser amplifier/compressor** for the next generation of high-power laser systems. Femtosecond seed pulses interacting with high-energy picosecond pump pulses in plasma produce electron plasma waves that exchange energy from the pump to the seed pulses. The research aims at developing a quantitative understanding of the wave–particle interactions and their effect on plasma-wave amplifiers that effectively "compress" the energy of the pump into the amplified seed. These experiments and modeling will lead to determining the optimum regime for plasma-wave amplification that will provide the foundation for designing a high-efficiency plasma-wave amplifier capable of exceeding current laser powers. Amplification in the nonlinear regime could provide the path for the laser systems to achieve a 100-PW laser by using multiple-kJ-picosecond pumps available from existing high-energy lasers.

---

33 D. A. Alessi *et al.*, "Active Cooling of Pulse Compression Diffraction Gratings for High Energy, High Average Power Ultrafast Lasers," Opt. Express **24** (26), 30,015–30,023 (2016).

Few-cycle pulses focused to $\lambda^2$-scale spots produce plasma mirrors that move relativistically in a direction normal and transverse to the original optical surface. These **relativistic plasma mirrors** can efficiently reflect and focus an isolated attosecond pulse with a pulse duration that scales inversely with normalized vector potential, $a_0$, which itself scales as the square root of the intensity.

**5.3.3.7 Thin-film compressors –** Thin-film compressor (TFC) technology, a novel compression scheme with the potential to compress high-energy pulses, has been proposed [ 34 ] with encouraging results experimentally demonstrated at low energy [35]. The scheme extends proven small-scale techniques making use of the optical Kerr effect in optical fibers, noble gas-filled hollow-core fibers and capillaries, and multipass gas cells to broaden the spectrum of a short pulse followed by compression with chirped mirrors. This approach has become a relatively conventional for few-cycle laser systems and has been proposed for nonlinear compression of high-energy picosecond Yb:YAG laser pulses into the femtosecond regime, as shown schematically in Fig. 5.3.2, to provide a path forward to generate Joule-level femtosecond pulses at high average powers. This approach depends on guided or cavity modes that significantly limit the energy of the pulse that can be compressed before the ionization limit is reached.

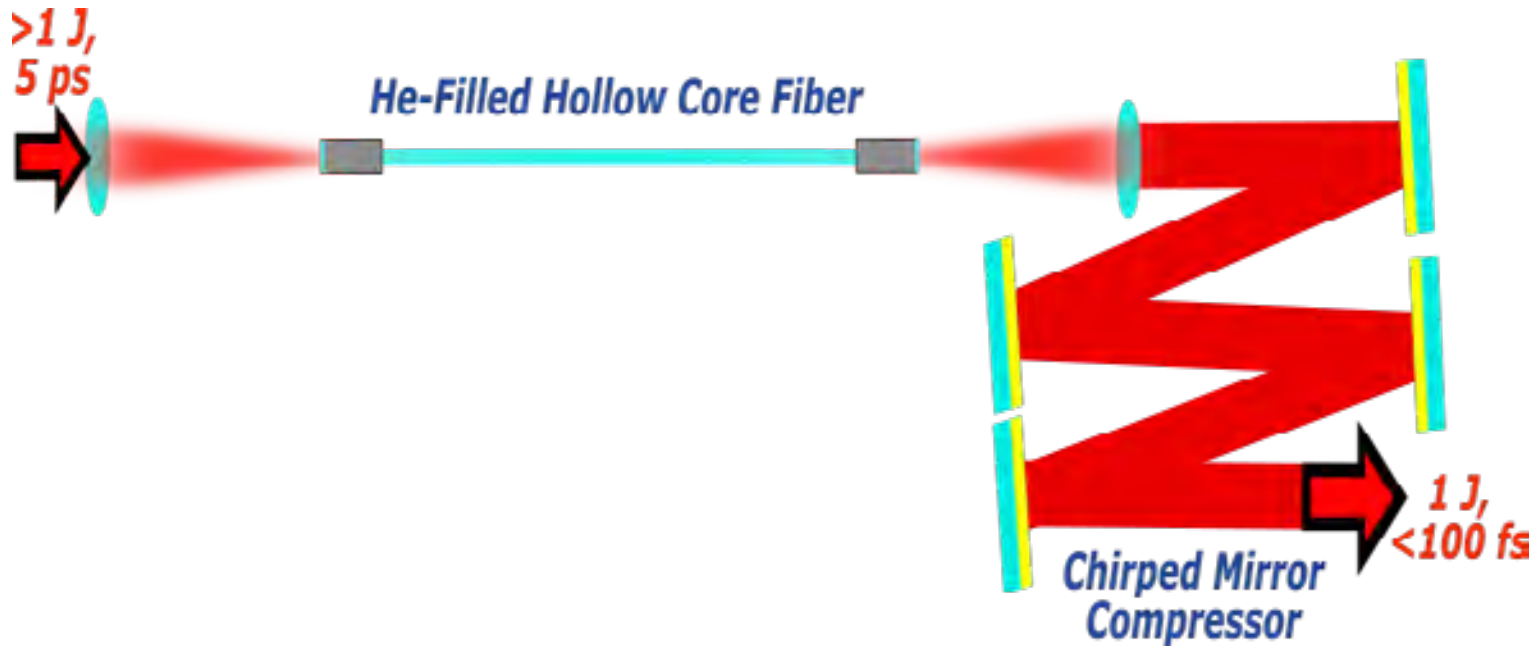

Figure 5.3.2. Conceptual diagram of hollow core fiber for spectrally broadening picosecond pulses and compressing them with chirped mirrors.

A TFC uses a thin film in a high-peak-power laser beam with a flattop spatial profile and bell-shaped temporal profile. Thin-plastic, glass, or crystalline substrates located in the transport path of a high-intensity laser beam self-phase modulate the pulse. This requires high-quality beam profiles and optical-quality substrates to uniformly broaden the spectrum. Spatial filtering can remove non-uniformities induced by small-scale self-focusing. Relatively low losses and pulse-compression factors up to 10× have been predicted in two-stage TFC systems. A first experimental demonstration of a single-stage TFC showed ~1.7× pulse compression near the Fourier transform

34 G. Mourou *et al.*, "Single Cycle Thin Film Compressor Opening the Door to Zeptosecond-Exawatt Physics," Eur. Phys. J. Spec. Top. **223** (6), 1181–1188 (2014).

35 D. M. Farinella *et al.*, "Focusability of Laser Pulses at Petawatt Transport Intensities in Thin-Film Compression," J. Opt. Soc. Am. B **36** (2), A28–A32 (2019).

limit, but some energy was redistributed out of the central portion of the focal spot, which compromised peak intensity.

Advances in high-precision laser control (Sec 5.2.3) to produce high-quality flattop beam profiles with negligible spatial intensity fluctuations and well-controlled temporal profiles in kJ-class ultrashort lasers could lead to exawatt-class performance that can be focused to ultrarelativistic intensities. Applying the output of such a system to a relativistic plasma mirror (Sec. 5.3.3.6) could produce attosecond pulses frequency upshifted to x-ray or gamma-ray photon energies and focused even more tightly to yield even higher peak intensities.

### 5.3.4 Research and Development Priorities

Key development programs to mature pulse compression technologies include:

1. Develop partnerships between research laboratories and industry to advance grating technology.
2. Develop and test ultra-broadband MLD and hybrid metal-MLD gratings for near-IR lasers that can be scaled to several-meter apertures.
3. Develop and test gratings for mid- and deep-IR (2- to 10-$\mu$m) operating wavelengths to enable applications that scale favorably with $I\lambda^2$.
4. Develop and test actively cooled gratings for compact, high-average-power pulse compressors.
5. Advance detailed understanding of the laser-induced damage processes in gratings and HR reflective optics via enhanced monitoring, observation and further developing predictive diagnostics for the precursors of optical damage in existing, operational systems.
6. Advance multi-aperture grating compressor technologies for sub-100-fs pulses using phased/tiled gratings or fold mirrors to increase effective grating size.
7. Support research in plasma amplification and compression of ultrashort pulses.
8. Develop spectral broadening and compression techniques based on thin-film compression technology.

## 5.4 Optics and Optical Materials for Ultrahigh-Intensity Lasers

### 5.4.1 Introduction

Similar to diffraction gratings for pulse compression, optics and optical materials can limit both high-energy/single-shot and high-average-power (HAP)/high-repetition-rate laser systems. Increasing laser-induced–damage thresholds and bandwidths of optical coatings would benefit both classes of lasers. Development of new optical architectures and material for fiber based beam combining and new active materials for new wavelengths regions would provide a larger selection of potential future laser systems for specific applications. Plasma optics can replace conventional optics in some cases to extend operating parameters.

*Future ultraintense lasers will require new optics and optical materials with higher laser-induced–damage thresholds and support larger bandwidths.*

### 5.4.2 Mapping to BLI Scientific Research Needs and Facility Capabilities

Advances in optics and optical material technology would benefit all scientific applications identified in Chapter 3.

### 5.4.3 Related BLI Technology R&D Areas

**5.4.3.1 Optical coatings and surface treatments –** The optical design of lasers and beam transport systems is dictated largely by the availability of suitable optical coatings. Several classes of broadband optical coatings are required for ultra-intense lasers: antireflection (AR), high reflection (HR), polarizers, dichroic mirrors for beam combination and separation, and partial reflectors for beam sampling.

**Broadband HR mirrors** are required to transport both chirped and compressed pulses. Ultra-broadband multilayer dielectric coatings for *s*-polarized light with high reflectivity (>99%), high damage threshold, and limited group-delay dispersion have been developed. All-dielectric coatings maximize laser-damage resistance while providing controlled, low-order dispersion, except that satisfactory solutions do not exist for *p* polarization so enhanced metal-coated mirrors must be developed. Dense MLD coatings that are insensitive to environmental changes, particularly relative humidity, ensure that dispersion properties are stable. Scaling coatings for meter-scale optics is a significant challenge that requires tight control of coating uniformity and coating stresses. Some combination of the following are required: new coating processes to yield lower compressive film stress; substrates with appropriate surface figures to precompensate for deflections resulting from the film stress; or depositing stress-compensation films on the second surface of substrates to balance bending forces.

**Broadband AR coatings** are needed for laser and nonlinear optical components. Sol-gel coatings provide good optical performance but they are fragile and susceptible to contamination.

Nanostructures on the surfaces of bulk optical materials have proven effective for a few common and commercially important materials, like YAG and fused silica. Extending this technology to ultrafast laser materials, like Ti:sapphire and nonlinear optical crystals, would be a significant advance.

**Large-aperture chirped mirrors** are required to realize the thin-film compressors discussed in Sec. 5.3.3.6. Current coating designs and technology need to be scaled to meter-class sizes with sufficient coating uniformity and stability to operate in vacuum.

**Curved surfaces** required for all-reflective image relays and focusing pose the final challenge. Changes in the height of the substrate surface (sag) can lead to significant film nonuniformity and operating angles of incidence. High-curvature optics require custom deposition processes. Additionally, large changes of incidence angle over a highly curved surface lead to compound angles and mixed polarization states that can lead to large variations in performance over the spectral bandwidth of the pulse. Enhanced or protected metal coatings provide performance over the spectrum that is quite similar for *s* and *p* polarizations, but typically with lower laser damage thresholds. Design and deposition processes for interference coatings must be developed to match spectral phase performance over large-aperture, curved components operating at a range of incidence angles.

High pulse energy and high repetition rate impose significant demands on the dielectric coatings in the laser amplifier crystals. Investigating the physical mechanisms that limit the laser damage resistance of coatings must proceed to understand and mitigate them. Understanding the roles that strain and material interfaces play in laser damage at ambient and cryogenic conditions need particular attention. Extending the most promising multilayer designs to longer wavelengths also promises new performance.

**5.4.3.2 Plasma optics** – Traditional optical components—solid mirrors, lenses, diffraction gratings, wave plates, polarizers, beam splitters, nonlinear crystals, filters, and many others—are subject to optical breakdown and material damage when handling beams with intensities exceeding approximately 1 TW/cm$^2$. Plasmas can exhibit light-manipulating properties similar to or even superior to those of solid-state materials at several orders-of-magnitude-higher intensities.

Transient plasmas created in solids, liquids, or gases by laser beams of ionizing intensities can efficiently redirect and focus [36, 37], diffract [38], split [39] and combine [40] light beams, and change their polarization [41]. Plasmas have been shown to temporally and spatially filter intense light, amplify and compress laser pulses, as well as generate coherent radiation with new frequencies spanning from x-rays to the terahertz range for applications that are beyond the reach of current light sources. Novel plasma-based components may yield the paths to more compact, controllable, and affordable architectures for the highest power laser systems offering an exciting scientific and technological opportunity to greatly expand the frontiers of achievable laser intensities. Plasma-based optical components, such as wave plates, beam combiners, and diffraction gratings, need to advance beyond limited theoretical studies and proof-of-principle experiments so they can be deployed as practical optical elements.

---

36 H. M. Milchberg, R. R. Freeman, and S. C. Davey, "Resistivity of a Simple Metal from Room Temperature to $10^6$ K," Phys. Rev. Lett. **61** (20), 2364–2367 (1988); M. M. Murnane, H. C. Kapteyn, and R. W. Falcone, "High-Density Plasmas Produced by Ultrafast Laser Pulses," Phys. Rev. Lett. **62** (2), 155–158 (1989).

37 A. Kon *et al.*, "Geometrical Optimization of an Ellipsoidal Plasma Mirror Toward Tight Focusing of Ultra-Intense Laser Pulse," J. Phys.: Conf. Ser. **244** (3), 032008 (2010).

38 H.-C. Wu *et al.*, "Manipulating Ultrashort Intense Laser Pulses by Plasma Bragg Gratings," Phys. Plasmas **12** (11), 113103 (2005).

39 S. Palaniyappan *et al.*, "Dynamics of Relativistic Transparency and Optical Shuttering in Expanding Overdense Plasmas," Nat. Phys. **8** (10), 763–769 (2012).

40 R. K. Kirkwood *et al.*, "Plasma-Based Beam Combiner for Very High Fluence and Energy," Nat. Phys. **14** (1), 80–84 (2018).

### 5.4.4 Research and Development Priorities

Key development programs to mature optical technologies include the following:

1. Develop and commercialize optical coating technology for meter-scale optics with dense and durable coatings and high laser-induced damage thresholds;
2. Develop and commercialize nanostructured AR surfaces for ultrafast laser materials;
3. Develop and commercialize large-aperture chirped mirrors; and
4. Develop plasma-based optics to extend operating intensities by an order of magnitude.

## 5.5 Experimental Technologies

### 5.5.1 Introduction

Frontier science with high-intensity laser systems will require advances in experimental technology. The workshop identified several key areas for research and development: high-repetition-rate targets, mitigating debris from targets, and managing electro-magnetic pulse (EMP) radiation. These issues, discussed below, will become even more acute with high-repetition-rate lasers discussed in Sec. 5.1. Additionally, experimental diagnostics originally developed for other disciplines, like high-energy physics, nuclear physics, and accelerator science, will need to be adapted for use in next-generation laser facilities that can advance these same fields.

*Frontier science using ultraintense lasers will also depend on new experimental technologies, like handling targets and mitigating target debris at high-repetition rates, as well as understanding and managing new threats from electromagnetic pulse radiation.*

41 P. Michel *et al.*, “Dynamic Control of the Polarization of Intense Laser Beams via Optical Wave Mixing in Plasmas,” Phys. Rev. Lett. **113** (20), 205001 (2014).

### 5.5.2 Mapping to BLI Scientific Research Needs and Facility Capabilities

Advances in high-repetition-rate targets and target-debris mitigation would benefit all scientific applications identified in Chap. 3, while electromagnetic pulse radiation requires particular attention for ultrahigh-fields associated with SRNs described in Secs. 3.1, 3.2, and 3.5–3.7.

### 5.5.3 Related BLI Technology R&D Areas

**5.5.3.1 High-repetition-rate targets –** Most high-shot-rate laser experiments will consume large numbers of targets that must be manufactured and positioned accurately, and the laser target shot interaction will typically generate considerable threats to final optics of the laser, such as x-ray, ions, plasma, and contamination by target debris.

Targets based on gas jets, and liquid jets or sheets are options for high-repetition-rate experiments. They are continuously present in the target chamber, except for the time it takes to reform the flow pattern after a shot, which can be quite short using high-velocity flows. For example, crossed liquid jets forming a thin film have been demonstrated that support shot rates >10 kHz for joule-class lasers. Nozzles forming the target can be eroded by output threats, so more robust nozzles must be developed that are formed from refractory materials or diamond. Nozzles require thermal management, especially for cryogenic fluids like liquid hydrogen, where heat can both vaporize the liquid, and deflect the jet position out of the laser focus. Active cooling, optimizing nozzle configurations, and implementing active feedback offer promising solutions.

Trains of droplets can reduce the mass of debris produced and the mass of material that must be pumped from the target chamber at a low vacuum pressure. This type of target introduces the additional challenge of timing the laser shot to hit the droplet. Concepts developed and tested at a low power for inertial fusion energy reactors can be adapted.

**5.5.3.2 Target debris mitigation –** Debris mitigation is critical in studies with increasing repetition rate that use short focal-length optics, especially in solid-target experiments. In principle, the threat can be mitigated by interposing a transparent window or film as a debris shield between the target and the final optics. The debris shield can be used until its transmission is no longer acceptable due to contamination or damage from debris when the debris shield must be replaced. The debris shield can also degrade the focal spot due to self-phase modulation. High shot rates would require automating this process. Debris shield material options that pose advantages and disadvantages include:

- glass—high optical quality can be produced but costs can be prohibitive;
- polymer film—large quantities can be produced cheaply but maintaining optical quality poses a technical challenge. Compensating spatial and/or temporal phase distortions from the debris shield might be possible using adaptive optics, but this significantly increases system complexity; and

- liquid crystal film—renewable liquid crystal thin films can be deployed as plasma mirrors in configurations where a baffle or solid shield blocks target debris from hitting conventional focusing optics or other critical components. Promising proof-of-principle concepts have been demonstrated [42], but additional R&D is needed to prove them for high-repetition-rate experiments.

It may be possible with high electric and magnetic field and field gradients to deflect electrically charged debris away from final optics, although it provides no protection against charge neutral debris.

**5.5.3.3 Electromagnetic pulse mitigation –** The interaction of high-intensity lasers with matter generates intense electromagnetic radiation with multi-GHz bandwidths lasting up to hundreds of nanoseconds. These fields have been observed in experiments using femtosecond to nanosecond laser pulses. The fields scale with the laser energy and intensity. Electric fields up to MV/m have been observed that pose a serious threat to electronic equipment placed both inside and outside the experimental chamber. EMP environments expected from future laser facilities will exceed anything currently experienced worldwide, so EMP studies are a vital topic of research.

Collecting and comparing experimental EMP data from current laser facilities, using standardized measurement techniques, is an important and urgent first step to developing accurate measurement methods along with predictive theoretical models and numerical codes that can reproduce the existing data and extrapolate them to future laser parameters. These tools can be used to both develop methods to limit the generation of EMP and mitigate its impact. This research also applies to the generation of extreme laser-induced transient currents and ultrastrong quasi-static fields with applications described in Sec. 3.4. Several workshops in Europe have been dedicated to EMP-related issues [Bordeaux (2016), Warsaw (2017) and Frascati (2018)]. Coordination through a formal network that includes U.S. and European collaborators is strongly recommended to advance this research.

### 5.5.4 Research and Development Priorities

Key development programs to mature experimental technologies include:

1. Develop and commercialize high-repetition-rate target and debris mitigation technology; and
2. Establish international collaborations to study and develop EMP mitigation technologies.

42 P. L. Poole *et al.*, "Liquid Crystal Films as On-Demand, Variable Thickness (50–5000 nm) Targets for Intense Lasers," Phys. Plasmas **21** (6), 063109 (2014).

## 5.6 Engagement of Industry and Supply Chain Challenges

All four TRN panels noted that addressing many of the technology challenges discussed in this chapter will require significant participation by the optics and laser manufacturing industry, and funding this engagement will be critical to reestablishing capacity in these highly competitive fields where the U.S. previously dominated.

In the past, many of the advances in high-energy and high-peak-power lasers occurred at U.S. national labs and universities. This model no longer competes successfully with European nations that have developed successful approaches to establish teams combining industry and academic researchers, like Germany and France, as discussed in Sec. 4.3. Consequently, the dominant high-power, short-pulse laser manufacturers worldwide reside in Europe – Thales and Amplitude Technologies in France, and TRUMPF in Germany. The U.S. must seek ways to foster R&D at laser companies and optics manufacturers in the U.S., as well as to transfer technology between universities, national laboratories and industry.

*Any strategy to advance high-intensity laser science and applications in the U.S. must include a plan to support partnerships with industry and financial support for advanced R&D.*

The issues facing industry involvement in this field become particularly distressing when considering supply chain availability. A dwindling number of companies supply the components and integrated systems needed for the high-peak-power lasers considered in this report. In some areas there is a strong set of competitive suppliers, such as, laser diodes for pumping solid state lasers, but further R&D at companies will still likely require government investments in technology development. Other key areas face a very uncertain supply with high-quality manufacturers rapidly disappearing from the U.S. or even Europe. This is illustrated by three critical examples:

1. **Diffraction gratings** with large-aperture, high-damage-threshold, and high-diffraction efficiency are key elements in CPA laser systems. Currently, only two companies worldwide and one U.S. national laboratory can manufacture meter-scale gratings: Plymouth Grating Laboratory (PGL), a small company in Massachusetts; and HORIBA Jobin-Yvon, a well-established manufacturer in France and a member of the HORIBA Group, a Japanese conglomerate. Lawrence Livermore National Laboratory (LLNL) invented multi-layer dielectric grating technology and produced many of the large-aperture gratings installed in large CPA lasers worldwide. Linking the capabilities of LLNL and PGL to advance grating technology with federal investments would significantly boost capabilities and competitiveness.

2. Companies able to supply **Nd-doped laser glass** are disappearing. During construction of the National Ignition Facility (NIF) and Laser Megajoule (LMJ), two companies, Schott (a German company with manufacturing in Pennsylvania) and Hoya (a Japanese company with manufacturing in California) established considerable capacity that has been largely dismantled after fulfilling these two facilities. This presents a major challenge to any near-term high-energy CPA laser project, such as the MEC laser upgrade (4.2.2.1) or future multi-PW laser (4.3). This problem extends beyond Nd:glass to a whole host of other laser gain materials (Nd- or Yb-doped, or even Ti:sapphire). Development of new laser materials as well as improving the optical quality and aperture size of novel materials is a key aspect of propelling new laser technologies.
3. Only a few domestic suppliers exist for **large nonlinear optical crystals** used in frequency conversion and optical parametric amplification, and the number is decreasing. R&D investment in the U.S. for producing larger crystals or new novel nonlinear crystals is almost nonexistent.

In all three cases, China has made enormous investments and Chinese suppliers may become the sole supplier for these critical laser components and materials, which puts supply to any future major laser projects in U.S. at risk. Any strategy to advance high-intensity laser science and applications must include a plan to support partnerships with industry and financial support for advanced R&D.

## APPENDIX A - List of Acronyms

| | |
|---|---|
| AFM | Atomic Force Microscope |
| AGN | Active Galactic Nuclei |
| AI | Artificial Intelligence |
| BNCT | Boron Neutron Capture Therapy |
| CDI | Coherent Diffractive Imaging |
| CERN | Conseil Européen pour la Recherche Nucléaire |
| CPA | Chirped Pulse Amplification |
| DC | Direct Current |
| ELI | European Light Infrastructure |
| EM | Electro-Magnetic |
| EPa | Eta ($10^{18}$) Pascal |
| EPW | Electron Plasma (Langmuir) Wave |
| EUV | Extreme Ultraviolet lithography |
| FEL | Free Electron Laser |
| GeV | Giga ($10^{9}$) electron Volt |
| HEDP | High Energy Density Plasma |
| HHG | High Harmonic Generation |
| IAW | Ion Acoustic Wave |
| IR | Infrared |
| LEP | Large Electron–Positron Collider |
| LHC | Large Hadron Collider |
| LPA | Laser Plasma Accelerator |
| LPI | Laser Plasma Interactions |
| LCFA | Local Constant Field Approximation |
| LCLS | Linac Coherent Light Source |
| LWFA | Laser-Wake Field Acceleration |
| MeV | Mega ($10^{6}$) electron Volt |
| MHD | Magneto Hydro Dynamic |
| ML | Machine Learning |
| NIF | National Ignition Facility |
| NRF | Nuclear Resonance Fluorescence |
| NSCAR | Nearby Skeleton Constrained Accelerated Re-computing |
| OFI | Optical Field Ionization |
| OUFTS | Optical Ultra Fast Thomson Scattering |
| PIC | Particle In Cell |
| PW | Peta ($10^{15}$) Watt |
| QED | Quantum Electro-Dynamics |
| RIT | Relativistically Induced-Transparency |
| SF-QED | Strong-Field Quantum Electro-Dynamics |
| STUD | Spike Trains of Uneven Duration and delay |
| SBS | Stimulated Brillouin Scattering |
| SNM | Special Nuclear Material |
| SRN | Science Research Needs |
| SRS | Stimulated Raman Scattering |
| THz | Tera ($10^{12}$) Hertz |
| TNSA | Target Normal Sheath Acceleration |
| TPa | Tera ($10^{12}$) Pascal |
| trARPES | time-resolved and Angle-Resolved Photo-Electron Spectroscopy |
| TW | Tera ($10^{12}$) Watt |
| UV | Ultra Violet |
| W | Watt |
| WDM | Warm Dense Matter |
| XANES | X-ray Absorption Near Edge Structure |
| XAFS | X-ray Absorption Fine Structure |
| XRTS | X-Ray Thomson Scattering |
| XPCI | X-Ray Phase Contrast Imaging |